\documentclass[20pt, letterpaper, hyper]{JHEP3}
\usepackage[utf8]{inputenc}
\usepackage{epic,eepic}
\usepackage{amsmath,epsfig}
\usepackage{amssymb,amsfonts}
\usepackage{dsfont}
\usepackage{graphicx}
\usepackage{multirow,cite}
\usepackage{enumerate}

\title{Spacetime emergence via holographic RG flow from incompressible Navier-Stokes at the horizon}

\preprint{CPHT-RR053.0612}

\author{$\text{Stanislav Kuperstein}^{a}$ and $\text{Ayan Mukhopadhyay}^{a,b}$\\
\textsf{a} \textit{Institut de Physique Th\'{e}orique, CEA Saclay, CNRS URA 2306 \\F-91191 Gif-sur-Yvette, France}\\
\textsf{b} \textit{Centre de Physique Th\'{e}orique, \'{E}cole Polytechnique, CNRS, 91128 Palaiseau, France}

\texttt{e-mails}: \textsf{ayan.mukhopadhyay@cpht.polytechnique.fr
, stanislav.kuperstein@cea.fr}\\
}

\abstract
{We show that holographic RG flow can be defined precisely such that it corresponds to emergence of spacetime. We consider the case of pure Einstein's gravity with a negative cosmological constant in the dual hydrodynamic regime. The holographic RG flow is a system of first order differential equations for radial evolution of the energy-momentum tensor and the variables which parametrize it's phenomenological form on hypersurfaces in a foliation. The RG flow can be constructed without explicit knowledge of the bulk metric provided the hypersurface foliation is of a special kind. The bulk metric can be reconstructed once the RG flow equations are solved. We show that the full spacetime can be determined from the RG flow by requiring that the horizon fluid is a fixed point in a certain scaling limit leading to the non-relativistic incompressible Navier-Stokes dynamics. This restricts the near-horizon forms of all transport coefficients, which are thus determined independently of their asymptotic values and the RG flow can be solved uniquely. We are therefore able to recover the known boundary values of almost all transport coefficients at the first and second orders in the derivative expansion. We conjecture that the complete characterisation of the general holographic RG flow, including the choice of counterterms, might be determined from the hydrodynamic regime.}

\date
\maketitle
\begin{document}
\section{Introduction}

One of the least understood features of the holographic correspondence  \cite{Maldacena:1997re, Gubser:1998bc, Witten:1998qj, Aharony:1999ti} is the construction of the field-theoretic renormalization group flow from the classical theory of gravity. Indeed the radial coordinate of the one higher dimensional spacetime, in which the theory of classical gravity lives, has been readily identified as the scale of defining the dual effective field theory, since early days of holography. This identification of the radial direction with the scale of the dual field theory has been used to define holographic renormalization \cite{Henningson:1998ey, Balasubramanian:1999re, deBoer:1999xf, deHaro:2000xn, Bianchi:2001kw}, which has been instrumental in making the holographic correspondence precise. Nevertheless, it is not clear, what is the precise nature of the coarse graining at a given scale in the field theory that the radial direction corresponds to. It is certainly possible that a lot of notions of coarse graining or integrating out degrees of freedom can be defined in the field theory, each of which has a holographic dual.

In recent literature, there has been various suggestions of replicating Wilsonian renormalization group in holography (see for instance \cite{Swingle, Heemskerk:2010hk, Faulkner:2010jy,Douglas:2010rc,Sen:2012fc,Lee:2012xba,
Oh:2012bx,Swingle:2012wq,Balasubramanian:2012hb,
Zayas:2013qda, Nakayama:2013fha,Sachs:2013pca}).\footnote{An instance of an earlier work is \cite{Akhmedov:1998vf}.} It is not entirely clear if the coarse graining of the dual field theory is being done in momentum space or in real space, or with unitary operations like minimising entanglement amongst directly interacting parts added to the repertoire before the coarse graining is done. It is also not clear how to define the rules for constructing counterterms generally, and also what is the precise role of multi-trace operators in the holographic renormalization group (RG) flow. 

In this paper we will take a different approach. We will adopt the point of view that before interpreting holographic renormalization group flow in field-theoretic terms, we will need to understand first how Einstein equations can be equivalent to the \emph{first} order evolution of physical data with the scale of averaging.  This physical data lives on the hypersurfaces of a spacetime foliation and is typically constructed out of the explicit metric of the bulk spacetime. The question is if we reverse this procedure. Firstly, can we define a radial flow of the physical data along a hypersurface foliation without knowing the bulk spacetime explicitly? Secondly, can we reconstruct the bulk spacetime metric which will be a solution of classical gravity from this holographic RG flow? This also necessitates that we should determine the properties of the solutions of classical gravity, like absence of naked singularities, from the holographic RG flow itself, as without knowing the explicit bulk spacetime metric, we have no other way of analysing regularity.

Our approach is built on the usual notion that the holographic RG flow should be equivalent to emergence of spacetime, in particular the emergence of the radial direction. The non-trivial part apparently is that the equations of gravity are \emph{second} order in radial derivatives and involve the explicit metric of spacetime. On the other hand, the RG flow equations are first order and involve couplings and physically measurable quantities like transport coefficients built out of the couplings. Our aim will be to establish that solutions of Einstein's equations of pure gravity with negative cosmological constant are equivalent to first order radial flow of only physically measurable data. The latter should be constructed without any explicit knowledge of the bulk spacetime metric. In other words, we will show that we can construct the holographic RG flow without requiring to solve Einstein's equations first. Furthermore, we will show that the solutions of Einstein's equations of a given class can be reconstructed from the holographic RG flow. 

We will also demonstrate that this holographic RG flow can be constructed in a unique way up to trivial scale reparametrizations. We will require that: 
\begin{itemize}
\item the holographic RG flow can be constructed without knowing the bulk spacetime metric explicitly, and
\item the physical data in the infrared is sensible.
\end{itemize}
Based on these requirements, we will give evidence that:
\begin{itemize}
\item  only a certain kind of hypersurface foliation works, and also
\item the counter-terms can be determined uniquely. 
\end{itemize}

It is known that in the holographic correspondence, there is a special limit of long wavelength solutions on the gravity side, which corresponds to hydrodynamics in the dual systems \cite{Policastro:2002se, Policastro:2002tn, Janik:2005zt, Janik:2006ft, Baier:2007ix, Bhattacharyya:2008jc, Natsuume:2007ty}. We will call this limit of the holographic correspondence as the \emph{fluid/gravity limit}. 
In this paper we will be mainly concerned with what we can learn about the general repertoire of holographic RG flow on the gravity side from this limit. This limit poses a major challenge for showing emergence of spacetime from holographic RG flow. In this limit, we expect the full spacetime to be determined by horizon dynamics, \emph{i.e.} the nature of the horizon fluid alone. This horizon fluid should be close to the membrane paradigm fluid \cite{Damour1, Damour2, Damour:1978cg, Blandford:1977ds, Price:1986yy, Thorne:1986iy, Damour:2008ji}, and the relativistic boundary fluid with corrections to Navier-Stokes equations to all orders in the derivative expansion should also be reconstructed from the horizon fluid via RG flow.

We will show that in the fluid/gravity limit the horizon emerges as a natural endpoint of the holographic RG flow - it is the hypersurface where the temperature, pressure and speed of sound blow up at late time, thus the notion of a thermodynamic equation of state does not exist beyond the thermal scale corresponding to the radial location of the horizon.\footnote{At the horizon, the speed of sound diverges giving rise to incompressibility. We will see later this makes the equation of state ill defined, because the pressure instead of becoming determined by the local temperature, becomes a non-local functional of the velocity fields.} It is important to remark that by the horizon we strictly mean the late-time horizon. We will elaborate on this issue later.

Under a certain rescaling of the radial coordinate (corresponding to the renormalization scale), time and the hydrodynamic variables, the horizon fluid turns out to be a fixed point of the RG flow at the leading order in the derivative expansion. This fixed point is precisely described by the non-relativistic incompressible Navier-Stokes equations, if the shear and bulk viscosities are finite at the horizon. Demanding that this fixed point is not altered by higher derivative corrections is what allows us to restrict near-horizon behaviour of the higher order transport coefficients. We find sufficient evidence by explicit calculations that these restrictions lead us to solve the RG flow uniquely, thus allowing us to determine the boundary values of the shear viscosity and higher order transport coefficients.

Our RG flow indeed reproduces the known values of first and second order transport coefficients at the boundary. The latter have been derived earlier in the literature by constructing solutions of Einstein's equations corresponding to long wavelength perturbations of black branes and then requiring existence of smooth late time (future) horizons in these solutions (the most complete method is as in \cite{Bhattacharyya:2008jc}). 

Our procedure typically simplifies the calculations involved in determining these boundary transport coefficients. However in some cases, our procedure is slightly more complex than the usual procedure, as we need to go to higher orders in derivative expansion to fix values of a few lower order transport coefficients, like in the case of the shear viscosity $\eta$. We will see this complexity is unavoidable in the RG flow approach. Nevertheless we believe that our results give us sufficient evidence that the RG flow determines the values of all transport coefficients at the boundary to all orders in derivative expansion, which can equivalently be determined from regularity of the late time (future) horizon of the explicit bulk spacetime metric. 

The fact that the horizon originally appears as an end point rather than a fixed point of the RG flow is quite reminiscent of the situation in multi-scale entanglement renormalization Ansatz (MERA) \cite{Vidal:2007, Vidal:2008}. In the latter case, the RG flow ends at the thermal scale or the scale of mass gap simply because the operations involved cannot be defined any more. In fact, similarities between MERA and holographic RG have also been pointed out in the literature from the Euclidean point of view \cite{Swingle, Swingle:2012wq}. In our case, we additionally learn the virtue of Lorentzian holographic RG flow is immense - we can determine all the information at the boundary from the dynamics at the infra-red scale in which the RG flow terminates. The end point of the RG flow is also a fixed point in disguise. Demanding that the dynamics at the fixed point is of a special kind allows us to fix the data in the ultraviolet.

We will find out that constructing holographic RG flow without a priori knowledge of the bulk spacetime metric implies the choice of Fefferman-Graham foliation. In the fluid/gravity limit, this foliation has a natural end-point where the Fefferman-Graham coordinates have a coordinate singularity, which coincides with the horizon at late time in the fluid/gravity limit.  It is precisely here where we expect the fluid to follow incompressible Navier-Stokes equations. The Fefferman-Graham foliation can be defined for any asymptotically anti de-Sitter spacetime. Therefore, even generally beyond the fluid/gravity limit, we should impose good behaviour of the physical data on the infrared screen. This should determine the RG flow uniquely. We will have more to say on this in the Discussion Section.

Our results and discussion here will indicate that the fluid/gravity limit itself will be good enough to define a unique holographic RG flow (up to trivial scale re-parametrisation) which will satisfy the requirements mentioned earlier. Though we need to generalize our methods beyond the hydrodynamic limit, we will argue that the choice of foliation and the counter-terms will be generally valid. We will also comment later that our considerations can be extended beyond pure gravity as well.

In this paper, we will not attempt to interpret the holographic RG flow in field-theoretic terms. Nevertheless, let us point out here that the holographic RG flow achieves something remarkable. The couplings of a field theory, and also the transport coefficients (as functions of thermodynamic variables and coupling constants) are defined independently of the background metric in which the field theory lives. The averaging procedure described by the holographic RG flow defines scale dependent transport coefficients in a background independent manner. In particular, there is no assumption about the boundary metric other than it is weakly curved. The behaviour of the fluid variables at any scale will depend on the choice of boundary metric. Still the radial flow of transport coefficients will be independent of this choice. 

Let us see a bit closely what happens if we coarse grain the velocity field $u^\mu$ over a scale. It is clear that the coarse grained velocity field $\langle u^\mu \rangle$ will not satisfy $\langle u^\mu \rangle\langle u^\nu \rangle g_{\mu\nu} = -1$, with $g_{\mu\nu}$ being the fixed background metric unless the background metric $g_{\mu\nu}$ is also defined in a coarse grained manner. Our holographic RG procedure also averages $g_{\mu\nu}$ in a simple and consistent manner - by identifying it with the physical hypersurface metric at any scale.

In a way, in the fluid/gravity limit, the holographic RG flow is close to the RG flow of partial differential equations with singular perturbations as defined by Barenblatt, Chen, Goldenfeld and Oono \cite{Chen:1994, Chen:1995ena, Barenblatt}.\footnote{For a similar perspective see \cite{Nakayama:2013fha}.} In their method, a specific construction of RG flow is used to derive generic long term behaviour of solutions and this is applicable to forced non-relativistic incompressible Navier-Stokes equation. The goal is to define a consistent averaging of a generic solution, such that the end point of this averaging is the so-called unique scaling solution which determines the long-term behaviour. This procedure has successfully reproduced the drag coefficient on a body moving in the fluid at low values of Reynold's number, though it has been difficult to implement it for large values of Reynold's number which is the realm of turbulence \cite{GoldenfeldReview}. 

The holographic RG flow seems to be the natural choice of RG for relativistic fluids as it can average both $u^\mu$ and the metric $g_{\mu\nu}$ in a consistent way maintaining the norm of $u^\mu$. Morally it is similar to Barenblatt, Chen, Goldenfeld and Oono approach as the endpoint of the holographic RG is the horizon fluid, and the horizon is indeed expected to control the dynamics at large time scales. In the future, we will like to explore whether our construction of holographic RG flow leads to a better understanding of turbulence. We will have more to say about this in the Discussion Section.

The explicit metric has a coordinate singularity at the horizon in the Fefferman-Graham coordinates, but this does not affect the RG flow procedure. This is because the transport coefficients on hypersurfaces of the foliation depend only on choice of the foliation up to trivial redefinitions of the radial coordinate as a function of itself, and are independent of the choice of the bulk coordinate system. The reason is that the hypersurface energy-momentum tensor is after all uniquely determined by the embedding of the hypersurface. The Landau-Lifshitz definition uniquely determine the fluid variables $u^\mu$ and $T$ at each hypersurface and hence the transport coefficients. Thus the hypersurface transport coefficients can be evaluated in any bulk coordinate system if the explicit spacetime metric is known. Since the foliation we choose naturally involves the Fefferman-Graham radial coordinate for reasons described before, we will stick to this coordinate system for our convenience.

The hydrodynamic derivative expansion parameter is the ratio of the typical scale of variation of fluid variables to the mean-free path. It flows radially with the RG flow because the hydrodynamic variables and also their covariant derivatives are defined at each hypersurface individually.  The derivative expansion, when defined in this scale dependent manner, is expected to become better as we flow to the horizon. The fluid approximation should work better in the infra-red. Unfortunately we cannot make this more precise because we cannot determine the scale dependent mean free path from hydrodynamic considerations alone. At a given scale, this should be a function of the effective temperature and the scale as well. It will be interesting to determine the scale dependent mean-free path precisely in the future. Regardless, such considerations do not affect our results here.
\newpage

\emph{Comparison with other approaches}
\newline

The Wilsonian approach to fluid/gravity correspondence was initiated by Bredberg, Keeler, Lysov and Strominger \cite{Bredberg:2010ky} (see also \cite{Bredberg:2011jq, Lysov:2011xx, Compere:2011dx, Cai:2011xv, Kuperstein:2011fn, Brattan:2011my, Eling:2011ct}). They adopted the procedure of cutting off the geometries along hypersurface foliation of Eddington-Finkelstein coordinates and solved for the metric in the region bounding the horizon and the cut-off hypersurface. The interpretation was that the ultraviolet part of the geometry is removed. Then Dirichlet boundary conditions were imposed on the cut-off hypersurface such that the induced metric there remained flat to all orders in the derivative expansion. The bulk metric was found in the long-wavelength approximation. This generated a hydrodynamic energy-momentum tensor on the cut-off hypersurface with cut-off dependent equation of state and transport coefficients. Clearly this gave a method to investigate fluid/gravity correspondence with many possible asymptotic conditions in a cut-off dependent way.

In our previous work \cite{ Kuperstein:2011fn}, we showed that one can obtain the velocity and temperature fields on the cut-off hypersurface by cut-off dependent field redefinitions of the boundary velocity and temperature fields. We assumed asymptotically AdS boundary conditions. The cut-off dependent bulk solutions are related by field redefinitions of the fluid velocity and temperature fields. These field redefinitions are uniquely fixed by the Landau-Lifshitz definitions of the velocity and temperature fields applied to the hypersurface energy-momentum tensor which is covariant with respect to the induced metric. This directly leads to the cut-off dependent transport coefficients. We implemented this in the Fefferman-Graham coordinates and a similar approach was later developed in Eddington-Finkelstein coordinates \cite{Brattan:2011my}.

We also noticed that having a flat induced metric at the cut-off was not necessary for fixing the cut-off dependent transport coefficients. The regularity at the horizon determined the RG flow uniquely. This result actually forms the basis of our present work. If transport coefficients at the cut-off depend only on the regularity of the horizon but not on the precise nature of the cut-off metric, then somehow we should be able to obtain the flow of transport coefficients from first order equations. 

In a way, our work here makes the Wilsonian RG program for holographic fluids precise. We construct a first order system of equations for evolution of all physical variables in asymptotically AdS spaces. 

It should be also possible to generalize to other asymptotic spacetimes provided we find the right choice of hypersurface foliation which is crucial to our construction. For asymptotically Ricci-flat spaces one can also try to use methods of \cite{Caldarelli:2012hy} which map asymptotically AdS solutions to Ricci-flat solutions. If the equations can also be mapped, then we can map the RG flow as well. We leave this investigation for the future.

The other related approach is due to Iqbal and Liu \cite{Iqbal:2008by} which defines an RG flow of response functions. One can show by this approach that the ratio of shear viscosity $\eta$ to the entropy density $s$, \emph{i.e.} $\eta/s$ does not flow and is determined by the regularity of the horizon only. Crucially the RG flow equation for the response function is non-linear but first order in $r$-derivative. The drawback of this approach is that we can only define the RG flow in the equilibrium black brane geometry - thus we can only probe limited transport coefficients. Knowing the RG flow of the response functions of non-equilibrium geometries, or of multi-point correlators in equilibrium geometries, we can define the RG flow of all transport coefficients. In our approach we reproduce the basic result that $\eta/s$ does not run radially. In the future, we would like to explore the RG flow of response functions in non-equilibrium geometries.
 
Finally we would like to mention that the black-fold approach of generating approximate solutions of Einstein's equation \cite{Emparan:2009cs, Emparan:2009at,Armas:2011uf} has a similar spirit to our RG flow approach. Just like we construct the spacetime from the horizon fluid, the black-fold approach generates the (approximate) solution from long wavelength perturbations of membrane like sources. These long wavelength perturbations have fluid interpretation in the extended direction but also elastic shear interpretation in the compact directions. In the future we would like to investigate if we would be able to reconstruct the spacetime from horizon dynamics via RG flow in the hydro-elastic approximation by adding compact directions to the hypersurfaces. Once we understand how to choose hypersurface foliation for various asymptotic conditions, it will be possible to make the black-fold approach more rigorous via construction of the RG flow.\footnote{For a recent work connecting the holographic RG flow with the blackfold approach see \cite{Emparan:2013ila}.}
\newline

\emph{Organization of the paper}
\newline\newline
The organization of the paper is as follows. In Section 2, we review fluid mechanics briefly and introduce the systematics of the holographic RG flow Ansatz. In Section 3, we formulate our basic Ansatz for the RG flow of the holographic fluid. In Section 4, we show why we require to choose Fefferman-Graham foliation and also determine the method for renormalization of energy-momentum tensor from general principles. In Section 5, we give the general algorithm for construction of holographic RG flow order by order in the derivative expansion. In Section 6, we discuss how should transport coefficients behave near the horizon for the horizon fluid to be  governed by non-relativistic incompressible Navier-Stokes equations. In Section 7, we present the results up to second order in the derivative expansion explicitly, demonstrating how the RG flow recovers known values of  boundary transport coefficients. Finally, in Section 8 we discuss the broader lessons we can learn from our results and possible future developments. The appendices contain details of our calculations corroborating with the main line of development of the paper.

\section{A brief review of fluid mechanics}

In this section we will first briefly review fluid mechanics in an arbitrary weakly curved background metric. Our presentation will mostly follow \cite{Romatschke:2009kr}, though we will emphasize some points which may not be completely familiar.

The basic data of uncharged fluid mechanics are:
\begin{equation}
g_{\mu\nu} \, , \quad u^\mu \, \, \quad  \textrm{and} \quad T \, ,
\end{equation}
namely the background space-time metric $g_{\mu\nu}$, the velocity field $u^\mu$ which is of norm $-1$ with respect to the background metric (\emph{i.e.} $u^\mu u^\nu g_{\mu\nu} = -1$) and the temperature $T$. The velocity and temperature fields are directly measurable. 

We also require the background metric $g_{\mu\nu}$ to be weakly curved, \emph{i.e.} the typical curvature radius to be larger than the mean free path. In such a background, the hydrodynamic approximation is valid near equilibrium. The fluid energy-momentum tensor takes a covariant form with respect to the background metric. This form is a functional of $g_{\mu\nu}, u^\mu, T$, covariant derivatives of $u^\mu$ and  $T$, and the Riemann curvature $R^\mu_{\phantom{\mu}\nu\rho\sigma}$ constructed out of $g_{\mu\nu}$ and it's covariant derivatives. This functional form can be expanded phenomenologically in the derivative expansion, where the expansion parameter is the typical length scale of variation of the curvature, $u^\mu$ and $T$ with respect to the mean free path.

The equations of fluid mechanics are simply given by the conservation of this hydrodynamic energy-momentum tensor, \emph{i.e.} $\nabla^\mu t_{\mu\nu} =0$. These determine the $d$ independent variables in $T$ and $u^\mu$.

Thermodynamics is an important input in construction of fluid mechanics. The equation of state is used locally to define pressure $P(T)$ and the energy density $\epsilon(T)$ as functions of temperature. Conversely, we can also use thermodynamics to define the temperature $T$ and the entropy density $s$ from $P$ and $\epsilon$ locally using the following thermodynamic identities: 
\begin{equation}
\label{Thermodynamics}
\frac{\textrm{d} \epsilon}{\textrm{d} s} = T, \quad \epsilon+P = T s \, .
\end{equation}
The speed of sound is given by:
\begin{equation}
\label{TheSpeedOfSound}
c_s^2 = \frac{\textrm{d} P}{\textrm{d} \epsilon} = \frac{\textrm{d} \ln T}{\textrm{d} \ln s} \, ,
\end{equation}
where in the second equality we used the thermodynamic identities \eqref{Thermodynamics}.

The equilibrium energy-momentum tensor $t_{\mu\nu}^{\text{eq}}$ takes the form:
\begin{equation}
\label{equlibrium}
t_{\mu\nu}^{\text{eq}} = \epsilon u_\mu u_\nu + P \Delta_{\mu\nu}, \quad \text{with}\quad \Delta_{\mu\nu} = u_\mu u_\nu + g_{\mu\nu}
\end{equation}
being the projection tensor on the spatial plane orthogonal to $u^\mu$. Here $u^\mu$ plays the r\^{o}le of an arbitrary relativistic boost. Obviously this is the most general form of the stress tensor that does not involve space-ime derivatives of $u^\mu$ and $T$.
The conservation of the equilibrium energy-momentum tensor $\nabla^\mu t_{\mu\nu}^{\text{eq}} = 0$ gives the covariant Euler equations which can be written as:
\begin{equation}
\label{Euler-T}
D \ln T = -c_s^2 \nabla\cdot u, \quad D u^\mu = - \nabla_{\perp}^{\mu} \ln T 
\end{equation}
or equivalently,
\begin{equation}
\label{Euler-s}
D \ln s = - \nabla\cdot u, \quad D u^\mu = - c_s^2\nabla_{\perp}^{\mu} \ln s \, ,
\end{equation}
where
\begin{equation}
D \equiv u\cdot \nabla, \quad \nabla_{\perp\mu} = \Delta_\mu^{\phantom{\mu}\nu}\nabla_\nu
\end{equation}
are the covariant derivatives along the vector $u^\mu$ and orthogonal to it.

In the full hydrodynamic energy-momentum tensor, $T$ and $u^\mu$ are space-time dependent and the equilibrium form is corrected by non-equilibrium contribution $t_{\mu\nu}^{\text{non-eq}}$. The full energy-momentum tensor is then  $t_{\mu\nu}^{\text{eq}}+t_{\mu\nu}^{\text{non-eq}}$. At a given $n$ order in the derivative expansion each term in $t_{\mu\nu}^{\text{non-eq}}$ will have exactly $n$ space-time derivatives acting on $u^\mu$, $T$ and the metric.

In any phenomenological description it is useful to eliminate quantities which are equivalent on shell. In hydrodynamics one may use Euler equations of motion \eqref{Euler-T} in order to reduce the number of possible  scalars, vectors and tensors terms in $t_{\mu\nu}^{\text{non-eq}}$ at the same order in the derivative expansion. The non-equilibrium corrections to the equilibrium Euler equations are relevant to relate quantities at different orders.
 
We should, however, emphasize that neither fluid mechanics, nor the fluid/gravity correspondence, nor our holographic RG flow construction depends on the removal of the phenomenological redundancy. In fact the full construction of the energy-momentum tensor, the explicit space-time metric and the holographic RG flow can all be constructed completely off-shell. The removal of redundancy using Euler equations order by order in derivative expansion merely reduces our labour.

We will denote independent scalars (\emph{i.e.} those unrelated by Euler equations) as $\mathcal{S}_i^{(n)}$, where superscript $(n)$ will denote the order in the derivative expansion and the subscript $i$ will be a counting index. Note we can readily construct a vector out of a scalar at the same order in the form of $\mathcal{S}_i^{(n)}u^\mu$. We will therefore consider only transverse vectors $\mathcal{V}_{i\mu}^{(n)}$, which are orthogonal to $u^\mu$, namely those satisfying $\mathcal{V}_{i\mu}^{(n)}u^\mu = 0$, with the same meaning for the superscript $(n)$. Furthermore, we can built a symmetric tensor of the form $\mathcal{S}_i^{(n)}u_\mu u_\nu$, $\mathcal{S}_i^{(n)}\Delta_{\mu\nu}$ or $u_\mu \mathcal{V}_{i\nu}^{(n)} + u_\nu \mathcal{V}_{i\mu}^{(n)}$ out of the same order scalars and vectors. Therefore we are left only with traceless and transverse symmetric tensors $\mathcal{T}^{(n)}_{i\mu\nu}$ that satisfy $\mathcal{T}_{\mu\nu}g^{\mu\nu}= 0$ and $u^\mu\mathcal{T}_{\mu\nu} =0$.

Before presenting independent scalars, vectors and tensors at the first and the second orders we would like to make the following remarks:
\begin{itemize}
\item It is useful to eliminate local time-derivatives (denoted by $D$) in favour of local spatial derivatives (given by $\nabla_{\perp\mu}$). It can always be achieved using Euler equations.
\item Despite the fact that we treat the temperature, and not the entropy density, as the ``fundamental" thermodynamic variable, we will prefer $\nabla_\mu \ln s$ over $\nabla_\mu \ln T $ as a building block for the scalars, vectors and tensors. This choice will make many expressions in the paper significantly shorter. One can always go the $\nabla_\mu \ln T$ basis using the thermodynamical relation $\nabla_\mu \ln s = c_s^2 \nabla_\mu \ln  T$.
\item Throughout this paper we will consider only parity even contributions to the energy momentum tensor. One example of such an odd parity contribution in $d=3$ is $(\varepsilon^{\alpha \beta \gamma} u_\alpha \nabla_\beta u_\gamma )\Delta_{\mu\nu} $. 
\end{itemize}

We will denote the number of independent scalars, transverse vectors, and symmetric, traceless and transverse tensors at $n$-th order in derivative expansion as $m_s^{(n)}$, $m_v^{(n)}$ and $m_t^{(n)}$ respectively.
At the first order in the derivative expansion we find that $\left( m_s^{(1)}, m_v^{(1)}, m_t^{(1)} \right) = (1,1,1)$. In other words, there is only one independent scalar, one independent vector and one independent tensor. They are:\footnote{Notice that the definition of $\sigma_{\mu\nu}$ varies in the literature by a factor of two.}
\begin{equation}
\label{ListOfFirstOrderSVT}
\nabla\cdot u, \quad \nabla_{\perp\mu}\ln s, \quad \text{and}\quad \sigma_{\mu\nu} = \langle  \nabla_{\perp\mu} u_\nu \rangle
\end{equation}
respectively.
 We will use $\langle  A_{\mu\nu} \rangle$ to denote the symmetric, traceless and transverse (orthogonal to $u^\mu$) part of $A_{\mu\nu}$ which can be obtained by using:
\begin{equation}
\label{brackets}
\langle  A_{\mu\nu} \rangle = \frac{1}{2}\Delta_\mu^{\phantom{\mu}\rho}
\Delta_\nu^{\phantom{\mu}\sigma}\Big(A_{\rho\sigma}+
A_{\sigma\rho}\Big) - \frac{1}{d-1}\Delta_{\mu\nu}\Delta^{\rho\sigma}A_{\rho\sigma}.
\end{equation}
We note that we can also construct a transverse anti-symmetric tensor at the first order in derivative expansion, which is the vorticity tensor $\omega_{\mu\nu}$ which is given by
\begin{equation}
\omega_{\mu\nu} = \frac{1}{2}\Big(\nabla_{\perp\mu}u_\nu - \nabla_{\perp\nu}u_\mu\Big).
\end{equation}

At second order in derivative expansion, there are seven independent scalars $\mathcal{S}^{(2)}_i$:
\begin{eqnarray}
\label{ListOfSecondOrderScalars}
&& \qquad\qquad
\mathcal{S}_1^{(2)} = R \, , \quad
\mathcal{S}_2^{(2)} = u_\mu R^{\mu}_{\,\, \nu} u^\nu \, , \quad
\mathcal{S}_3^{(2)} = \left( \nabla \cdot u\right)^2 \, ,
\\ 
&& 
\mathcal{S}_4^{(2)} = {\nabla_\bot}^\mu {\nabla_\bot}_\mu \ln s \, , \quad
\mathcal{S}_5^{(2)} = {\nabla_\bot}^\mu \ln s {\nabla_\bot}_\mu \ln s \, , \quad 
\mathcal{S}_6^{(2)} = \sigma^\mu_{\,\, \nu} \sigma^{\nu}_{\,\,  \mu} \, , \quad
\mathcal{S}_7^{(2)} = \omega^\mu_{\,\, \nu} \omega^{\nu}_{\,\, \mu} \, .
\nonumber 
\end{eqnarray}
Notice that our definition of $\mathcal{S}_7 = \omega^2$ is different from conventions adopted in some papers, where $\omega^2$ is defined as $\omega^{\mu \nu} \omega_{\mu \nu}$. Since $\omega_{\mu \nu}$ is antisymmetric, one picks up a minus sign compared to our definition.

As an illustration of the use of Euler equations to eliminate redundant quantities, we may consider the scalar $D(\nabla\cdot u)$. In Appendix \ref{VariousUsefulIdentities} we list various off-shell identities involving covariant derivatives of $u^\mu$. Starting from \eqref{D-Nabla-u} and using equations of motion to eliminate $D u^\mu$ we get:
\begin{equation}
\label{D-Nabla-u}
D(\nabla\cdot u) = - \mathcal{S}_2^{(2)} - \frac{1}{d-1}\mathcal{S}_3^{(2)} - c_s^2 \mathcal{S}_4^{(2)} + \left( c_s^4 - \dfrac{\partial c_s^2}{\partial \ln s}\right) \mathcal{S}_5^{(2)} 
	- \mathcal{S}_6^{(2)} - \mathcal{S}_7^{(2)}\, .
\end{equation}

There are six independent transverse vectors $\mathcal{V}^{(2)\mu}_i$ at the second order in derivative expansion, namely:
\begin{equation}
\label{ListOfSecondOrderVectors}
\nabla_{\perp\alpha}\sigma^{\alpha\mu} - u^\mu \sigma^2, \quad \nabla_{\perp\alpha}\omega^{\alpha\mu} - u^\mu \omega^2, \quad \sigma^{\mu\nu}\nabla_{\perp\nu}\ln s, \quad \omega^{\mu\nu}\nabla_{\perp\nu}\ln s, \quad
(\nabla\cdot u)\nabla_\perp^\mu \ln s, \quad \Delta^{\mu\alpha}u^\beta R_{\alpha\beta} \, .
\end{equation}

Finally there are eight independent symmetric, traceless and transverse tensors $\mathcal{T}^{(n)}_{i\mu\nu}$ at second order, namely
\begin{eqnarray}
\label{ListOfSecondOrderTensors}
 && 
 {\mathcal{T}_1}^\mu_{\,\,\,\nu} = \left< R^\mu_{\phantom{\mu}\nu} \right> , \quad 
 {\mathcal{T}_2}^\mu_{\,\,\,\nu} = \left< u^\alpha R^{\phantom{\alpha} \mu \phantom{\nu} \beta}_{\alpha \phantom{\mu} \nu \phantom{\beta}} u_\beta  \right> , \quad 
 {\mathcal{T}_3}^\mu_{\,\,\,\nu} = (\nabla\cdot u)\sigma^\mu_{\,\,\,\nu},  \quad
 {\mathcal{T}_4}^\mu_{\,\,\,\nu} =  \left<\nabla_\bot^\mu {\nabla_\bot}_\nu\ln s \right>, \quad
 \\
 && 
 {\mathcal{T}_5}^\mu_{\,\,\,\nu} =  \left<{\nabla_\bot}^\mu\ln s \nabla_{\perp\nu}\ln s \right> ,  \quad
 {\mathcal{T}_6}^\mu_{\,\,\,\nu} =  \left< \sigma^{\mu}_{\,\,\, \tau} \sigma^{\tau}_{\,\,\,\nu}  \right> , \quad
 {\mathcal{T}_7}^\mu_{\,\,\,\nu} = \left< \omega^{\mu}_{\,\,\, \tau} \omega^{\tau}_{\,\,\,\nu}  \right>, \quad
 {\mathcal{T}_8}^\mu_{\,\,\,\nu} = \left<\sigma^{\mu}_{\,\,\, \tau} \omega^{\tau}_{\,\,\,\nu}  \right> \, .
 \nonumber
\end{eqnarray}
Similarly to the scalars, it is necessary to use Euler equations and the tensor identities from Appendix \ref{VariousUsefulIdentities} to express terms like, for instance, $\left< D \sigma^\mu_{\,\,\, \nu } \right>$ in terms of ${\mathcal{T}_i}^\mu_{\,\,\,\nu}$'s.

To summarize, we see that at the second order we have $\left( m_s^{(2)}, m_v^{(2)}, m_t^{(2)} \right) = (7,6,8)$.

Our goal now is to express the non-equilibrium correction to the energy-momentum tensor, $t_{\mu\nu}^{\text{non-eq}}$, in terms of scalars $\mathcal{S}_i^{(n)}$, vectors $\mathcal{V}_i^{(n)}$ and tensors $\mathcal{T}_i^{(n)}$. Before doing so we have to adopt a new definition of $u^\mu$ since away from the equilibrium, it is not anymore a relativistic boost. It is customary to use the Landau-Lifshitz field definitions of $u^\mu$ and $T$. In these field definitions, $u^\mu$ is the local velocity of energy transport, and the local energy density $\epsilon(T)$ is then given by:
\begin{equation}
\label{LandauGauge}
 t^\mu_{\phantom{\mu}\nu}u^\nu = - \epsilon(T) u^\mu \, .
\end{equation}
The temperature $T$ is then defined locally via the equation of state. Since \eqref{LandauGauge} trivially holds for the equilibrium tensor \eqref{equlibrium}, the non-equilibrium part of the hydrodynamic tensor satisfies $ t_{\mu\nu}^{\text{non-eq}} u^\mu= 0$. This means that the $\mathcal{V}_i^{(n)}$'s don't appear in the full hydrodynamic energy-momentum tensor, which can be written now in the following form:
\begin{equation}
\label{HydrodynamicEMTensor}
t_{\mu\nu} = \epsilon u_\mu u_\nu + P\Delta_{\mu\nu}+ \sum_{n=1}^\infty \left( \sum_{i=1}^{m_t^{(n)}} \gamma^{(n)}_{i} \mathcal{T}^{(n)}_{i\mu\nu} + \left( \sum_{i=1}^{m_s^{(n)}} \delta^{(n)}_{i} \mathcal{S}^{(n)}_i \right) \cdot \Delta_{\mu\nu}  \right) \, .
\end{equation}
Above the phenomenological coefficients $\delta_{(n)i}$ are the scalar transport coefficients. At the first order there is only one such coefficient ($m_s^{(1)}=1$), the bulk viscosity $\zeta$ which is given by $\delta^{(1)}_{1} \equiv -\zeta $. The phenomenological coefficients $\gamma^{(n)i}$ are the tensor transport coefficients. At the first order, there is one such coefficient ($m_t^{(1)}=1$), namely the shear viscosity $\eta$ given by $\gamma^{(1)}_{1} \equiv -2\eta$. All scalar and tensor transport coefficients are functions of the entropy density $s$ or alternatively the temperature $T$, which like the equation of state can be obtained from the underlying quantum field theory.

In this paper we will not go beyond the second order in the derivative expansion. We will, therefore, omit  in the rest of the paper the ${}^{(2)}$ superscript used in \eqref{ListOfSecondOrderScalars}, \eqref{ListOfSecondOrderVectors} and \eqref{ListOfSecondOrderTensors}.

In a conformal fluid, most of the scalar transport coefficients vanish, except those required for satisfying the conformal anomaly.  Furthermore, the energy-momentum tensor except for the anomalous part should be Weyl covariant. At each order in derivative expansion, only a certain combination of symmetric, traceless and transverse tensors are Weyl covariant. At the first order $\sigma_{\mu\nu}$ is Weyl covariant, therefore the shear viscosity $\eta$ will be non-vanishing in a conformal field theory. At second order, only five linear combinations of the eight independent symmetric, traceless and transverse tensors are Weyl covariant. These are:
\begin{eqnarray}
\label{ListOfConformalTensors}
{\mathcal{T}_2}^\mu_{\,\,\,\nu} -\frac{1}{d-2} {\mathcal{T}_1}^\mu_{\,\,\,\nu} \, , \qquad
{\mathcal{T}_2}^\mu_{\,\,\,\nu} -\frac{{\mathcal{T}_3}^\mu_{\,\,\,\nu}}{d-1}  - \dfrac{{\mathcal{T}_4}^\mu_{\,\,\,\nu}}{d-1} + \dfrac{{\mathcal{T}_5}^\mu_{\,\,\,\nu}}{(d-1)^2} \, , \qquad
{\mathcal{T}_6}^\mu_{\,\,\,\nu}  \, , \qquad
{\mathcal{T}_7}^\mu_{\,\,\,\nu}  \, , \quad
\textrm{and} \quad
{\mathcal{T}_8}^\mu_{\,\,\,\nu}  \, .
\end{eqnarray}
Thus, a conformal fluid has only five second order transport coefficients, one corresponding to each of these Weyl covariant tensors.

Also in a conformal fluid, the temperature is the only scale for hydrodynamics. Therefore all the non-vanishing tensor coefficients at $n$-th order will be of the form $k_{(n)i}T^{d-n}$ or $k_{(n)i}s^{\frac{d-n}{d-1}}$, where $k_{(n)i}$ are numerical constants.

The conformal fluid relevant for our discussion is the one dual to Einstein gravity with a negative cosmological constant. It has the following stress tensor \cite{Baier:2007ix}, \cite{Bhattacharyya:2008jc}:
\begin{eqnarray}
\label{BoundaryFluid}
t^\mu_{\,\,\, \nu} &=& \epsilon_\textrm{b} u^\mu u_\nu + P_\textrm{b} \Delta ^\mu_{\,\,\, \nu} -
		2 \eta_\textrm{b} \sigma^\mu_{\,\,\, \nu} - 2 \eta_\textrm{b} b \cdot 
			\left[{\mathcal{T}_2}^\mu_{\,\,\, \nu} - \dfrac{1}{d-2} {\mathcal{T}_1}^\mu_{\,\,\, \nu} \right]  +
\\		
 &+&    2 \eta_\textrm{b} \left(b-\tau_\omega \right) \left[ 
		{\mathcal{T}_2}^\mu_{\,\,\,\nu} -\frac{{\mathcal{T}_3}^\mu_{\,\,\,\nu}}{d-1}  - \dfrac{{\mathcal{T}_4}^\mu_{\,\,\,\nu}}{d-1} + \dfrac{{\mathcal{T}_5}^\mu_{\,\,\,\nu}}{(d-1)^2}
		\right] +
		2 \eta_\textrm{b} \tau_\omega {\mathcal{T}_6}^\mu_{\,\,\, \nu} + 
		2 \eta_\textrm{b} \left(\tau_\omega - b\right) \eta_\textrm{b} \tau_\omega  {\mathcal{T}_7}^\mu_{\,\,\, \nu} + 
		4 \eta_\textrm{b} \tau_\omega  {\mathcal{T}_8}^\mu_{\,\,\, \nu} \, .
\nonumber		
\end{eqnarray}
Here we intentionally added the ${}_{b}$ subscript in order to distinguish the boundary conformal fluid and a cut-off fluid we will introduce further in the paper. The constants in (\ref{BoundaryFluid}) are \cite{Bhattacharyya:2008mz}:
\begin{eqnarray}
\label{BoundaryConstants}
&&
\epsilon_\textrm{b} = (d-1) P_\textrm{b} \, , \quad 
P_\textrm{b} = \kappa_\textrm{AdS} \left( \dfrac{d}{4 \pi T_\textrm{b}} \right)^{-d} \, , \quad
\eta_\textrm{b} = \kappa_\textrm{AdS} \left( \dfrac{d}{4 \pi T_\textrm{b}} \right)^{1-d} \, , 
\nonumber
\\ 
&& \quad
\tau_\omega = 
    b \int_1^\infty \dfrac{y^{d-2} - 1}{y(y^d-1)} \textrm{d} y \, , \qquad
b = \dfrac{d}{4 \pi T_\textrm{b}} \, , \qquad
\kappa_\textrm{AdS} = \frac{16\pi G_N}{l_\textrm{AdS}^{d-1}} \, ,
\end{eqnarray}
where $l_\textrm{AdS}$ is the AdS radius and $T_\textrm{b}$ is the Hawking temperature.

\section{A general formulation of the Ansatz}

We will now formulate our holographic RG flow Ansatz in the fluid/gravity limit. This Ansatz will make an assumption about how $u^\mu$ and $T$ flow radially along the hypersurface foliation, along with our basic assumption that the energy-momentum tensor $t_{\mu\nu}$ is purely hydrodynamic on each hypersurface of the foliation. We will impose no restrictions on the induced metric on each hypersurface other than that it is weakly curved.

As we will explain in the next section, gravity equations of motion provide a first order differential equation for the radial evolution of the hydrodynamical stress tensor $t_{\mu\nu}$. At the moment we will need neither the general form of these equations nor even the definition of the radial foliation. For the aim of this section it is sufficient to realize that $u^\mu$ and $T$ both necessarily have non-trivial radial evolution, for otherwise we cannot preserve the Landau-Lifshitz gauge of $t_{\mu\nu}$ and the norm of $u^\mu$ with respect to the induced metric ($u^\mu g_{\mu\nu} u^\nu=-1$). In other words, $u^\mu$ and $T$ have to be properly redefined at each radial slice. Moreover, for given boundary values of $u^\mu$ and $T$ we should be able to reproduce their values at any cut-off surface. 

The main observation that drastically facilitates our calculations is that instead of focusing on the explicit solution for $u^\mu=u^\mu(r)$ and $T=T(r)$ in terms of their boundary values, it suffices to determine the first order differential equations for their radial evolution.  Furthermore, these two equations should express ${u^\mu}^\prime(r)$ and $T^\prime(r)$ in terms of $u^\mu(r)$ and $T(r)$ (and their space-time derivatives) on the same hypersurface. 

Our Ansatz, therefore, will be similar in spirit to the form of the hydrodynamical stress tensor in \eqref{HydrodynamicEMTensor}: 
\begin{eqnarray}
\label{Ansatz-uT}
{u^\mu}^\prime &=& \alpha_0 u^\mu + \sum_{n=1}^{\infty} \left(\sum_{i=1}^{m^{(n)}_s} \alpha^{(n)}_i \mathcal{S}^{(n)}_i u^\mu+ \sum_{i=1}^{m^{(n)}_v} \beta^{(n)}_i {\mathcal{V}^{(n)}}^\mu_i \right) \, ,
\nonumber\\
\dfrac{T^{\prime}}{T} &=&  \lambda_0 + \sum_{n=1}^{\infty} \sum_{i=1}^{m^{(n)}_s} \lambda^{(n)}_i \mathcal{S}^{(n)}_i \, .
\end{eqnarray}
Clearly this is the most general way to express ${u^\mu}^\prime$ and $T^{\prime}$ in terms of the induced metric, $u^\mu$ and $T$ and their derivatives (the overall $T$ factor on the right hand side of the second equation is introduced to make $\lambda^{(n)}_i$'s dimensionless). In general it is not immediately clear that the differential equations for $u^\mu(r)$ and $T(r)$ are of the first order and so \emph{a priori} our minimalistic Ansatz may not be sufficient. We will see that this is not the case.

The coefficients in \eqref{Ansatz-uT} are not the only unknown parameters in the problem. We also have transport coefficients $\gamma_i$ and $\delta_i$ in \eqref{HydrodynamicEMTensor}, which is just the general form of the Landau-Lifshitz hydrodynamic energy-momentum tensor presented in the previous section. Overall we see that at a given order there are $3m^{(n)}_s + m^{(n)}_v+ m^{(n)}_t$ parameters, namely:
\begin{itemize}
\item $m^{(n)}_s$ of  $\alpha^{(n)}_i$'s and $\lambda^{(n)}_i$'s each,  needed for defining ${u^\mu}^\prime$ and $T^\prime$ respectively;
\item $m^{(n)}_v$ of $\beta^{(n)}_i$'s, also needed for defining ${u^\mu}^\prime$;
\item $m^{(n)}_s$ of scalar transport coefficients $\delta^{(n)}_i$;
\item $m^{(n)}_t$ of tensor transport coefficients $\gamma^{(n)}_i$.
\end{itemize}
Of these $\alpha^{(n)}_i$, $\beta^{(n)}_i$ and $\lambda^{(n)}_i$ are auxiliary variables which are required for scale dependent field definitions. These are analogues of redundant couplings in field theory. Therefore they should be determined by algebraic equations.

\setlength{\unitlength}{0.9cm}
\FIGURE[t]{
\begin{picture}(15,8)
\put(1.0,0){\vector(0,1){8}}
\put(1.5,7.5){\Large{$r$}}
\put(0.8,0.1){\line(1,0){0.4}}
\put(0.1,0.0){\Large{$0$}}
\put(0.8,6.5){\line(1,0){0.4}}
\put(0.0,6.4){\Large{$r_\textrm{H}$}}
\put(6.5,1.0){\centering{\Large{Boundary}}}
\put(6.5,7.0){\centering{\Large{Horizon}}}
\put(5.0,4.0){\centering{\Large{$u^\mu(r)$, \quad $T(r)$, \quad $g_{\mu\nu}(r)$}}}
\linethickness{0.05mm}
\qbezier(9.5,6.4)(11.5,6.0)(13.5,6.4)
\qbezier(5.5,6.4)(7.5,6.8)(9.5,6.4)
\qbezier(1.5,6.4)(3.5,6.0)(5.5,6.4)
\linethickness{0.05mm}
\qbezier(9.5,3.3)(11.5,3.1)(13.5,3.3)
\qbezier(5.5,3.3)(7.5,3.4)(9.5,3.3)
\qbezier(1.5,3.3)(3.5,3.1)(5.5,3.3)
\linethickness{0.5mm}
\qbezier(9.5,0.3)(11.5,0.1)(13.5,0.3)
\qbezier(5.5,0.3)(7.5,0.4)(9.5,0.3)
\qbezier(1.5,0.3)(3.5,0.1)(5.5,0.3)
\end{picture}
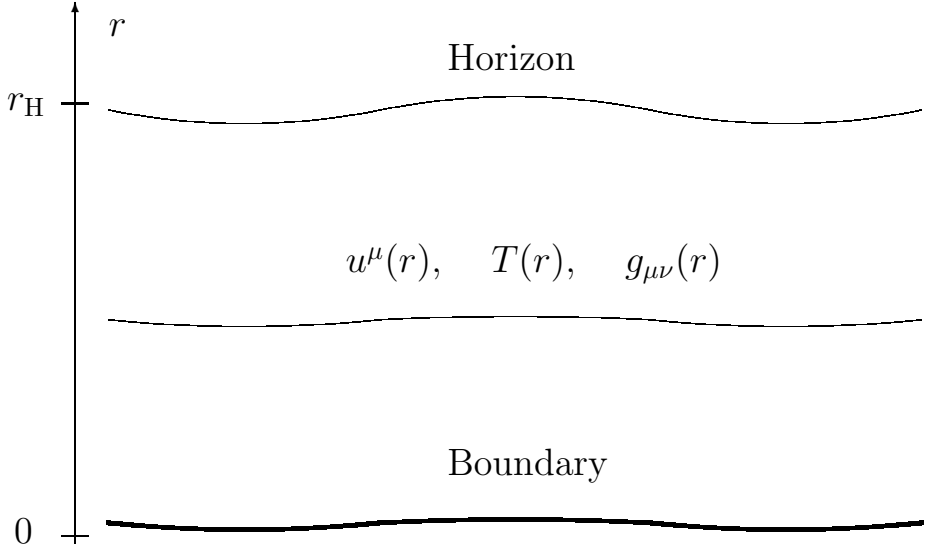
\caption{The induced metric and the hydrodynamic tensor on a fixed hypersurface change as we move from the conformal boundary to the horizon. Similarly the vector $u^\mu$ and the temperature $T$ have to be properly redefined on each cut-off surface. Their radial evolutions are governed by \eqref{Ansatz-uT}, while the hydrodynamical stress tensor $t_{\mu\nu}$ takes the form \eqref{HydrodynamicEMTensor}.}
}

The physical parameters are the scalar and the tensor transport coefficients,  $\delta^{(n)}_i$ and $\gamma^{(n)}_i$, which are directly measurable. They should follow first order ordinary differential equations, exactly like physical couplings in field-theoretic RG flow follow corresponding $\beta$-function equations.

We will see that if we substitute the Ansatz \eqref{Ansatz-uT} in Einstein's equations we will have exactly $2m^{(n)}_s + m^{(n)}_v+ m^{(n)}_t$ equations for these variables at the $n$-th order in derivative expansion, of which $m^{(n)}_s + m^{(n)}_v$ will be algebraic and $m^{(n)}_s + m^{(n)}_t$ will be first order. There will be additionally a set of $m^{(n)}_s$ algebraic equations which will be equivalent to satisfying $u^\mu u^\nu g_{\mu\nu} = -1$, but without using the metric explicitly. Thus we will have $3m^{(n)}_s + m^{(n)}_v+ m^{(n)}_t$ equations determining $3m^{(n)}_s + m^{(n)}_v+ m^{(n)}_t$ variables in \eqref{Ansatz-uT} and \eqref{HydrodynamicEMTensor} at each order in the derivative expansion. Moreover, we will see that the equations determining the redundant variables $\alpha^{(n)}_i$, $\beta^{(n)}_i$ and $\lambda^{(n)}_i$ will be indeed by algebraic.

\section{Preliminary issues}

In this section, we begin with the details considerations for defining holographic RG flow. Let us outline the main issue we will address:
\begin{itemize}
\item \emph{The proper hypersurface foliation}: We will show that the choice of the Fefferman-Graham foliation is necessary to construct the holographic RG flow without needing to solve for the bulk spacetime metric explicitly. Later in the paper we will see how to find this metric out of the flow. The radial coordinate in this foliation directly corresponds to the energy scale in the problem. We will also address the coordinate singularity at the horizon.
\item \emph{The stress tensor counterterms}: We will argue that the nature of the counter-terms can be determined from very general considerations without specifying boundary condition on the metric at the hypersurface where the transport coefficients are evaluated.
\item \emph{Einstein's equations of motion}: We will write these equations of in terms of useful variables which will help us to eliminate the metric.
\end{itemize}

\subsection{Which hypersurface foliation to choose?}

In order to develop a holographic RG scheme, we need to choose a hypersurface foliation of spacetime first. We also demand that we should be able to obtain the evolution of data along the hypersurface foliation, \emph{i.e.} construct the holographic RG flow, directly from Einstein's equations, without knowing the spacetime metric explicitly. We will show that this requires a specific choice of foliation.

A choice of foliation is equivalent to a choice of coordinate system. Once a coordinate system is chosen, we can separate the coordinates into those coordinates which span the boundary (which we collectively denote as $x$) and the radial coordinate (which we denote as $r$). Any hypersurface foliation is given by $r = r_{\text{c}}$.

As Einstein's equation has general covariance, we need an appropriate gauge to choose a coordinate system. The most obvious choice for our purpose is to choose a gauge of the following kind:\footnote{Throughout the paper $G_{MN}$ stands for the $d+1$ dimensional metric.}
\begin{equation}
\label{GaugeChoice}
G_{rr} = f(r/l_\textrm{AdS}) 
\qquad \textrm{and} \qquad G_{r\mu} = 0 
\end{equation}
with a definite function $f(r/l_\textrm{AdS})$. With such a choice of gauge, $G_{rr}$ and $G_{r\mu}$ do not involve any hydrodynamic variables. This will allow us to write $G_{\mu\nu}^\prime$, (with ${}^\prime$ denoting the radial derivative) in terms of $t_{\mu\nu}$ and subsequently derive an equation for $t_{\mu\nu}$ directly from Einstein's equations without using the metric $G_{\mu\nu}$ explicitly. In due course of the paper, this procedure will be laid bare. 

The most well known choice in this class of gauges is the Fefferman-Graham gauge $f(r/l_\textrm{AdS}) = l_\textrm{AdS}^2/r^2$, where $l$ denotes the AdS radius. Any other gauge choice in this class is related to the Fefferman-Graham gauge by a trivial reparametrization of the radial coordinate as a function of itself.

Let us illustrate the difficulty in implementing the above procedure in other choices of gauge fixing with the example of the ingoing Eddington-Finkelstein gauge. The latter is fixed by requiring :
\begin{equation}
G_{rr} = 0 \, , \quad G_{r\mu} = -u^{\text{b}}_\mu (x),
\end{equation}
where $u^{\text{b}}_\mu (x)$ is the velocity field of the boundary fluid. If we are solving for the metric explicitly, this is the canonical choice for doing perturbation theory in the derivative expansion \cite{}. One can readily see the manifest regularity at the late time (future) horizon order by order in the perturbation expansion. Nevertheless, for direct construction of holographic renormalization group flow, this is not an appropriate choice. This is because when the energy-momentum tensor $t_{\mu\nu}$ is constructed on hypersurfaces $r=r_{\text{c}}$ from the full space-time metric $G_{MN}$, one can readily see that it involves $u^{\text{b}}_\mu (x)$. On the other hand, the physical velocity field $u_{\mu}(x,r)$ out of which the hypersurface hydrodynamic energy-momentum tensor is constructed, in a manner which is covariant with respect to the induced metric on the hypersurface, is different from $u^{\text{b}}_\mu (x)$. Therefore, in order to eliminate $G_{\mu\nu}'$ in favour of $t_{\mu\nu}$, one should eliminate $u^{\text{b}}_\mu (x)$ in favour of $u_{\mu}(x,r)$.

Let us argue that the latter cannot be achieved without knowing the metric explicitly. Indeed, in order to write $u_\mu(x,r)$ in terms of $u_{\mu}^{\text{b}}(x)$, we need to directly integrate the right hand side of the ${u^\mu}^\prime$ equation in \eqref{Ansatz-uT}. However, it cannot can accomplished without knowing the explicit metric because the covariant derivatives and raising/lowering operations, \emph{etc.} are defined using the explicit induced hypersurface metric. As for illustration:
\begin{equation}
\int d r^\prime (\nabla\cdot u)  \neq \nabla_\mu \int  u^\mu d r^\prime \, .
\end{equation}
To summarize, we cannot construct the RG flow without determining the metric explicitly first.

In order to remove this difficulty one may attempt an alternative gauge choice :
\begin{equation}
G_{rr} = 0, \quad G_{r\mu} = -u_\mu (x, r) \, ,
\end{equation}
where $u_\mu (x,r)$ is the hypersurface velocity field. This will require computation of $u_\mu^{\prime\prime} (x,r)$ as this will appear in Einstein's equations explicitly through the Ricci tensor. As a result the auxiliary variables $\alpha_i^{(n)}$, $\beta_i^{(n)}$ and $\lambda_i^{(n)}$ will not follow algebraic equations, but rather first order ordinary differential equations in $r$-derivative. Eventually in order to fix integration constants for these auxiliary variables, we will need to know the relation between $u^{\text{b}}_\mu (x)$ and $u_{\mu}^\prime(x,r)$ at $r=0$. 
This in turn will require us to also solve for the metric at least near the boundary.

Also, this gauge choice is not good even to solve the explicit metric perturbatively. We will see that $u_\mu(x,r)$ is singular at the horizon due to the red shift, as a result the metric is not invertible at the horizon. This is unlike the situation in the ingoing Eddington-Finkelstein coordinates. 

As far as we have checked there are similar problems with any other gauge choice, other than that given by \eqref{GaugeChoice}.

There is also a pragmatic argument in favour of the Fefferman-Graham gauge. If one solves for the metric explicitly in the derivative expansion in Eddington-Finkelstein coordinates, up to say the $n$-th order, the regularity at horizon is preserved at higher orders also, in the sense that all higher curvature invariants (of the $d+1$ dimensional metric $G_{MN}$) are finite at the horizon. However, the anti-de Sitter (boundary) asymptotics of the $n$-th order solution will be violated from the $(n+1)$-th order, as can be deduced from the fact that it can be translated to Fefferman-Graham coordinates only up to $n$-th order \cite{Gupta:2008th}. Conversely, if the metric is solved explicitly in the Fefferman-Graham coordinates perturbatively to the $n$-th order, the solution this time could be made regular at the horizon only up to this order in the expansion and already some of the order $n+1$ curvature invariants will diverge there even if the transport coefficients are chosen correctly \cite{Gupta:2008th}. Nevertheless, the metric is manifestly asymptotically anti-de Sitter to all orders in the expansion, because any asymptotically anti-de Sitter space must match the Fefferman Graham expansion at the boundary. Thus, solving up to $n$-th order in one of the two gauges renders the metric either smooth to all orders at the horizon or asymptotically AdS to all orders but not both. This is directly related to the general fact that $n$-th order solution in one gauge can be translated into the other only up to the $n$-th order itself.

In our construction of the holographic RG flow we want to ensure that we end up with the same UV fixed point at all orders of the perturbation theory. Even though we have solved up to a given order in perturbation theory we want the asymptotic behaviour to be maintained even at higher orders. More specifically the conformal structure of the boundary fluid is better to be established to all orders.

Before concluding this subsection let us emphasize that one should not be concerned about the Fefferman-Graham coordinate singularity at the horizon, simply because our prime goal is to find only the transport coefficients of the energy momentum tensor. Though the metric is singular at the horizon and the velocity vector $u^\mu$ and the temperature are both infinitely red-shifted there, the transport coefficients are uniquely fixed by the Landau gauge and so should be the same for a given hypersurface independently of the coordinate choice.

\subsection{Einstein's equations of motion}

In the Fefferman-Graham coordinates, any asymptotically anti de-Sitter metric in $d+1$ dimensions takes the form :
\begin{equation}\label{ads}
ds^2 = \frac{l_\textrm{AdS}^2}{r^2}\Big(dr^2 + g_{\mu\nu}(x, r) dx^\mu dx^\nu\Big).
\end{equation}
As we discussed in the previous subsection, these coordinates are always valid in a finite patch away from the boundary, $r=0$, for any asymptotically anti-de Sitter spacetime. For the metrics of fluid/gravity correspondence, these are good till the location of the horizon in the far future at each order in perturbation theory. 

In this subsection, we will not specialize to fluid/gravity correspondence - our results will be valid for any asymptotically anti-de Sitter spacetime. For the rest of this paper, if not specified otherwise, $g_{\mu\nu}$ will denote the $d$ dimensional metric appearing in (\ref{ads}).

Einstein's equations with a negative cosmological constant $\Lambda = -d(d-1)/2l_\textrm{AdS}^2$ in $d+1$ dimensions are:
\begin{equation}
\label{Einstein}
R_{MN} = -\frac{d}{l_\textrm{AdS}^2}\, G_{MN} \, ,
\end{equation}
where $M$ and $N$ run over $r$ and the $d$ dimensional index.
Let us now introduce a new tensor:
\begin{equation}
\label{Def-z}
z^\mu_{\phantom{\mu}\nu} = g^{\mu\rho}g_{\rho\nu}^\prime \, ,
\end{equation}
where ${}^\prime$ denotes derivative with respect to the radial coordinate $r$. In terms of $z^\mu_{\phantom{\mu}\nu}$ Einstein's equations \eqref{Einstein} takes the following form in Fefferman-Graham coordinates:
\begin{eqnarray}
\label{EoM}
z^{\mu\prime}_{\phantom{\mu}\nu} - \frac{d-1}{r}
z^\mu_{\phantom{\mu}\nu} + \text{Tr}\,z\Bigg(\frac{z^\mu_{\phantom{\mu}\nu}}{2}-\frac{\delta^\mu_{\phantom{\mu}\nu}}{r}\Bigg) &=& 2 R^\mu_{\phantom{\mu}\nu} \, , \nonumber\\
\nabla^\mu \text{Tr}\,z - \nabla^\nu z^\mu_{\phantom{\mu}\nu} &=& 0 \, ,\\
\text{Tr}\,z' - \frac{1}{r}\text{Tr}\,z
+ \frac{1}{2} \text{Tr} \left( z^2 \right)  &=& 0 \, .\nonumber
\end{eqnarray}
Above both the covariant derivative $\nabla^\mu$ and the Ricci tensor $R^\mu_{\phantom{\mu}\nu}$ are constructed out of the $d$ dimensional $g_{\mu\nu}$, and all traces are taken by contractions with $\delta^\mu_{\phantom{\mu}\nu}$. Also, all indices are lowered (or raised) with $g_{\mu\nu}$ (or it's inverse $g^{\mu\nu}$). In particular, it implies that $z^{\mu\nu} = -{g^{\mu\nu}}^\prime$ and $z_{\mu\nu} = {g_{\mu\nu}}^\prime$.

The three equations in \eqref{EoM} come from the $(\mu,\nu)$, $(r,\mu)$ and the $(r,r)$ components of Einstein's equations \eqref{Einstein}. For reasons that will be immediately clear, we will refer to the first equation above as the dynamical equation, and the second and third equations as the vector and scalar constraints respectively. Note that all three equations are of second order, since there is already one $r$-derivative in the definition of $z^\mu_{\phantom{\mu}\nu}$.

The dynamical equation of motion determines the metric in the bulk with given data at $r=0$ boundary. This data is the boundary metric:
\begin{equation}
\label{BoundaryMetric}
g^{\text{b}}_{\mu\nu}(x) = g_{\mu\nu}(r=0,x)
\end{equation}
and the boundary energy-momentum tensor which we will define  precisely later in the section. Importantly there is a residual class of diffeomorphisms which keep the form of the metric (\ref{ads}) invariant. These diffeomorphisms are in one-to-one correspondence with conformal transformations of the boundary metric, and so \eqref{BoundaryMetric} can be defined only up to conformal transformation.

The vector and scalar equations merely constrain this boundary data. Once they are satisfied at $r=0$, the dynamical equation ensures they are satisfied at all radial slices \cite{Gupta:2008th}. The vector constraint amounts to conservation of the energy-momentum tensor, for any hypersurface $r=r_{\text{c}}$ and thus gives the equations of fluid mechanics, $\nabla^{\mu} t_{\mu\nu}=0$. The scalar constraint amounts to imposing a trace condition on the energy-momentum tensor at the boundary. For a flat metric the trace of the energy-momentum tensor should vanish at the boundary. If, on the other hand, the boundary metric is curved, the trace of the boundary  stress tensor is also completely fixed by conformal anomaly.

\subsection{The renormalized energy-momentum tensor on hypersurfaces}

\label{Section-Counterterms}

Traditionally, we define the hypersurface energy-momentum tensor as a functional derivative of the on-shell gravitational action with respect to the induced metric. The renormalized hypersurface energy-momentum tensor is the functional derivative of the renormalized on-shell gravitational action, where counterterms have been added to regulate infra-red divergences - divergences arising from infinite volume of space-time as the cut-off hypersurface 
is taken to asymptotic infinity. 

In order to have a well-defined action principle, we need to specify a boundary condition on the cut-off hypersurface, as well as add a variational counterterms, if necessary. The traditional choice is the Dirichlet boundary condition, which necessitates addition of Gibbons-Hawking counterterms.

The above procedure has been implemented in asymptotically anti-de Sitter spacetimes. Let us choose a coordinate system in the bulk which gives a hypersurface foliation of spacetime with $\rho$ denoting the radial coordinate. The renormalized energy-momentum tensor on each hypersurface $\rho=\rho_{\text{c}}$ in the foliation (for $d>2$) is \cite{Henningson:1998ey},\cite{Balasubramanian:1999re}:
\begin{eqnarray}
\label{GravityEMTensor}
t_{\mu\nu} = -\dfrac{l_\textrm{AdS}^{d-2}}{8 \pi G_N h(\rho_{\text{c}})} \left( K_{\mu\nu}-K \gamma_{\mu\nu}+\frac{d-1}{l_\textrm{AdS}} \gamma_{\mu\nu} - \frac{l_\textrm{AdS}}{d-2}\left( {{}^\gamma\!{R}}_{\mu\nu}-\frac{1}{2}{{}^\gamma\!{R}} \gamma_{\mu\nu}\right) - ...\right) \, ,
\end{eqnarray}
where $\gamma_{\mu\nu}$ is the induced metric on the hypersurface, $K_{\mu\nu}$ is the extrinsic curvature of the hypersurface, while ${{}^\gamma\!{R}}_{\mu\nu}$ and ${{}^\gamma\!{R}}$ is constructed from the intrinsic (Riemann) curvature of the hypersurface. The $...$ denotes other counterterms which may be necessary to correct infra-red divergences. 

Despite the appearance of the $\rho_c$ factor this form of the cut-off stress tensor is coordinate independent. The explicit form of the function $h(\rho_\text{c})$ is uniquely fixed from the overall regularization of the energy-momentum tensor at the boundary. Notice also that the Brown-York term and the Einstein tensor counter-term are separately conserved.

In the Fefferman-Graham foliation $\rho=r$ with $r$ given by \eqref{ads} and we also have:
\begin{equation}
h(r) = r^{d-2} \, , \qquad
\gamma_{\mu\nu} = \left(  \dfrac{l_\textrm{AdS}}{r} \right)^2 g_{\mu\nu} \, , \qquad
K_{\mu\nu} =\frac{l_\textrm{AdS}}{r} \left(  \frac{g_{\mu\nu}}{r} - \frac{g_{\mu\nu}^\prime }{2} \right) \, .
\end{equation}
The renormalized energy-momentum on each hypersurface then takes the following form (from now on we omit the ${}_\textrm{c}$ subscript in $r_\textrm{c}$):
\begin{equation}
\label{EMTensor-z}
t^{\mu}_{\phantom{\mu}\nu} = \frac{1}{\kappa_\textrm{AdS} r^{d-2}}\left( \frac{z^\mu_{\phantom{\mu}\nu}}{r} - \frac{\text{Tr} \, z}{r}\delta^\mu_{\phantom{\mu}\nu} + \frac{2}{d-2} \left(R^{\mu}_{\phantom{\mu}\nu} - \frac{1}{2}R \delta^\mu_{\phantom{\mu}\nu}\right)+ ...\right).
\end{equation}
Here $z_{\mu\nu}$ is defined in \eqref{Def-z}, the Ricci tensor and Ricci scalar are constructed out of the physical hypersurface metric $g_{\mu\nu}$.  Again, all raising and lowering is done with the inverse of $g_{\mu\nu}$. We call $g_{\mu\nu}$ the physical metric because it is finite in the limit $r\rightarrow 0$, \emph{i.e.} in the limit the hypersurface is taken to the boundary. We have raised the first index of $t_{\mu\nu}$ for later convenience and $\kappa_\textrm{AdS}$  is defined in \eqref{BoundaryConstants}.

The boundary energy-momentum tensor, which is the expectation value of the energy-momentum in the field-theoretic state dual to the specific bulk geometry, is:
\begin{equation}
t_{\mu\nu}^{\,\,\text{b}}(x) = \lim_{r \rightarrow 0}t_{\mu\nu}(r,x).
\end{equation}
This limit is indeed finite due  to the $r^{d-2}$ factor in \eqref{EMTensor-z}. One of the consistency checks can be done by using bulk diffeomorphisms which are in one-to-one correspondence with conformal transformations of the boundary metric. Under such bulk diffeomorphisms, which can be defined as those which preserve Fefferman-Graham form of the metric, the energy-momentum tensor also transforms Weyl covariantly up to conformal anomaly \cite{Henningson:1998ey, Balasubramanian:1999re, deHaro:2000xn}. The conformal anomaly is uniquely determined by central charges, which turn out to be fixed by the $d+1$ dimensional Newton's gravitational constant.

We have the following issue in light of the discussion above. We want to reformulate fluid/gravity correspondence as first order renormalization group flow equations, so that finiteness of transport coefficients at the horizon alone determines the flow uniquely. However, the hypersurface energy-momentum tensor implicitly uses a boundary condition for the induced metric - the Dirichlet boundary condition. The implicit use of the boundary condition on the metric is in conflict with the idea of an emergent spacetime (where the radial direction should be emergent).

This issue can be addressed as follows. The hypersurface energy-momentum tensor \eqref{GravityEMTensor} is defined by a variation of the gravitational action and according to the variational principle the induced metric is kept fixed \cite{Henningson:1998ey},\cite{de Haro:2000xn}. Nevertheless no specific form of the induced metric is assumed. Thus the energy-momentum tensor \eqref{GravityEMTensor} does not assume any specific boundary condition from the point of view of the bulk equations of motion, \emph{i.e.} Einstein's equation. This is in contrary to the previous approach \cite{Bredberg:2010ky}, where it was assumed that the induced metric on the cut-off surface remains flat to all orders in perturbation theory.

Assuming that there are no constraints on the induced metric other than it is weakly curved, the most general perturbation at the first order is $\delta\gamma_{\mu\nu} =(c_1/r^2) \pi T \sigma_{\mu\nu}$ , with an arbitrary $r$-independent $c_1$. In our previous work \cite{Kuperstein:2011fn} we have taken this issue into account and showed that for any choice of $c_1$ we get the same RG flow for the shear viscosity. Therefore, at the first order in the derivative expansion the RG flow indeed is independent of the specific choice of boundary conditions.

We can also ask if we can give up the variational principle to construct the cut-off energy momentum tensor \eqref{GravityEMTensor} and thus getting rid of even implicit boundary conditions at the cut-off.
We will demonstrate now that, indeed, the form of the energy-momentum tensor \eqref{GravityEMTensor} can be obtained without relying on the variational principle.

With no variational principle at hand what are the other possible limitations on the form of the energy momentum tensor? We will propose the following four conditions that will set the form of the energy-momentum tensor up to some numerical constants multiplying the counter-terms:
\begin{itemize}
\item The ``bare" energy momentum tensor should satisfy the junction condition. In other words, $t_{\mu\nu}^{\text{bare}}$ is the physical source localized on the hypersurface one should add to the right hand side of Einstein's equation - choosing it appropriately allows for two arbitrarily different solutions of vacuum Einstein's equation on the two sides of this hypersurface to be glued together. 
\item The form of the energy-momentum tensor should be universal meaning it is conserved for any solution of Einstein's equations.
\item The dependence of the full energy-momentum tensor on the anti-de Sitter radius $l$ comes via an overall dimensionless factor $\kappa_\textrm{AdS}^{-1}=l_\textrm{AdS}^{d-1}/(16 \pi G_N)$. This is because we don't consider here $l^{-1}$ corrections which translate to $\alpha^\prime$ or $1/\sqrt{\lambda}$ corrections in string theory or dual field theory respectively. 
\item The UV stress tensor is finite. 
\end{itemize}

The ``bare" hypersurface energy-momentum tensor $t_{\mu\nu}^{\text{bare}}$ in \eqref{GravityEMTensor} is the Brown-York stress tensor, and we will argue first why this should be so. We note that it is the only functional of the extrinsic curvature $K_{\mu\nu}$ which is covariantly conserved on-shell, \emph{i.e.} which satisfies
$\nabla^\mu t_{\mu\nu}^{\text{bare}} = 0$ in an arbitrary solution of Einstein's equation. Thus the ``bare" hypersurface stress-tensor should be proportional to the Brown-York energy-momentum tensor, with the proportionality factor exactly as in \eqref{GravityEMTensor} for the reasons explained above.

The universality principle makes the counter-terms be functionals only of the induced metric, the intrinsic and the extrinsic curvature, since other variables, like $u^\mu$ and $T$, can be ingredients in a conserved current only for specific non-generic solutions. Next, we observe that the extrinsic curvature can contribute only to the Brown-York tensor. Higher extrinsic curvature invariants will contain more than one $r$-derivative of the induced metric. They will over-constrain the hypersurface data and hence will not be conserved in a generic solution of the two-derivative Einstein's equations.

We are left only with terms built out of the induced metric and its intrinsic curvature. Therefore no $r$-derivatives appear except in the ``bare" Brown-York term. Constructing conserved tensors out of the metric, the Riemann curvature and its covariant derivatives is a standard well-known problem. There is a finite number of them at any order in the derivatives. For example, at the second order, the Einstein tensor is the only candidate. 

It then follows from the penultimate requirement that the $r$-dependence of the counter-term coefficients is necessarily of the form $r^{-d+k}$, where $k$ is the mass dimension of the counter-term.\footnote{Multiplying the counter-terms by function $f(r,T(x))$ will not only spoil universality but also the conservation of the energy-momentum tensor.} For example, the counter-term in \eqref{GravityEMTensor} (the Einstein tensor) has two space-time derivatives (dimension two) and so its coefficient goes like $r^{-d+2}$. Thus the coefficients of all these counter-terms are determined up to numerical constants.

The on-shell gravitational action has finite number of ultra-violet divergences as the cut-off moves to the boundary. Hence it can fix the numerical coefficients of a few leading counter-terms only. For instance, the counter-term proportional to the induced metric in \eqref{GravityEMTensor} cancels the volume divergence which goes like $r^{-d}$ and so its coefficient is fixed uniquely. The coefficient of the Einstein tensor in \eqref{GravityEMTensor} is also set exactly this way, since it cancels $r^{2-d}$ divergence in the Brown-York tensor. The last UV divergence cancelling counter-term in even dimensions will have $d$ derivatives. It will be related to the conformal anomaly and will cancel the $\log(r)$ divergence. In odd $d$ the last counter-term will have $d-1$ derivatives cancelling $r^{-1}$ divergence.

To summarize, even giving up the variational principle we can uniquely fix the stress tensor up to numerical constants that multiply counter-terms that do not cancel UV divergences. Later we will conjecture that all these numerical constants can be fixed by imposing constrains on the horizon fluid.

\section{Construction of the holographic RG flow}

In this section, we will construct the explicit algorithm for solving the holographic fluid RG flow order by order in the derivative expansion. We will first sketch the basic procedure, highlighting why it does not require explicit knowledge of the bulk spacetime metric. We will then construct the RG flow for equilibrium variables and see how we can define thermodynamics in a consistent way at each hypersurface in the foliation. We will then demonstrate the iterative procedure of solving the RG flow of the transport coefficients order by order in the derivative expansion. Finally, we will show that once the RG flow is solved, we can also explicitly reconstruct the bulk spacetime metric from the scale dependent physical quantities only. Therefore, we will establish that our construction of the holographic RG flow is equivalent to emergence of spacetime.

\subsection{How to eliminate the metric?}
\label{MetricElimination}

In order to show that holographic RG flow leads to emergence of spacetime, we should be able to construct it without requiring explicit knowledge of the spacetime metric. Here we will outline how this can be achieved in the Fefferman-Graham hypersurface foliation. In subsequent subsections, we will give the full systematics of the algorithm.

Firstly to simplify expressions, it is convenient to redefine the renormalized hypersurface energy-momentum tensor $t^{\mu}_{\phantom{\mu}\nu}$ as follows:
\begin{equation}
\label{t-rescaling}
t^{\mu}_{\phantom{\mu}\nu} \, \rightarrow \,
\kappa_\textrm{AdS} \cdot t^{\mu}_{\phantom{\mu}\nu} \, ,
\end{equation}
where $\kappa_{AdS}$ was defined in \eqref{BoundaryConstants}.
In the AdS/CFT correspondence, the above redefinition simply strips off an $N^2$ factor, so now $t^{\mu}_{\phantom{\mu}\nu}$ has a finite large $N$ limit.

With the above re-definition, we can now invert the relation \eqref{EMTensor-z} between $t^{\mu}_{\phantom{\mu}\nu}$ and $z^{\mu}_{\phantom{\mu}\nu}$ as follows:
\begin{equation}
\label{z}
z^{\mu}_{\phantom{\mu}\nu} = r^{d-1}\Bigg(t^\mu_{\phantom{\mu}\nu} - \frac{\text{Tr} \, t}{d-1}  \delta^\mu_{\phantom{\mu}\nu}\Bigg) - \frac{2r}{d-2} \Bigg(R^{\mu}_{\phantom{\mu}\nu} - \frac{R}{2(d-1)} \delta^\mu_{\phantom{\mu}\nu}\Bigg)+ ... \, .
\end{equation}
Above the $...$ denotes corrections relevant only at higher order in derivatives and which originate from the counterterms used to define the renormalized energy-momentum tensor. We also see from above that
\begin{equation}
\label{Trz}
\text{Tr}\, z = - \frac{r^{d-1}}{d-1} \text{Tr}\,t - \frac{r}{d-1} R + ... \, .
\end{equation}

The strategy is simple. We substitute \eqref{z} and \eqref{Trz} in Einstein's equations of motion as given in the form \eqref{EoM}. This gives the equation for evolution of $t^\mu_{\phantom{\mu}\nu}$ which is first order in radial derivative. We actually only need to use the first equation in \eqref{EoM} as the other two are merely non-dynamical constrains. Once they satisfied at the initial hypersurface, they are preserved along the flow. The dynamical equation of motion for $t^\mu_{\phantom{\mu}\nu}$ is explicitly as below:
\begin{eqnarray}
\label{EMTensorEoM-Full}
&&
{t^\mu_{\phantom{\mu} \nu}}^\prime - \frac{2 r^{2-d}}{d-2} {R^\mu_{\phantom{\mu} \nu}}^\prime - 
 \dfrac{r^{d-1}}{2(d-1)}\left( \textrm{Tr} t + r^{2-d} R \right) \left( t^\mu_{\phantom{\mu} \nu} - \frac{2 r^{2-d}}{d-2} R^\mu_{\phantom{\mu} \nu} \right) +
\\
&& 
\,
 + \frac{1}{d-1}\left( - \textrm{Tr} t^\prime + \dfrac{r^{2-d}}{d-2} R^\prime + \frac{\textrm{Tr} t}{r} + \frac{r^{d-1}}{2(d-1)} 
 \left( \textrm{Tr} t + r^{2-d} R \right) \left( \textrm{Tr} t - \dfrac{r^{2-d}}{d-2} R \right) \right) \delta^\mu_{\phantom{\mu} \nu} + \ldots
 = 0 \, .
\nonumber
\end{eqnarray}
Above $...$ again represents terms which are relevant only onwards from third order in derivative expansion.

Here we encounter a conceptual difficulty. Notice that $R^\mu_{\phantom{\mu}\nu}$ and $R$ appear in the dynamical equation for radial evolution of $t^\mu_{\phantom{\mu}\nu}$ - this is not yet the problem, because this does not involve any radial derivative of the metric. In fact, both $R^\mu_{\phantom{\mu}\nu}$ and $R$ can be readily represented in terms of the scalar and tensor transport terms we introduced in \eqref{ListOfSecondOrderScalars} and \eqref{ListOfSecondOrderTensors}. However, their radial derivatives, \emph{i.e.} ${R^{\mu}_{\phantom{\mu}\nu}}^\prime$ and $R^\prime$ also appear in \eqref{EMTensorEoM-Full}. Since we want to proceed without solving for the metric, we have to express these quantities in terms of $z^\mu_{\phantom{\mu}\nu}$, as the latter can be written in terms of $t^\mu_{\phantom{\mu}\nu}$
using \eqref{z}, and therefore the equation of motion \eqref{EMTensorEoM-Full} will eventually involve only $t^\mu_{\phantom{\mu}\nu}$, its $r$-derivative, $R^\mu_{\phantom{\mu}\nu}$ and $R$.

Let us show that the radial derivative of the Riemann curvature and other covariant quantities constructed out of $g_{\mu\nu}$ can be indeed written in terms of covariant derivatives of $z^\mu_{\phantom{\mu}\nu}$ only. Firstly, the Levi-Civita connection constructed out of $g_{\mu\nu}$ identically satisfies:
\begin{equation}
\label{Christoffel-prime}
\Gamma^{\mu\prime}_{\nu\rho} = \frac{1}{2}\left(\nabla_\nu z^\mu_{\phantom{\mu}\rho} + \nabla_\rho z^\mu_{\phantom{\mu}\nu}- \nabla^\mu z_{\nu\rho}\right).
\end{equation}
We note though the Levi-Civita connection is not covariant with respect to $g_{\mu\nu}$, it's radial derivative is. We recall that $\nabla$ denotes the covariant derivative constructed from $g_{\mu\nu}$, and this will be so for the rest of paper. Using the above, one can readily prove the following useful identity:
\begin{eqnarray}
\label{Riemann-prime}
{R^{\mu}_{\phantom{\mu}\nu\rho\sigma}}^\prime
= \frac{1}{2} \left(  \nabla_\rho \nabla_\nu z^\mu_{\phantom{\mu}\sigma}  - \nabla_\sigma \nabla_\nu z^\mu_{\phantom{\mu}\rho}- \nabla_\rho \nabla^\mu z_{\nu\sigma} + \nabla_\sigma \nabla^\mu z_{\nu\rho} \right)  + \frac{1}{2}\left(  R^{\mu}_{\phantom{\mu}\kappa\rho\sigma}
z^\kappa_{\phantom{\mu}\nu} -
R^{\kappa}_{\phantom{\mu}\nu\rho\sigma}
z^\mu_{\phantom{\mu}\kappa} \right) \, .
\end{eqnarray}
Tracing over (\ref{Riemann-prime}) and using the vector constraint in (\ref{EoM}) it follows that:
\begin{equation}
\label{Ricci-prime}
R^{\prime}_{\mu\nu} = \frac{1}{2}\left(\nabla_\rho\nabla_\mu z^\rho_{\phantom{\mu}\nu} + \nabla_\rho\nabla_\nu z^\rho_{\phantom{\mu}\mu}- \nabla_\alpha\nabla^\alpha z_{\mu\nu}- \nabla_\mu \nabla_\nu \text{Tr}\, z\right).
\end{equation}
Furthermore, contracting the above with $g^{\mu\nu}$ and using the vector constraint in (\ref{EoM}) again it also follows that:
\begin{equation}
\label{RicciScalar-prime}
R^\prime = -R_{\mu\nu}z^{\mu\nu}.
\end{equation}
The derivation of the four above formulae is completely general and does not rely on any specific property of $g_{\mu\nu}$.
One can also derive similar expressions for covariant derivatives of Riemann curvature tensor; however, we will not need these in this paper.

To recap, we can use the identities \eqref{Ricci-prime} and \eqref{RicciScalar-prime} to eliminate ${R^{\mu}_{\phantom{\mu}\nu}}^\prime$ and $R^\prime$ from Einstein's equation of motion \eqref{EMTensorEoM-Full}. Then we use \eqref{z} to rewrite \eqref{EMTensorEoM-Full} solely in terms of ${t^{\mu}_{\phantom{\mu}\nu}}^\prime$, $t^{\mu}_{\phantom{\mu}\nu}$, $R^{\mu}_{\phantom{\mu}\nu}$ and $R$. Schematically we obtain that:
\begin{equation}
t_{\mu\nu}^\prime = t_{\mu\nu}^\prime \left( t_{\mu\nu}, R_{\mu\nu\rho\sigma} \right) \, .
\end{equation}

In our discussion so far, we have not used any specific form of the energy-momentum tensor. In the hydrodynamic limit, the derivative expansion is the systematics on which we can construct an iterative algorithm. In each step of the iteration, we bring in more covariant derivatives of $t^\mu_{\phantom{\mu}\nu}$ and $R^\mu_{\phantom{\mu}\nu}$, and higher order polynomials of $R^\mu_{\phantom{\mu}\nu}$ as well. To see this explicitly, we now assume that $t^\mu_{\phantom{\mu}\nu}$ is hydrodynamic on each hypersurface of the foliation, namely takes the form \eqref{HydrodynamicEMTensor}.

It what follows in this section it will be useful to have the leading order term in the explicit hydrodynamical form of $z^\mu_{\phantom{\mu} \nu}$ and $\textrm{Tr} z$.
Substituting \eqref{HydrodynamicEMTensor} into \eqref{z} and \eqref{Trz} we obtain that:
\begin{eqnarray}
\label{z-ZerothOrder}
z^\mu_{\phantom{\mu}\nu} &=& r^{d-1} \Bigg(\left( \frac{d-2}{d-1}\epsilon + P \right) u^\mu u_\nu + \frac{\epsilon}{d-1}\Delta^\mu_{\phantom{\mu}\nu}\Bigg)
+ ...\, ,\nonumber\\
\text{Tr}\, z &=& r^{d-1}\left(\frac{\epsilon}{d-1} - P  \right)+ \ldots\,.
\end{eqnarray}
Above $\ldots$ denotes the first and higher order derivative corrections. 

Once we have eliminated ${R^{\mu}_{\phantom{\mu}\nu}}^\prime$ and $R^\prime$ in favour of $t^{\mu}_{\phantom{\mu}\nu}$ and its covariant derivatives, we can substitute the hydrodynamical Ansatz \eqref{HydrodynamicEMTensor} into Einstein's equation of motion \eqref{EMTensorEoM-Full}. Any time the radial derivative hits $u^{\mu}$ and $T$ we have to further use our Ansatz for $u^{\mu\prime}$ and $T^\prime$ as given in \eqref{Ansatz-uT}. On the other hand, $r$-derivatives of $\nabla_\mu$ need be treated with \eqref{Christoffel-prime}. This way the only $r$-derivatives in the equation will be those of the transport coefficients $\gamma_i$ and $\delta_i$. This will give ordinary differential equations for all the coefficients in \eqref{HydrodynamicEMTensor}. We will later show that indeed there will be as many equations (both algebraic and differential) as there are parameters, and these can be solved iteratively order by order in the derivative expansion.

Before we give the details of the algorithm, it is worth to illustrate our approach with a simple example. The hydrodynamic stress tensor \eqref{HydrodynamicEMTensor} has a bulk viscosity term $-\zeta \left( \nabla \cdot u \right)$. Plugging it into \eqref{EMTensorEoM-Full} produces a term $\left( \nabla \cdot u \right)^\prime$. Let us show how this term can be rewritten:
\begin{eqnarray}
\label{nabla-u-Example}
(\nabla\cdot u)^\prime &=& \nabla_\mu {u^{\mu}}^\prime + {\Gamma^{\mu}_{\nu\mu}}^\prime u^\nu = \alpha_0 (\nabla\cdot u) + \frac{d\alpha_0}{d \ln s} D\ln s  + \frac{1}{2}D \text{Tr}\, z + ... \nonumber\\
&=& \Big(\alpha_0 - \frac{d\alpha_0}{d \ln s}-\frac{1}{2}\Big(c_s^2 - \frac{1}{d-1}\Big)(\epsilon+P)\Big)(\nabla\cdot u) + ...\, .
\end{eqnarray}
In the first equality we have simply substituted the definition of the covariant derivative. In the second we have used the Anstaz \eqref{Ansatz-uT} (keeping only the leading term in the derivative expansion), \eqref{Christoffel-prime}\footnote{Notice that ${\Gamma^{\mu}_{\nu\mu}}^\prime = \dfrac{1}{2} \nabla_\nu \textrm{Tr} z$.} and the definition of $D=u^\mu\nabla_\mu$. Finally,
in the second line we have substituted \eqref{z-ZerothOrder} and have used the thermodynamic identities \eqref{Thermodynamics} and the Euler equations in the form \eqref{Euler-s}. 

Here we have truncated up to first order in derivatives. As $\nabla\cdot u$ is the only independent scalar at the first order, it's radial derivative can only be proportional to itself at the leading order. It is clear from the above example we can systematically go to higher orders in the derivative expansion.

We note that just like in the example above, if we consider any scalar $\mathcal{S}^{(n)}_i$, or a transverse vector  $\mathcal{V}^{(n)\mu}_i$ or a symmetric, traceless transverse tensor  $\mathcal{T}^{(n)}_{i\mu\nu}$ at the $n$-th order in derivatives, it's radial derivative will involve $n$-th but also higher order scalars, transverse vectors or symmetric, traceless and transverse tensors at higher orders. 

This is in exact analogy with RG flow in quantum field theory. Operators of lower dimension mixes with operators of lower dimensions in the RG flow. If we think of $\mathcal{S}^{(n)}_i$, $\mathcal{V}^{(n)\mu}_i$ and $\mathcal{T}^{(n)}_{i\mu\nu}$ as operators, then indeed their mixing with operators at higher order in derivative expansion along the radial evolution is analogous to an operator mixing with other operators of same dimension and also higher dimensional operators in RG flow in field theory. 

On the other hand, the parameters $\alpha^{(n)}_{i}$, $\beta^{(n)}_i$, $\lambda^{(n)}_i$, $\gamma^{(n)}_i$ and $\delta^{(n)}_i$ in \eqref{HydrodynamicEMTensor} and \eqref{Ansatz-uT} are analogous to couplings in field theory. Due to the nature of operator mixing, the RG flow of couplings of a given degree of relevance, namely of a given mass dimension can be calculated without involving couplings of lesser relevance, \emph{i.e.} lesser mass dimension. Similarly, we will find that the RG flow of the transport coefficients $\gamma^{(n)}_i$ and $\delta^{(n)}_i$, and the field redefinition parameters $\alpha^{(n)}_i$, $\beta^{(n)}_i$ and $\lambda^{(n)}_i$ relevant at the $n$-th order in derivatives can be calculated without involving those relevant at higher order in derivatives. We also recall that the parameters of field redefinitions $\alpha^{(n)}_i$, $\beta^{(n)}_i$ and $\lambda^{(n)}_i$ satisfy algebraic equations, and are analogous to redundant couplings in field theory  which can be algebraically eliminated as well by field redefinitions.

\subsection{RG flow of thermodynamic data}

In any iterative construction, we should understand the zeroth order first. Therefore, we should begin by understanding how the holographic RG flow works for the thermal equilibrium geometry, corresponding to the unperturbed black brane. Solving for the black brane geometry should be equivalent to the thermodynamic data RG flow. We will need to address the conceptual issue of how to define thermodynamic quantities at each hypersurface of the foliation without knowing the bulk metric explicitly. Also the RG flow should determine the radial location of the horizon without requiring to know the metric explicitly.

The RG flow Ansatz of the black brane geometry is given by \eqref{Ansatz-uT} together with \eqref{HydrodynamicEMTensor}. We first show how $\alpha_0$ which gives the cut-off dependent field definition of $u^\mu$ can be solved without knowing the explicit form of the metric. 

Differentiating $u^\mu u^\nu g_{\mu\nu} = -1$ with respect to $r$ we obtain:
\begin{equation}
\label{norm}
u^{\mu}u^\nu z_{\mu\nu} + 2 u^{\mu\prime}u_\nu = 0.
\end{equation}
We recall that in equilibrium we can ignore higher order terms in $z_{\mu\nu}$ leaving only the zeroth order term as in \eqref{z-ZerothOrder}. Substituting \eqref{z-ZerothOrder} into \eqref{norm} we find a linear algebraic equation for $\alpha_0$. Solving it yields:
\begin{equation}
\label{alpha_0}
\alpha_0 = \frac{r^{d-1}}{2}\left( P + \frac{d-2}{d-1}\epsilon \right) \, .
\end{equation}
It is worth mentioning here again that raising/lowering indices does not commute with the $r$-derivative, because the metric is itself $r$-dependent. For example we note that: 
\begin{equation}
u_\mu^\prime = u^{\mu\prime} g_{\mu\nu} + u^\mu z_{\mu\nu} = - \alpha_0 u_\mu.
\end{equation}

Substituting $t_{\mu\nu}^\text{eq}=\epsilon u^\mu u_\nu + P \Delta^{\mu}_{\phantom{\mu}\nu}$ into the radial evolution equation of $t_{\mu\nu}$ given by \eqref{EMTensorEoM-Full}, truncating to zeroth order in derivatives and using \eqref{alpha_0} we obtain radial evolution of the energy density $\epsilon$ and the pressure $P$. Essentially we get two coupled ordinary differential equations that can be conveniently written as follows:
\begin{eqnarray}
 \textrm{Tr}\, t^\prime &=& \left( \frac{r^{d-1}}{2(d-1)} \textrm{Tr}\, t + \frac{d}{r}\right) \textrm{Tr}\, t \, ,
 \nonumber \\
 \epsilon^\prime  &=& \left( \frac{r^{d-1}}{2 (d-1)} \epsilon - \frac{1}{r}\right) \textrm{Tr} \, t \, ,       
\end{eqnarray}
where $\text{Tr}\, t = (d-1) P - \epsilon$. The general solution of the first equation is:
\begin{equation}
 \textrm{Tr} \,t = 4 d (d-1)\, \frac{r^d}{r_\textrm{H}^{2d} - r^{2d}} \, ,
\end{equation} 
where $r_\textrm{H}$ is a constant of integration. Notice that one can reproduce a traceless thermal stress tensor asymptotically at $r =0$ as expected. On the other hand, $\text{Tr} \, t$ blows up at $r = r_{\textrm{H}}$. We will see immediately that $r_{\textrm{H}}$ is the radial location of the horizon. 

The $\epsilon$-equation then yields:
\begin{equation}
\epsilon = 4 (d-1)\,  \frac{\mathrm{const} - r^d}{r_\textrm{H}^{2d} - r^{2d}} \, .
\end{equation}
In order to get a finite $\epsilon$ at the horizon $r = r_{\textrm{H}}$ we have to put $\mathrm{const} =  r_\textrm{H}^{d}$.

Finally we get:
\begin{equation}
\label{e(r)-and-P(r)}
\epsilon = \frac{4 (d-1) }{r_\textrm{H}^{d} + r^{d}}
\quad \textrm{and} \quad
P = 4\,  \frac{r_\textrm{H}^{d} + (d-1) r^d}{r_\textrm{H}^{2d} - r^{2d}} \, .
\end{equation}
We see above that $\epsilon$ is regular at the horizon and it's horizon value is half of the boundary value. The pressure $P$ blows up at the horizon. We will see later this corresponds to the fact that the fluid at the horizon is non-relativistic and incompressible. We will also see that consequently the effective temperature and the speed of sound also blows up at the horizon.

We now address the question how to define thermodynamics consistently at each hypersurface in the foliation. We note that the parameter $r_{\textrm{H}}$ which gives the location of the horizon actually parametrizes the curve $P=P(\epsilon , r)$ which gives the equation of state at a fixed value of $r$, \emph{i.e.} on a given hypersurface. Using the thermodynamic identities \eqref{Thermodynamics}, we get:
\begin{equation}
\int \textrm{d} r_{\textrm{H}}\Bigg( \frac{\partial \epsilon}{\partial r_{\textrm{H}}}\Bigg)\frac{1}{\epsilon+P} = \int  \textrm{d} \ln\,s.
\end{equation}
The above allows us to determine $s$ up to an overall constant, as given by
\begin{equation}
\label{s}
s = C \, \frac{r_{\textrm{H}}^{d-1}}{(r_{\textrm{H}}^d + r^d)^{2 \left(1 - 1/d \right)}}.
\end{equation}
We note that thermodynamic identities alone do not determine the $x$-independent constant $C$.

\setlength{\unitlength}{0.9cm}
\FIGURE[t]{
\begin{picture}(12,9)
\put(1.0,0){\vector(0,1){7}}
\put(1.0,0){\vector(1,0){8}}
\put(1.5,6.5){\Large{$P$}}
\put(9.0,0.5){\Large{$\epsilon$}}
\put(5.6,5.5){\centering{\Large{$r_\textrm{H}$  or  $T$}}}
%\linethickness{0.7mm}
%\put(8.2,5.3){\vector(1,0){0.6}}
\linethickness{0.5mm}
\qbezier(8.5,6.8)(6.5,2.3)(1.5,1.7)
\put(5.85,3.8){\line(1,0){0.5}}
\put(6.35,3.3){\line(0,1){0.5}}
\end{picture}
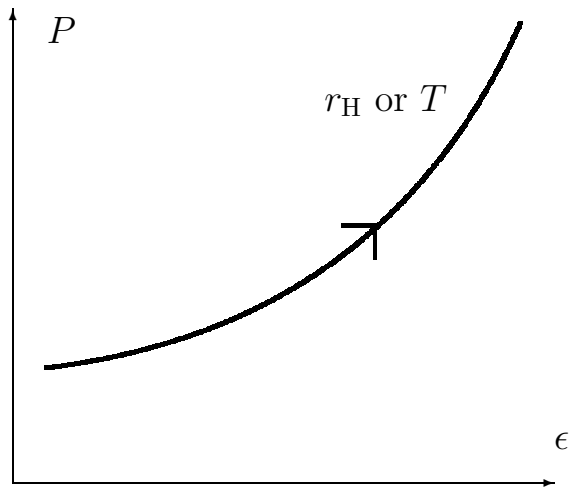
\caption{Solving \eqref{e(r)-and-P(r)} we find that $\epsilon$ and $P$ lie on a curve parametrized by $r_\textrm{H}$ at a fixed $r$. This allows us to find the temperature $T$ using the thermodynamic identities \protect\eqref{Thermodynamics}. Eliminating $r_\textrm{H}$ in favour of $T$ we arrive at $\epsilon=\epsilon(T,r)$ and $P=P(T,r)$.}
\label{Thermodynamics-Figure}
}

We can now determine $T$ from $T = (\epsilon+P)/s$. We get
\begin{equation}
\label{T}
T = \frac{4d}{C} \cdot r_{\textrm{H}} \dfrac{\left( r_{\textrm{H}}^d+ r^d \right)^{1-\frac{2}{d}}}{ r_{\textrm{H}}^d - r^d} \, .
\end{equation}
From the above we can determine $\lambda_0$ in our equilibrium Anstaz \eqref{Ansatz-uT} by calculating $T^\prime/T$. We will, however, see later that for actual computations, the explicit form of $\lambda_0$ will not be necessary.

It will also not be necessary to determine the constant $C$ appearing in $s$ and $T$. This constant is related to the scaling symmetry $\left( T,s\right) \to \left( C^{-1} \cdot T, C \cdot s\right)$ that does not affect the thermodynamic relations.

Nevertheless, we would like to fix $C$, merely because it will be useful later for the comparison of our results with those in the literature. We can so by identifying $T_\textrm{b}$, the value of $T$ at the boundary $r=0$, with the Hawking temperature $T_\textrm{Hawking}$ of the unperturbed black brane. The relation is:
\begin{equation}
\label{Hawking} 
T_\textrm{Hawking} = \dfrac{2^{2/d-2} d}{\pi r_\textrm{H}}  
\qquad \textrm{and so} \qquad
C = 2^{4-2/d} \pi \, .
\end{equation}
Again, this value of $C$ can be set only knowing the black brane geometry, but we only need this identification for thee future reference to the previous works on the subject.
With this relation we now have:
\begin{equation}
\label{s(r)-T(r)}
s(r) = 2^{-2/d + 4} \pi \cdot \frac{r_{\textrm{H}}^{d-1}}{(r_{\textrm{H}}^d + r^d)^{2 \left(1 - 1/d \right)}}
\qquad \textrm{and} \qquad
T(r) = \dfrac{2^{-2 + 2/d} d}{\pi} \cdot \dfrac{ r_\textrm{H} \left( r_{\textrm{H}}^d+ r^d \right)^{1-2/d}}{ r_{\textrm{H}}^d - r^d} \, . 
\end{equation}
We can also derive the speed of sound $c_s$:
\begin{equation}
\label{SpeedOfSound}
c_s^2 = \frac{\partial p/\partial r_{\textrm{H}}}{\partial \epsilon/\partial r_{\textrm{H}}} = \frac{r_{\textrm{H}}^{2d}+ 2(d-1)r_{\textrm{H}}^d r^d + r^{2d}}{(d-1) (r_{\textrm{H}}^{d} -r^{d})^2}.
\end{equation}

In the above equations $r_{\textrm{H}}$ is a constant. In such a case, the fluid is in static equilibrium at all hypersurfaces in the foliation with an equation of state that depends on the radial location of the hypersurface. This is an exact solution of the RG flow. To go to non-equilibrium, we need to promote $r_\textrm{H}$ to a function of the hypersurface coordinates $x$. Then $\epsilon$, $P$ and $u^\mu$ will also depend on the spacetime coordinates through $r_\textrm{H}=r_\textrm{H}(x)$. In such a case, when we substitute the spacetime dependent $t^\mu_{\phantom{\mu}\nu}$ in it's equation for radial evolution \eqref{EMTensorEoM-Full}, there will be higher derivative corrections. Thus we need to go to higher orders to find the full solution of the RG flow. The systematic iterative algorithm for solving the RG flow will be presented in the next subsection.

At zeroth order, $r_{\rm{H}}$ or $T$ are interchangeable variables. At higher orders, $T$ will be a fundamental thermodynamic variable, and we will eliminate $r_{\rm{H}}$ in favor of $T$. Nevertheless, the equation of state given by $\epsilon(T,r)$ and $P(T,r)$ will remain unchanged to all orders. We illustrate our approach in Figure \ref{Thermodynamics-Figure}.

There is no reason to believe that the $r_\textrm{H} = r$ hypersurface is the event horizon of the full non-equilibrium geometry, but it does coincide with the event horizon at late time, when the geometry settles down to the static brane with constant $u^\mu$ and $T$.

Before going to higher orders, it is useful to check the consistency of our construction of thermodynamic variables at each hypersurface of the foliation. We note that once $\epsilon$, $P$ and $u^\mu$ become functions of the hypersurface coordinates, at the leading order they follow Euler equations given by \eqref{Euler-T}, or equivalently by \eqref{Euler-s}. If we take the radial derivatives of the Euler equations in the form of \eqref{Euler-s}, we readily get:
\begin{eqnarray}
\label{s,c_s-prime}
s^\prime &=& -\frac{r^{d-1}}{2}\,\epsilon \,s, \nonumber\\
\left( c_s^2 \right)^\prime  &=& \frac{r^{d-1}}{2}(\epsilon+P) \Big( 2 c_s^2 + \frac{d-2}{d-1}\Big).
\end{eqnarray}
We can check that $s$ and $c_s^2$ as given by \eqref{s} and \eqref{SpeedOfSound} indeed satisfy the above identities.

As we have mentioned earlier, since the equations of fluid mechanics follow from the constraints of Einstein's equations, they are automatically satisfied for all hypersurfaces provided they hold at the initial hypersurface. The radial data evolution given by the dynamical equation \eqref{EMTensorEoM-Full} preserves these constraints. Thus it is not necessary to check that the radial evolution of the equations of fluid mechanics is consistent with the RG flow obtained from \eqref{EMTensorEoM-Full}. Nevertheless, the above check of the consistency of the radial flow of Euler equations with the thermodynamics at each hypersurface shows that there is a unique way of constructing the hypersurface thermodynamic variables, namely $s$, $T$ and $c_s^2$. 

We also note that the total equilibrium entropy does not change along our RG flow. From \eqref{s,c_s-prime}, we find $(\ln \, s)'= -r^{d-1}\epsilon/2$. It is easy to see that the spatial volume flows radially as $(\ln \,\sqrt{g_\perp})' = (1/2) z^\mu_{\phantom{\mu}\nu}
\Delta^\nu_{\phantom{\nu}\mu}$. From \eqref{z-ZerothOrder}, we find that $(\ln \,\sqrt{g_\perp})' = r^{d-1}\epsilon/2$. Thus the rate of change of entropy density is compensated by the rate of change of the spatial volume in static equilibrium, making total entropy invariant along the RG flow. A similar feature was also found in \cite{Bredberg:2010ky}.

\subsection{The iterative algorithm and some simplifications}
\label{algo}

We now present a systematic algorithm for solving for the $3 m^{(n)}_s + m^{(n)}_v + m^{(n)}_t$ parameters in our holographic fluid RG flow Ansatz given by \eqref{Ansatz-uT} order by order in the derivative expansion. We will also see that determining these variables will completely exhaust the norm condition $u^\mu g_{\mu\nu} u^\nu = -1$ and  Einstein's equations, or equivalently the equation for radial evolution of $t^\mu_{\phantom{\mu}\nu}$, which up to second order in derivatives is given explicitly by \eqref{EMTensorEoM-Full}. 

Let us assume we have solved the RG flow up to $(n-1)$-th order, meaning that we have found all the transport coefficients $\gamma^{(m)}_{i}$ and
$\delta^{(m)}_{i}$ from \eqref{HydrodynamicEMTensor}, as well as the reparametrization parameters $\alpha^{(m)}_{i}$, $\beta^{(m)}_{i}$ and $\lambda^{(m)}_{i}$ from \eqref{Ansatz-uT} for all $m<n$. In the previous subsection we have solved the system at the zeroth order, so with an iterative algorithm at hand one will be able to compute these coefficients for any $n$.

In subsection \ref{MetricElimination} we demonstrated how one can eliminate $z^\mu_{\phantom{\mu}\nu}$, and therefore the metric, from the calculations. As an example we have calculated $\left( \nabla \cdot u \right)^\prime$ up to the first order in the derivative expansion. It is worth emphasizing again that the $r$-derivative does not commute with the $d$-dimensional covariant derivative, since the latter is built form an $r$-dependent induced metric.

Notice that in our Ansatz \eqref{Ansatz-uT} as the $r$-derivatives of the zeroth order functions $u^\mu(r,x)$ and $T(r,x)$ have infinite expansion in terms of the space-time derivatives. In order to calculate the first order term in $\left( \nabla \cdot u \right)^\prime$, we needed only the zeroth order coefficients, $\epsilon$, $P$ and $\alpha_0$. From our derivation it is clear that the second order contribution will be spanned by $\mathcal{S}^{(2)}_i$'s from \eqref{ListOfSecondOrderScalars} and this time it will require also the first order coefficients $\gamma^{(1)}_{i}=-\eta$,  $\delta^{(1)}_{i}=-\zeta$ from \eqref{HydrodynamicEMTensor} together with $\alpha^{(1)}_{i}$, $\beta^{(1)}_{i}$, $\lambda^{(1)}_{i}$. More generally, solving the RG flow up to the $m$-th order provides enough information for expanding $\mathcal{S}^{(k)}_i$ up to the order $(k+m)$ in the derivative expansion.

The same observation holds also for the $r$-derivatives of the vectors ${\mathcal{V}^{(k)}_i}^\mu$ , although this time the higher term expansion might include scalar terms of the form $\mathcal{S}^{(l)} u^\mu$ with $l > k$. Similarly, acting with $\partial/\partial r$ on the tensors ${\mathcal{T}^{(k)}_i}_{\mu\nu}$ we can \emph{a priori} get terms of the form $u_\mu {\mathcal{V}^{(l)}_i}_\nu + u_\nu {\mathcal{V}^{(l)}_i}_\mu$,  $\mathcal{S}^{(l)} u_\mu u_\nu$
and $\mathcal{S}^{(l)} \Delta_{\mu\nu}$.

Note, however, that at the leading order $\left( {\mathcal{V}^{(k)}_i}^\mu \right)^\prime$ is still orthogonal to $u^\mu$, because at the zeroth order $u_\mu$ is parallel to itself, ${u_\mu}^\prime = -\alpha_0 u_\mu + \ldots$.\footnote{Indeed $\left( {\mathcal{V}^{(k)}_i}^\mu \right)^\prime u_\mu = \left( {\mathcal{V}^{(k)}_i}^\mu  u_\mu \right)^\prime - {\mathcal{V}^{(k)}_i}^\mu u_\mu^\prime$ and both terms on the right hand side are zero.} This implies that:
\begin{equation}
\label{Vector_n-prime}
\left({{\mathcal{V}_i^{(k)}}^\mu} \right)^\prime = \sum_{ij}^{m^{(k)}_v} B_{ij}^{(k)} {\mathcal{V}_i^{(k)}}^\mu + \mathcal{O}\left( \nabla^{k+1}\right)  \, .
\end{equation}
Remarkably, we can achieve a similar simplification for the tensors ${\mathcal{T}^{(n)}_i}^\mu_{\phantom{\mu}\nu}$ just by keeping one of the their indices lowered and another raised. 
If we take a radial derivative of such a tensor, it will remain traceless to all orders in the derivative expansion, ${{\mathcal{T}^{(n)}}^\mu_{\phantom{\mu}\nu}}^\prime
\delta_{\,\mu}^\nu = 0$. This can be easily seen from the fact that as $\delta_{\,\mu}^{\nu}$ is a constant, contraction with $\delta_{\,\mu}^{\nu}$ commutes with the radial derivative operation, which is definitely not true for $g_{\mu\nu}$ and $g^{\mu\nu}$. Precisely as ${{\mathcal{V}_i^{(k)}}^\mu}^\prime$ the tensor ${{\mathcal{T}^{(k)}_i}^\mu_{\phantom{\mu}\nu}}^\prime$ is also orthogonal to $u^\mu$ at the leading $k$th order. Therefore:
\begin{equation}
\label{Tensor_n-prime}
\left( {{\mathcal{T}_i^{(k)}}^\mu_{\phantom{\mu}\nu}} \right)^\prime = \sum_{j=1}^{m^{(k)}_t} L_{ij}^{(k)} {{\mathcal{T}_j^{(k)}}^\mu_{\phantom{\mu}\nu}} + \mathcal{O}\left( \nabla^{k+1}\right) \, .
\end{equation}
Again, the higher order terms are still traceless, but not necessarily orthogonal to $u^\mu$. They will be of the form $u^\mu {\mathcal{V}^{(l)}_i}_\nu + u_\nu {\mathcal{V}^{(l)}_i}^\mu$ or $\mathcal{S}^{(l)} \left( (d-1) u^\mu u_\nu + \Delta^\mu_{\phantom{\mu} \nu} \right)$.

There is yet another issue we need to draw our attention to. Despite the fact that $\epsilon$ and $P$ are zero order quantities their radial derivatives contribute to the RG flow at higher orders via the radial derivative of $T$. Indeed, by construction $\epsilon$ and $P$ depend on two variables, the radial coordinate $r$ and the temperature $T$, and this dependence is fixed by the equation of state we have derived in the previous subsection.  More explicitly, we can use \eqref{e(r)-and-P(r)} and \eqref{T}, eliminate $r_\textrm{H}$ and construct $\epsilon(T,r)$ and $P(T,r)$. Note that these relations are fixed by thermodynamics once and for all, and do not change at higher orders by the method of phenomenological construction of the hydrodynamic equations. Therefore:
\begin{equation}
\label{de/dr-dP/dr}
\frac{\textrm{d} \epsilon}{\textrm{d} r} = \frac{\partial \epsilon}{\partial r} + \frac{\partial \epsilon}{\partial T} T^\prime
\qquad \textrm{and} \qquad
\frac{\textrm{d} P}{\textrm{d} r} = \frac{\partial P}{\partial r} + c_s^2(T,r) \frac{\partial \epsilon}{\partial T} T^\prime \, ,
\end{equation}
where in the second identity we used the thermodynamics definition of $c_s^2$ that also holds at any order in the derivative expansion and can be obtained from \eqref{T} and \eqref{SpeedOfSound} after eliminating $r_\textrm{H}$. The partial $r$-derivatives in (\ref{de/dr-dP/dr}) are zero order quantities, therefore higher order contributions come only through $T^\prime$ via our Ansatz \eqref{Ansatz-uT}.  Therefore, for $n \geqslant 1$:
\begin{equation}
(\epsilon^{\prime})^{(n)} = \frac{\partial \epsilon}{\partial T} (T^\prime)^{(n)} \, , \qquad (P^{\prime})^{(n)} = c_s^2 \frac{\partial \epsilon}{\partial T} (T^\prime)^{(n)}.
\end{equation}

There is a similar issue with the transport coefficients $\gamma^{(m)}_i$ and $\delta^{(m)}_i$. At each value of $r$ they are definite functions of temperature. To be more precise, solving their ODEs for radial evolution, we would obtain them as functions of $r_\textrm{H}$ and $r$. Eliminating $r_\textrm{H}$ in favour of $T$, we arrive at $\gamma^{(m)}_i=\gamma^{(m)}_i(T,r)$ and $\delta^{(m)}_i=\delta^{(m)}_i(T,r)$. These dependences will not alter at higher order, exactly as in the case of the equation of state. Therefore, analogously to $\epsilon(T,r)$ and $P(T,r)$, for $k > m$:
\begin{equation}
   \left( {\gamma^{(m)}_i}^\prime \right)^{(k)} = \frac{\partial \gamma^{(m)}_i}{\partial T} (T^\prime)^{(k)} 
 \qquad
   \left( {\delta^{(m)}_i}^\prime \right)^{(k)} = \frac{\partial \delta^{(m)}_i}{\partial T} (T^\prime)^{(k)} 
\end{equation}

Let us now show in details how starting from the norm condition and Einstein's equation one can derive the first order ODEs for $\gamma^{(n)}_i$ and $\delta^{(n)}_i$ and algebraic equations for $\alpha^{(n)}_i$, $\beta^{(n)}_i$ and $\lambda^{(n)}_i$ assuming that we have already determined the lower order coefficients.

We will start with the tensor transport coefficients. For this goal we consider the spatial traceless component of the radial evolution equation of $t^\mu_{\phantom{\mu}\nu}$. The corresponding projection is defined by the $\langle \rangle$ operator \eqref{brackets}. Projecting Einstein's equation $E^\mu_{\phantom{\mu}\nu}$ \eqref{EMTensorEoM-Full} we find that up to the second order in derivatives:
\begin{eqnarray}
\label{EMTensorEoM-Tensor}
\langle{t^{ \mu}_{\phantom{\mu}\nu} }^\prime\rangle - \frac{2 r^{2-d}}{d-2} \langle {R^{\mu }_{\phantom{\mu}\nu }}^\prime\rangle -    \frac{r^{d-1}}{2(d-1)} \textrm{Tr} t 
   \left( \langle t^{ \mu }_{\phantom{\mu}\nu } \rangle- \frac{2 r^{2-d}}{d-2}\langle R^{\mu}_{\phantom{\mu}\nu } \rangle \right) = 0.
\end{eqnarray}
Here we dropped third and higher order curvature terms, since we will not consider them in this paper. Our iterative algorithm, though, still goes through with these terms.

Next, for a fixed $i=1, \dots, m^{(n)}_t$, we consider the coefficients of $\mathcal{T}^{(n)\mu}_{i\phantom{(n)\mu}\nu}$ coming from various terms in \eqref{EMTensorEoM-Tensor}. At the $k$-th order:
\begin{equation}
 \langle t^\mu_{\phantom{\mu}\nu}\rangle^{(k)}= \sum_{i= 1}^{m^{(k)}_t}\gamma^{(k)}_i
{\mathcal{T}^{(k)}_i}^\mu_{\phantom{\mu}\nu} \, .
\end{equation}
Let us study separately the contributions of each of the four terms in  \eqref{EMTensorEoM-Tensor}.
\begin{itemize}
 \item
 From our discussion it follows that $\langle{t^{\mu}_{\phantom{\mu}\nu} }^\prime\rangle$ has two different contributions to the ${\mathcal{T}^{(n)}_i}^\mu_{\phantom{\mu}\nu}$ term. The first one
 is ${\gamma^{(m)}_i}^\prime$ and the second comes from the derivative expansion of $\left( {{\mathcal{T}^{(m)}_i}^\mu_{\phantom{\mu}\nu}} \right)^\prime$ for $m \leqslant n$. For example, the $r$-derivative of the shear viscosity term $-\eta \sigma^\mu_{\,\,\nu}$ will contribute to the ODEs of the second order transport coefficients. It is important to remember that both contributions depend on $\delta^{(k)}_i$, $\alpha^{(k)}_i$, $\beta^{(k)}_i$ or $\lambda^{(k)}_i$ only for $k<n$.   
 \item
 The $\langle {R^{\mu }_{\phantom{\mu}\nu }}^\prime\rangle$ terms contributes only at the second and higher orders. This contribution will involve only $k \leqslant n-2$ order RG parameters.
 \item The third term in \eqref{EMTensorEoM-Tensor} gives a ${\gamma^{(n)}_i}$ contribution for any $i$. 
 \item The last term in \eqref{EMTensorEoM-Tensor} provides a source term only for the ${\gamma^{(2)}_1}$ equation.
\end{itemize}
To summarize, at the $n$th order we obtain overall $m^{(n)}_t$ first order differential equations for each $\gamma^{(n)}_i$ with $i=1,\ldots,m^{(n)}_t$. Again, these equations will explicitly depend on the lower order coefficients, but nevertheless will completely decouple from the $n$th order coefficients $\delta^{(n)}_i$, $\alpha^{(n)}_i$, $\beta^{(n)}_i$ and $\lambda^{(n)}_i$.

We now proceed to the field reparametrization variables of $u^\mu$ multiplying transverse vector like terms in \eqref{Ansatz-uT}, namely $\beta^{(n)}_i$'s. We can determine these paramedters by taking the vector projection of the equation of motion $E^\mu_{\phantom{\mu}\nu}$ in \eqref{EMTensor-z}, \emph{i.e.} by considering $\Delta^{\mu\beta}u_\alpha E^\alpha_{\phantom{\alpha}\beta}$, and extracting $n$-th order terms. Up to the second order the relevant equation is:
\begin{equation}
\label{vfste}
\Delta^{\mu\beta} u_\alpha \left({t^\alpha_{\,\, \beta}}^\prime - \frac{2 r^{2-d}}{d-2} {R^\alpha_{\,\, \beta}}^\prime +
 \dfrac{r}{(d-1)(d-2)}  \textrm{Tr} t R^\alpha_{\,\, \beta} \right) = 0 \, .
\end{equation}  
Similar to the tensor analysis above we are interested here in the overall coefficient of ${\mathcal{V}^{(n)}_i}^\mu$ for fixed $n$ and $i=1, \ldots, m^{(n)}_v$. We will show that this projection leads to an algebraic equation for $\beta^{(n)}_{i}$. First, the contraction of 
$\Delta^{\mu\beta} u_\alpha$ with the two curvature terms in \eqref{vfste} produces only terms depending on $\gamma^{(m)}_i$, $\delta^{(m)}_i$, $\alpha^{(m)}_i$, $\beta^{(m)}_i$ and $\lambda^{(m)}_i$ with $m \leqslant n-2$. Secondly, the contraction with the first term involves relevant $\beta^{(n)}_i$ terms:
\begin{equation}
\left( \Delta^{\mu\beta}u_\alpha {t^\alpha_{\,\, \beta}}^\prime \right)^{(n)} = - (\epsilon+P) \sum_{i=1}^{m^{(n)}_v}
\beta^{(n)}_i \mathcal{V}^{(n)\mu}_{i} + \textrm{terms with $\alpha^{(m)}_i$, $\beta^{(m)}_i$, $\lambda^{(m)}_i$, $\gamma^{(m)}_i$ and $\delta^{(m)}_i$} \quad \text{for all $m<n$} \, .
\end{equation}
Here we used the Landau-Lifshitz form of the energy-momentum tensor \eqref{HydrodynamicEMTensor} and \eqref{Tensor_n-prime}. The $\beta^{(n)}_i$ term comes solely from the equilibrium part of the energy momentum tensor \eqref{equlibrium}, while all the other contributions follow from the $\mathcal{O}\left( \nabla^{k+1}\right)$ terms in \eqref{Tensor_n-prime}.

We see that \eqref{vfste} naturally leads to $m^{(n)}_v$ algebraic equations expressing $\beta^{(n)}_i$ with $i=1, \ldots, m^{(n)}_v$ in terms of $\alpha^{(l)}_i$'s, $\beta^{(l)}_i$'s, $\lambda^{(l)}_i$'s, $\gamma^{(l)}_i$'s and $\delta^{(l)}_i$'s with $l<n$.

In the final step of the algorithm, we will determine the scalar transport coefficients $\delta^{(n)}_i$'s and the auxiliary reparametrization variables $\alpha^{(n)}_i$'s and $\lambda^{(n)}_i$'s multiplying scalar like terms. 

For our convenience, let us define the non-equilibrium trace part of $t_{\mu\nu}$ as $\mathcal{S}$, so that
\begin{equation}
\mathcal{S} =\sum_{n=1}^\infty \sum_{i=1}^{n_s}\delta^{(n)}_i \mathcal{S}^{(n)i}.
\end{equation} 
Then
\begin{equation}
\label{Trace-t}
\textrm{Tr} t = - \epsilon + (d-1) \left( P + \mathcal{S} \right) \, .
\end{equation}
We also note that the Landau-Lifshitz form of the energy-momentum tensor gives:
\begin{equation}
\label{epsilon-prime}
 {t^{\mu}_{\,\, \nu}}^\prime u_\mu  u^\nu  = 
 \left(  {t^{\mu}_{\,\, \nu}} u^\mu u_\nu \right)^\prime - t^{\mu}_{\,\, \nu} \left( u_\mu^\prime u^\nu + u_\mu {u^\nu}^\prime  \right) = 
 \epsilon^\prime + \epsilon \left( u_\mu^\prime u^\mu + u_\mu {u^\mu}^\prime  \right) = 
 \epsilon^\prime \, ,
\end{equation}
where in the last identity we used the norm condition.

We can now take the projection of $E^\mu_{\phantom{\mu}\nu}$, the equation for radial evolution of $t^\mu_{\phantom{\mu}\nu}$ with $u_\mu u^\nu$ and $\delta_\mu^\nu$ to obtain two independent scalar equations. Nest we need to calculate the coefficients of $\mathcal{S}^{(n)}_i$ in these two scalar equations.

Due to \eqref{Trace-t} and \eqref{epsilon-prime}, among the (leading) $n$-th order RG flow parameters, only $\lambda^{(n)}_i$'s and $\delta^{(n)}_i$'s appear in these scalar equations which thus decouple from $\alpha^{(n)}_i$'s, $\beta^{(n)}_i$'s and $\gamma^{(n)}_i$'s. To see that, notice the only possible place of appearance is $\left( \textrm{Tr} t\right)^\prime$ and ${t^{\mu}_{\,\, \nu}}^\prime u_\mu  u^\nu$. Again, one needs to determine all these RG flow parameters at lower orders to define the scalar equations at the $n$-th order. Also it is easy to see that $\lambda_{(n)i}^\prime$'s never appear in  the equations simply because $T^{\prime\prime}$ never appears.

These scalar equations involve $\lambda^{(n)}_i$'s only through $T^\prime$ (which, in turn, come from $\epsilon^\prime$ and $P^\prime$) and $\delta^{(n)}_i$'s through $\mathcal{S}$. We can eliminate $T^\prime$ from the two scalar equations to obtain the equation for $\mathcal{S}$ only which will determine the first order differential equations for $\delta^{(n)}_i$'s. This way we decouple the $\delta^{(n)}_i$'s from the $\lambda^{(n)}_i$'s. The suitable linear combination of the scalar equations which determine the scalar transport coefficients $\delta_{(n)i}$'s up to the second order is:
\begin{eqnarray}
\label{EMTensorEoM-Scalar1}
&&
\mathcal{S}^\prime + 
\dfrac{\left( c_s ^2 (d-1) + (d-3) \right)}{(d-1)(d-2)}  r^{2-d} R^\prime + 
\frac{2 r^{2-d}}{(d-1)(d-2)} \left( c_s ^2 (d-1) - 1 \right) \left( u_\mu {R^{\mu}_{\,\, \nu}} u^\nu \right)^\prime =
\nonumber \\
&& \qquad
 = \left( \dfrac{d-1}{r} (c_s^2+1) + r^{d-1} \left( P - \dfrac{1}{2} \left( c_s^2 + \dfrac{1}{d-1} \right) \epsilon \right)\right) \, \mathcal{S} +
\\
&& \qquad \qquad
 + \dfrac{1}{2(d-2)}\left( \left( c_s^2 + \dfrac{2d-5}{d-1}\right) P - \left( c_s^2 + \dfrac{d-3}{(d-1)^2} \right) \epsilon\right) r R +
\nonumber \\
&& \qquad \qquad \,
+ \dfrac{( c_s ^2 (d-1) - 1)}{(d-1)(d-2)}\left( P - \dfrac{ \epsilon}{(d-1)} \right) \left( u_\mu {R^{\mu}_{\,\, \nu}} u^\nu \right) \, .
\nonumber 
\end{eqnarray}  

Once the $\delta^{(n)}_i$'s are determined, we can use the second scalar equation obtained by contraction with $\delta_\mu^\nu$ below (showing terms relevant only up to second order) to derive the algebraic equation which determines $\lambda^{(n)}_i$'s:
\begin{equation}
\label{EMTensorEoM-Scalar2}
(\textrm{Tr}\, t)^\prime + r^{2-d} R^\prime - \dfrac{d}{r} \textrm{Tr}\, t - \dfrac{r^{d-1}}{2(d-1)} \left( (\textrm{Tr}\, t)^2 +2 r^{2-d} (\textrm{Tr}\, t)R \right) = 0 \, .
\end{equation}
We thus have exactly $2 m^{(n)}_s$ equations to determine the $2 m^{(n)}_s$ variables $\lambda_{(n)i}$'s and $\delta_{(n)i}$'s.

The $\alpha_{(n)i}$'s can now be determined algebraically from the norm condition obtained by radially differentiating $u^\mu u^\nu g_{\mu\nu} = -1$, which yields $2u^{\mu\prime}u_\mu + u^\mu u^\nu z_{\mu\nu} = 0$. We also need to substitute the form of $z^\mu_{\phantom{\mu}\nu}$ in terms of $t^\mu_{\phantom{\mu}\nu}$, polynomials of Riemann curvature and their covariant derivatives. Up to the second order in the derivative expansion, the relevant equation is:
\begin{equation}
\label{NormCondition}
2u^{\mu\prime}u_\mu = -r^{d-1}\Bigg(\epsilon + \frac{\text{Tr} \, t}{d-1}  \Bigg) +  \frac{2r}{d-2} \Bigg(R^{\mu}_{\phantom{\mu}\nu} u_\mu u^\nu + \frac{R}{2(d-1)} \Bigg).
\end{equation}
Clearly $\alpha^{(n)}_i$'s appear only via $u^{\mu\prime}u_\mu$ and we thus get exactly $m^{{n}}_s$ equations to determine the $m^{{n}}_s$ number of $\alpha^{(n)}_i$'s. We can readily see that in order to determine them, we need to solve for the scalar transport coefficients $\delta^{(n)}_i$'s which appear via $\text{Tr}\, t$. Again, $\alpha^{(n)}_i$'s decouple from all other RG flow parameters at $n$-th order.

Thus we can determine all the parameters in the RG flow Ansatz given by \eqref{Ansatz-uT} at the $n$-th order.
To summarize:
\begin{itemize}
\item The equations giving radial evolution of the tensor transport coefficients $\gamma^{(n)}_i$'s and scalar transport coefficients $\delta^{(n)}_i$'s decouple from each other and also from the other parameters of the RG flow, namely the auxiliary variables $\alpha^{(n)}_i$'s, $\beta^{(n)}_i$'s and $\lambda^{(n)}_i$'s. 

\item The algebraic equations determining the auxiliary variables $\beta^{(n)}_i$'s multiplying vector like terms in ${u^\mu}^\prime$ decouple from the auxiliary variables $\alpha^{(n)}_i$'s and $\lambda^{(n)}_i$'s multiplying scalar like terms. The algebraic equations for $\beta^{(n)}_i$'s decouple from all other $n$-th order RG flow parameters. 

\item The algebraic equations determining the auxiliary variables $\alpha^{(n)}_i$'s and $\lambda^{(n)}_i$'s multiplying scalar like terms in ${u^\mu}^\prime$ and $T^\prime$ respectively decouple from each other and involve only the scalar transport coefficients $\delta^{(n)}_i$'s.

\item To solve the RG flow for radial evolution of the $n$-th order RG flow parameters, one needs to solve it completely at lower orders.

\item At each order, we have exactly as many equations as there are parameters in the RG flow which need to be solved. Thus determining the RG flow exhaust the norm condition and the full dynamical content of Einstein's equations.
\end{itemize}

\subsection{Reconstruction of the metric}

So far, we have presented the algorithm for construction of the holographic RG flow in the fluid/gravity limit which requires no explicit knowledge of the bulk spacetime metric. We have also seen that solving for the RG flow exhausts the full dynamical content of Einstein's equations. Therefore, it is intuitively clear, that solving the RG flow should also be equivalent to solving for the explicit spacetime metric order by order in the derivative expansion.

Nevertheless, the RG flow equations are first order in radial derivatives, while Einstein's equations are second order in radial derivatives. Therefore, to determine the metric we need additional integration constants - these are nothing but the boundary metric ${g_\textrm{b}}_{\mu\nu}$ and the boundary velocity and temperature fields, namely $u_\textrm{b}^\mu$ and $T_\textrm{b}$, amounting to a specific choice of solutions of the equations of fluid mechanics at the boundary.

We recall that though the metric in Eddington-Finkelstein metric and Fefferman-Graham coordinates are gauge equivalent at each order in derivative expansion, they differ at orders higher than the order to which the bulk spacetime metric is explicitly solved. In the former coordinate system, the metric is manifestly regular even at higher orders, but is not asymptotically AdS. In the latter coordinate system, the metric is not regular at higher orders even if the transport coefficients are chosen correctly up to the given order, but is asymptotically AdS even at higher orders.

In the following sections, we will show that RG flow equations are sufficient to ensure regularity of the solution at the late time (future) horizon. It is therefore useful to obtain the bulk spacetime metric in a form which preserves the asymptotic AdS boundary condition. Therefore, to demonstrate that the bulk spacetime metric can be reconstructed from the RG flow, we will reconstruct the bulk metric in Fefferman-Graham coordinates which preserve the asymptotic AdS boundary condition to all orders manifestly.

The explicit method is as follows. First, we have to make an Ansatz for the expressions of $u^\mu(x,r)$, $T(x,r)$ and $g_{\mu\nu}(r,x)$, which appears in the bulk Fefferman-Graham metric as given in \eqref{ads},  in terms  the boundary metric ${g_\textrm{b}}_{\mu\nu}(x)$ and the boundary fluid variables, ${u_\textrm{b}}^\mu (x)$ and $T_\textrm{b}(x)$. For the induced metric we have:
\begin{eqnarray}
\label{BulkAnsatz-metric}
g_{\mu\nu}(r,x) &=& A \cdot {g_\textrm{b}}_{\mu\nu} + B \cdot {u_\textrm{b}}_\mu {u_\textrm{b}}_\nu  
	+ \sum_{n=1}^{\infty} \left( T_\textrm{b} \right)^n 
		\Bigg( \sum_{i=1}^{m^{(n)}_t} C_i^{(n)} {\mathcal{T}_\textrm{b}}^{(n)}_{\mu\nu} + 
\\
&& \qquad \qquad		
		  + \sum_{i=1}^{m^{(n)}_v} E_i^{(n)}  \left( {u_\textrm{b}}_\mu {{{\mathcal{V}_\textrm{b}}}_i^{(n)}}_\nu  + 
		                               {u_\textrm{b}}_\nu {{\mathcal{V}_\textrm{b}}_i^{(n)}}_\mu \right)  +
				  \sum_{i=1}^{m^{(n)}_s} \left( D_i^{(n)} {\Delta_\textrm{b}}_{\mu\nu}   + 
		                               \widetilde{D}_i^{(n)} {u_\textrm{b}}_\mu {u_\textrm{b}}_\nu \right)  {\mathcal{S}_\textrm{b}}^{(n)} 
	\Bigg) \, .
\nonumber 	 
\end{eqnarray}
Let us explain the notations in this formula. The traceless transverse $n$-th order tensors ${\mathcal{T}_\textrm{b}}^{(n)}_{\mu\nu}$, the transverse vectors ${\mathcal{V}_\textrm{b}}^{(n)}_\mu$ and the scalars ${\mathcal{S}_\textrm{b}}^{(n)}$ are all built from the boundary variables ${g_\textrm{b}}_{\mu\nu}(x)$, $u_\textrm{b}^\mu (x)$ and $T_\textrm{b} (x)$. Accordingly ${\Delta_\textrm{b}}_{\mu\nu} = {g_\textrm{b}}_{\mu\nu} + {u_\textrm{b}}_\mu {u_\textrm{b}}_\nu$.
The temperature factor $\left( T_\textrm{b} \right)^n$ is introduced in order to keep the parameters $C_i^{(n)}$, $E_i^{(n)}$, $D_i^{(n)}$ and $\widetilde{D}_i^{(n)}$ dimensionless. It also follows from the dimensional analysis that all of them depend on the dimensionless combination $r T_\textrm{b} (x)$:
\begin{eqnarray}
&& \qquad \qquad \qquad \qquad
A = A \left( r T_\textrm{b} (x) \right) \, , \quad
B = B \left( r T_\textrm{b} (x) \right) \, , 
\\
&&
C_i^{(n)} = C_i^{(n)} \left( r T_\textrm{b} (x) \right) \, , \quad
E_i^{(n)} = E_i^{(n)} \left( r T_\textrm{b} (x) \right) \, , \quad
D_i^{(n)} = D_i^{(n)} \left( r T_\textrm{b} (x) \right) \, , \quad
\widetilde{D}_i^{(n)} = \widetilde{D}_i^{(n)} \left( r T_\textrm{b} (x) \right) \, .
\nonumber
\end{eqnarray}
This metric \eqref{BulkAnsatz-metric} is the most general 2-tensor one can construct out of ${g_\textrm{b}}_{\mu\nu}(x)$, $u_\textrm{b}^\mu (x)$ and $T_\textrm{b} (x)$ (see our discussion shortly after \eqref{Euler-s}). The  $C_i^{(n)}$ and $D_i^{(n)}$
terms are analogous to the  $\gamma_i^{(n)}$ and $\delta_i^{(n)}$ terms in the hydrodynamic energy-momentum tensor \eqref{HydrodynamicEMTensor}, while the $E_i^{(n)}$ and $\widetilde{D}_i^{(n)}$ terms do not appear in \eqref{HydrodynamicEMTensor} due to the Landau-Lifshitz gauge.

It is straightforward to write the Ansatz for ${u_\textrm{b}}^\mu(x)$ and $T_{\textrm{b}}(x)$:
\begin{eqnarray}
\label{BulkAnsatz-uT}
u^\mu(x,r) &=& a_0 {u_\textrm{b}}^\mu + \sum_{n=1}^{\infty}   T_\textrm{b}^n  \left( \sum_{i=1}^{m^{(n)}_s} a^{(n)}_i  {\mathcal{S}_\textrm{b}}^{(n)}_i {u_\textrm{b}}^\mu + \sum_{i=1}^{m^{(n)}_v} b^{(n)}_i {{\mathcal{V}_\textrm{b}}^{(n)}_i}^\mu \right) \, ,
\nonumber\\
T(x,r) &=& l_0 T_\textrm{b} + \sum_{n=1}^{\infty} T_\textrm{b}^n  \sum_{i=1}^{m^{(n)}_s} l^{(n)}_i {\mathcal{S}_\textrm{b}}^{(n)}_i  \, .
\end{eqnarray}
Again, all the parameters are dimensionless and depend on the combination $r T_\textrm{b}$ only.

At the next step, using \eqref{BulkAnsatz-metric} and 
\eqref{BulkAnsatz-uT} one has to express $\mathcal{S}^{(n)}_i$, ${\mathcal{V}_i^{(n)}}_\mu$ and ${\mathcal{T}_i^{(n)}}_{\mu\nu}$ in terms of their boundary counterparts ${\mathcal{S}_\textrm{b}}^{(n)}_i$, ${{\mathcal{V}_\textrm{b}}_i}^{(n)}_\mu$ and ${{\mathcal{T}_\textrm{b}}_i}^{(n)}_{\mu\nu}$. This should be done systematically in the derivative expansion parameter at the boundary. Obviously, the final expression will be covariant with respect to the boundary metric ${g_\textrm{b}}_{\mu\nu}$. These algebraic manipulations are straightforward and (up to the first order in derivatives) are exactly as those in our previous work \cite{Kuperstein:2011fn}. For example, one finds that:
\begin{equation}
 \sigma_{\mu\nu} = \dfrac{A}{( A-B )^{1/2}} {\sigma_\textrm{b}}_{\mu\nu} + \mathcal{O} \left( \nabla^2 \right) \, .
\end{equation}

Once the above step is completed, we can substitute \eqref{BulkAnsatz-metric} and \eqref{BulkAnsatz-uT} in  \eqref{Ansatz-uT} and the expression for $z^\mu_{\phantom{\mu}\nu}= g^{\mu\alpha}g_{\alpha\nu}^\prime$ as given by \eqref{z}. Knowing the reparametrization parameters in \eqref{Ansatz-uT} and the transport coefficients in \eqref{HydrodynamicEMTensor} we may obtain a full set of first order differential equations for all the parameters in \eqref{BulkAnsatz-metric} and \eqref{BulkAnsatz-uT}. It is straightforward to check that we have the same number of equations as the number of the functions. Contrary to the $\alpha$, $\beta$ and $\lambda$-parameters in the Ansatz \eqref{Ansatz-uT}, now the equations for $a^{(n)}_i$,  $b^{(n)}_i$ and  $l^{(n)}_i$ will be differential and not algebraic.

Solving these differential equations, we also need to satisfy the boundary conditions at $r=0$. By construction we need:
\begin{equation}
A (0)=1 \, , \qquad
a_0 (0)=1 \, , \qquad
l_0 (0)=1 \, 
\end{equation}
with all other parameters vanishing for at the boundary. Notice that the norm condition of $u_\textrm{b}^\mu$, namely $u_\textrm{b}^\mu {g_\textrm{b}}_{\mu\nu}u_\textrm{b}^\nu=-1$ then guaranteed by the norm condition we have already imposed on $u^\mu$ while solving the RG flow.

One can wonder whether these UV boundary conditions will determine the functions uniquely, as the corresponding differential equations may appear singular at $r=0$. Let us argue that this cannot be the case. We note that the first order RG flow equations already determine the transport coefficients at the boundary from the constraints on their near horizon forms which will be discussed in the following section. Having the transport coefficients together with $u_\textrm{b}^\mu$ and $T_\textrm{b}$ completely determines ${t_\textrm{b}}_{\mu\nu}$, the energy-momentum tensor at the boundary. For fixed ${t_\textrm{b}}_{\mu\nu}$ and ${g_\textrm{b}}_{\mu\nu}$ there is a unique solution of Einstein's equations in the bulk. This was shown explicitly in the power expansion near the boundary \cite{Henningson:1998ey}. Furthermore, as fixing the transport coefficients ensures regularity, the metric thus reconstructed from the RG flow will indeed have a regular future horizon up to the given order. This can be readily checked by translating to Eddington-Finkelstein coordinates order by order in the derivative expansion following the method of  \cite{Gupta:2008th}.

We will not further elaborate on these straightforward calculations as the explicit metric does not contain additional physical information which will be relevant for us. Nevertheless, the metric up to first order in Fefferman-Graham coordinates is explicitly known, and is as in \cite{Gupta:2008th}\footnote{Here it was assumed that the boundary metric is flat, however one can readily check that the solution is valid up to first order replacing $\eta_{\mu\nu}$ by the boundary metric in the solution.}. The reader can check via explicit computations that it will follow from our results for the solution of the RG flow which will be discussed in Section 7, that the first order vector and scalar perturbations of the metric will vanish. This was also the case in the explicit solution in \cite{Gupta:2008th}.

There is an alternative algorithm for reconstructing the bulk metric from the RG flow, which we would like to present in a future work. Inspite of this alternative algorithm being simpler, we omit it here because it is not as illustraitive as the one presented above.

\section{The nature of the horizon fluid}
\label{hf}

The parameters of the RG flow are the transport coefficients and the auxiliary variables determining scale dependent reparametrization of the fluid variables. The latter satisfy algebraic equations and do not require any boundary condition for their determination. The physical variables, namely the scalar and tensor transport coefficients, satisfy first order differential equations, whose boundary conditions should be determined from physical principles. We will show that these boundary conditions are simply provided by the requirement that the fluid at the horizon is a well defined non-relativistic fluid.

It is quite clear from our construction that at each hypersurface, the fluid is relativistic. Nevertheless, as we move towards the horizon, the energy density stays finite, while the pressure and the speed of sound blow up. The horizon thus is an end point of the RG flow because the notion of thermodynamics breaks down at the horizon. For proper physical interpretation of the nature of the RG flow near the horizon, we need to define an appropriate scaling of coordinates and variables such that the near-horizon fluid behaves non-relativistically. After this rescaling, the horizon fluid will turn out to be a fixed point of the RG flow. We may anticipate that the dynamics at the fixed point will be governed by non-relativistic incompressible Navier-Stokes equations.

Our strategy will be as follows :
\begin{itemize}
\item We will define a near-horizon non-relativistic scaling that regularizes the otherwise ill-defined equations of motion.
\item We will show that at the leading order in the scaling and up to first order in the derivative expansion, the horizon fluid is a fixed point of the RG flow, provided the shear and the bulk viscosities do not blow up on the horizon. The dynamics at the fixed point will indeed be given by non-relativistic incompressible Navier-Stokes equations.
\item We will demand that higher derivative corrections will not destroy this horizon fixed point, thus we will derive constraints on the near-horizon behaviour of all higher order transport coefficients.
\item We will later claim that these constraints on the near-horizon behavior of transport coefficients lead to a unique solution of the RG flow.
\end{itemize}

In the limit of linearised perturbations about equilibrium, a near-horizon scaling limit was proposed in \cite{Bredberg:2010ky}. It was shown that if the radial and time coordinates are scaled as follows:
\begin{equation}\label{reco}
r_{\rm{H}} - r = \xi \cdot \tilde{r}, \quad t = \frac{\tau}{\xi},
\end{equation}
with $\xi$ as the scaling parameter, then in the limit $\xi \rightarrow 0$, one gets the non-relativistic incompressible Navier-Stokes at the horizon as a fixed point up to first order in derivative expansion. The spatial coordinates are not re-scaled. We will do something similar but also consider the full non-linear fluctuations about equilibrium and go beyond the first order in the derivative expansion.

There is a crucial point here that should not be overlooked. We have already seen that in order to consider non-equilibrium non-linear corrections to the RG flow we need $r_{\rm{H}}$ to be dependent on $\mathbf{x}$ and $t$. Replacing $t$ by $\tau/\xi$ and setting $\xi$ to zero means that we take the long time limit where the fluid approaches equilibrium. Therefore, in this limit the temperature becomes independent of hypersurface coordinates implying that $r_{\rm{H}}$ becomes constant as can be seen from \eqref{s(r)-T(r)}.

As in our construction of the RG flow, we did not need to know the spacetime metric explicitly, we should also not need the explicit form of the metric to define the near-horizon scaling of the RG flow. We will introduce, therefore, the scaling limit of the RG flow in the local inertial frame of a radially infalling observer in the near equilibrium geometry. This infalling observer should also start with zero velocity at infinity with respect to the final static equilibrium configuration before starting to infall. As this observer will be infalling in the near-equilibrium geometry, the derivative expansion should also make sense for this observer throughout the infall, provided she redefines the fluid variables as we have done in the RG flow. Note this observer is fictional - she is not doing any real experiment on the hypothetical holographic fluid at each hypersurface. We are merely using her local frame to define a scaling limit of the RG flow.

We note in order to reproduce non-relativistic limit of a relativistic system, we indeed need to choose a specific frame. In the case of fluid dynamics, the natural choice is the inertial frame co-moving with final static equilibrium configuration. This is why when we consider the scaling of the relativistic RG flow, we also choose the frame of an inertial (or in other words infalling) bulk observer in the near-equilibrium geometry who is co-moving with the final static equilibrium configuration. 

We will use the re-scaled radial and time coordinates as in \eqref{reco} in the frame of the radially infalling observer as described above. We will not re-scale the spatial coordinates. As this observer is infalling, the metric in her frame should be $\eta_{\mu\nu}$ along her trajectory. Also the first derivatives of the metric should vanish along her trajectory.

The fluid velocity field along the trajectory of the observer should be close $u^\mu = (1,0, \dots, 0)$ as she is co-moving with the final static equilibrium configuration. We note this velocity vector has norm $-1$ with respect to the metric $\eta_{\mu\nu}$. Nevertheless, the geometry will not be in exact equilibrium, so there will be small deviations characterized by a spatial velocity vector $v^i$. This spatial vector $v^i$ should be $\mathcal{O}(\xi)$ as we have scaled the time-coordinate like $1/\xi$, while not scaling the spatial coordinates. Thus the fluid velocity field in this frame will take the form:
\begin{equation}\label{reu}
u^\mu(r,\mathbf{x}, t) = \Big(1, \xi v^i \left(r_{\rm{H}} + \xi \tilde{r}, \mathbf{x}, \tau/\xi\right)\Big) + \mathcal{O} \left( \xi^2 \right).
\end{equation}
Essentially this forces non-relativistic dynamics as the above amounts to taking the velocity fields small compared to the speed of light. Furthermore, this also signals that the deviations from static equilibrium are small. To be consistent, we should also scale the fluctuations of the temperature about a static equilibrium.

As the temperature blows up at the horizon along the RG flow as $(r_{\rm{H}}-r)^{-1}$, in order to get a finite limit at the horizon we should re-scale the temperature at static equilibrium as $\xi^{-1}$. We therefore scale the temperature field as follows:
\begin{equation}\label{tre}
T(r, \mathbf{x},t) = \frac{T^{\rm{eq}}(\tilde{r})}{\xi} +   \widetilde{T}\left(r_{\rm{H}} + \xi \tilde{r},\mathbf{x},\tau/\xi\right).
\end{equation}
We also note that $T^{\rm{eq}}$ will be  $1/(2\pi\tilde{r})$ as one can see directly from \eqref{s(r)-T(r)}. As the energy density is a function of $T$ and $\tilde{r}$, the scaling of $T$ implies that the energy density should scale as follows:
\begin{eqnarray}
\epsilon(\tilde{r}, \mathbf{x},t) &=& \epsilon_0 + \xi \cdot \tilde{\epsilon}(\tilde{r},\mathbf{x},\tau/\xi).
\end{eqnarray}
Above $\epsilon_0$ is a constant, namely $2(d-1)/ r_{\rm{H}}^d$, as can be seen from \eqref{e(r)-and-P(r)}.

The fluctuation of pressure about equilibrium in the incompressible non-relativistic limit should not depend on the temperature but rather only on the velocity field. We will show later that this is rightfully so. Thus the scaling of only the equilibrium part of pressure will be consistent with it's temperature dependence, the consistent scaling of the fluctuation will be $\mathcal{O}(\xi)$ instead of $\mathcal{O}(1)$. Since both the equilibrium temperature and the equilibrium pressure will blow up at the horizon like $(r_{\rm{H}}-r)^{-1}$, the equilibrium pressure will be proportional to the equilibrium temperature at the horizon and it should also scale like $1/\xi$. Overall the pressure should behave near the horizon like:
\begin{eqnarray}
P(r, \mathbf{x},t) &=& \frac{\rho_0(\tilde{r})}{\xi} + \xi \cdot  P^{\textrm{non-rel}}(r_{\rm{H}}+ \xi \tilde{r}, \mathbf{x},\tau/\xi).
\end{eqnarray}
Above $\rho_0=2r_{\rm{H}}^{1-d}/\tilde{r}$ (see \eqref{e(r)-and-P(r)}) has no hypersurface coordinate dependence. In fact, $\rho_0$ will become the constant mass density of the fluid at the horizon. This ties up with the fact that the fluid will be incompressible at the horizon, so that mass density cannot depend on hypersurface coordinates. Also, $P^{\textrm{non-rel}}$ will be the pressure of the fluid\footnote{It might seem bizarre that the mass density of the fluid comes from the relativistic pressure, but this is related to the fact that the near-horizon geometry is Rindler space and it is known that the holographic fluid dual to the Rindler space has the bizarre equation of state $\epsilon = 0$ with $P$ non-zero \cite{Bredberg:2011jq}. We note that the non-relativistic mass density is nothing but the $\epsilon + P$ of the relativistic fluid. Therefore, owing to the peculiar thermodynamics of Rindler space, the mass density of the non-relativistic fluid receives contribution from $P$ only.}. 

We also recall that the transport coefficients are functions of $T$ and $r$ - therefore if a tensor transport coefficient $\gamma_i^{(n)}$ and a scalar transport coefficient $\delta_j^{(m)}$ blow up at the horizon like $(r_{\rm{H}}-r)^{-k}$ and $(r_{\rm{H}}-r)^{-l}$ with $k, l >0$ respectively, then their near-horion scaling should be:
\begin{eqnarray}
\label{trre}
\gamma_i^{(n)}(r, \mathbf{x},t) &=& \frac{\gamma_{i\, \rm{H}}^{(n)}(\tilde{r})}{\xi^{k}} + \mathcal{O}(\xi^{k+1}) \, , \nonumber\\
\delta_j^{(m)}(r, \mathbf{x},t) &=& \frac{\delta_{j\, \rm{H}}^{(m)}(\tilde{r})}{\xi^{l}} + \mathcal{O}(\xi^{l+1}) \, .
\end{eqnarray}
We note that $\gamma_{i\, \rm{H}}^{(n)}$ and $\delta_{j\, \rm{H}}^{(m)}$ will be parameters of the fluid and have no dependence on hypersurface coordinates. Also they will constant times $1/\tilde{r}^k$ and $1/\tilde{r}^l$ respectively, the constants of proportionality being functions of $r_{\rm{H}}$.

We can now readily work out the near-horizon scaling limit of the RG flow up to the first order in the derivative expansion. If the shear and bulk viscosities are constant on the horizon, then, indeed, at the leading order we find that the horizon fluid obeys non-relativistic incompressible Navier-Stokes equations. The near horizon scaling limit of the equations $\nabla^\mu t_{\mu\nu} = 0$ projected along $u^\nu$ and $\Delta^\mu_{\phantom{\rho}\nu}$ gives us the following equations:  
\begin{eqnarray}
\label{NS}
&&  \qquad \qquad \qquad \qquad \qquad \qquad
\rho_0\,(\partial_i v_i) + \mathcal{O}(\xi) = 0 \, ,
\\
&&
\xi
\Bigg(\rho_0 (\partial_\tau  + v_j \partial_j )v_i + \partial_i P^{\textrm{non-rel}} - \eta_{\textrm{H}}\partial_j (\partial_i v_j + \partial_j v_i) 
- \left(\zeta_{\textrm{H}} - \frac{\eta_{\textrm{H}}}{d-1}\right)\partial_i(\partial_j v_j)  \Bigg)+\mathcal{O}(\xi^2) = 0 \, .
\nonumber
\end{eqnarray}
One should be cautious while performing the scaling of the RG flow. The incompressibility condition $\partial_i v_i = 0$ should be used only after obtaining the fixed point and not before. This is precisely why we can conclude that the bulk viscosity at the horizon $\zeta_{\rm{H}}$ should be finite in order to have the Navier-Stokes limit.

It is also easy to see that under the transformations $\tau \rightarrow \tau/\xi$ (implying $v_i \rightarrow \xi v_i$) and $\tilde{r}\rightarrow\xi\tilde{r}$ (implying $\rho_0 \rightarrow \rho_0/ \xi$) and $P^{\rm{non-rel}} \rightarrow \xi P^{\rm{non-rel}}$, the non-relativistic incompressible Navier-Stokes equations scales homogeneously. This is precisely how we scale the coordinates and variables in the RG flow. Thus after rescaling the coordinates and the variables, the horizon fluid is indeed a fixed point of the RG flow at leading order in the derivative expansion. 

We note that one can do a non-relativistic scaling to obtain incompressible non-relativistic Navier-Stokes from any relativistic fluid \cite{Fouxon:2008tb, Bhattacharyya:2008kq} (see also \cite{Eling:2009pb}). This non-relativistic scaling is also a symmetry of the incompressible Navier-Stokes equations. However, this is different from our near-horizon scaling and it is not the scaling of any RG flow. Additionally this scaling assumes finite values of transport coefficients at finite temperature. On the other hand our near horizon scaling will be able to constrain the near-horizon forms of all transport coefficients. At the first order in derivatives, this non-relativistic scaling is equivalent to the near-horizon scaling up to some forcing terms in the Navier-Stokes equations \cite{Bredberg:2010ky, Bredberg:2011jq}.

Before moving to higher order in derivative expansion, we note a peculiar feature of incompressible non-relativistic Navier-Stokes equations. The pressure $P^{\textrm{non-rel}}$ is the only additional variable (because $\rho_0$, $\eta_{\textrm{H}}$ and $\zeta_{\textrm{H}}$ are constant parameters), but actually it is not an independent variable in the equation. This is expected as the time derivative of $P^{\textrm{non-rel}}$ does not appear in the equations, while we have an additional constraint on the velocity fields only - namely the incompressibility condition - $\partial_i v_i = 0$. This incompressibility condition can be used to eliminate $P^{\textrm{non-rel}}$ from the equations of motion. It is easy to see that after taking the partial derivative $\partial_i$ of the Navier-Stokes equations and using the incompressibility condition we get:
\begin{equation}\label{pnr}
\partial_j \partial_j P^{\textrm{non-rel}} = - \rho_0 \partial_i\partial_j(v_i v_j) + \eta_{\textrm{H}}\partial_i\partial_j(\partial_i v_j+ \partial_j v_i ) 
\end{equation}
at the leading order. We can use the above equation to eliminate $P^{\textrm{non-rel}}$ from the Navier-Stokes' equations at the cost of getting an integro-differential equation for the velocity fields. Thus the only dynamical variable in non-relativistic incompressible Navier-Stokes  equation is the velocity field $v_i$.

We will demand the fixed point of the RG flow after the scaling limit given by the incompressible Navier-Stokes at the horizon will be unaltered even after considering higher derivative corrections. Let us consider for example the case of the $\gamma_6$ transport coefficient - the corresponding tensor is $\langle \sigma_\mu^{\phantom{\mu}\rho}\sigma_{\rho\nu}\rangle$.
We can readily see that,
\begin{equation}
\langle\sigma_\mu^{\phantom{\mu}\rho}\sigma_{\rho\nu}\rangle
= \xi^2 \delta_\mu^i \delta_\nu^j \sigma_{ik}\sigma_{kj} + \mathcal{O}(\xi^3),
\end{equation} 
where $\sigma_{ij}=(1/2)(\partial_i v_j + \partial_j v_i) - (1/(d-1)) \delta_{ij} (\partial_k v_k)$. If $\gamma_6$ blows up like $(r_{\rm{H}}-r)^{-k}$ at the horizon, then it's contribution to Navier-Stokes equation will be 
\begin{equation}
\xi^{2-k}\gamma_{6\, \rm{H}}\partial_j(\sigma_{ik}\sigma_{kj}) 
\end{equation}
and it's contribution to the incompressibility condition is
\begin{equation}
\xi^{3-k}\gamma_{6\, \rm{H}}(\sigma_{ik}\sigma_{kj}) \partial_i v_j.
\end{equation}
In order to retain the same fixed point at the horizon, the correction to Navier-Stokes equation should be subleading compared to $\mathcal{O}(\xi)$ and the correction to the incompressibility condition should be subleading compared to $\mathcal{O}(1)$. Thus we need $\gamma_6$ to behave weaker than $(r_{\rm{H}}-r)^{-1}$ near the horizon. 

We can similarly constrain the behaviour of other second order transport coefficients near the horizon, except for $\gamma_1$ and $\gamma_2$ which involve the curvatures. The near horizon scaling of $\nabla_{\perp\mu} \ln \, s$ is obtained as follows. We use the leading order equations of motion \eqref{Euler-s} to relate it to $c_s^2 Du^\mu$. Thus, we obtain that at leading order, it scales like $\xi^4$. 
The final result for the constraints on the near-horizon behavior of the second order tensor transport coefficients $\gamma_3$, $\gamma_4$, $\gamma_5$, $\gamma_7$ and $\gamma_8$ are as in the Table \ref{gamma345678-div}.
\begin{table}
\begin{center}
    \begin{tabular}{| c | c | c |}
    \hline
    Transport coefficient & Corresponding tensor & Near-horizon behaviour will be weaker than     \\
    \hline  
 	   
    	$\phantom{\Big]}\gamma_3 \phantom{\Big]}$ & 
    	$\left( \nabla \cdot u\right) \sigma^\mu_{\,\, \nu}$ & 
    	$(r_{\rm{H}}-r)^{-1}$
    \\
    \hline
    	$\phantom{\Big]}\gamma_4 \phantom{\Big]}$ & 
    	$\left< {\nabla_\bot}^\mu {\nabla_\bot}_\nu \ln s \right>$ & 
    	$(r_{\rm{H}}-r)^{-3}$ 
    \\
    \hline
    	$\phantom{\Big]}\gamma_5 \phantom{\Big]}$ & $\left< {\nabla_\bot}^\mu \ln s {\nabla_\bot}_\nu \ln s \right>$ & 
    	$(r_{\rm{H}}-r)^{-7}$   
    	\\
    \hline
    	$\phantom{\Big]}\gamma_6 \phantom{\Big]}$ & 
    	$\left< \sigma^{\mu}_{\,\,\, \tau} \sigma^{\tau}_{\,\,\,\nu}  \right>$ & 
    	$(r_{\rm{H}}-r)^{-1}$
    \\
    \hline
    	$\phantom{\Big]}\gamma_7 \phantom{\Big]}$ & 
    	$\left< \omega^{\mu}_{\,\,\, \tau} \omega^{\tau}_{\,\,\,\nu}  \right>$ & 
    	$(r_{\rm{H}}-r)^{-1}$
    \\
    \hline
    	$\phantom{\Big]}\gamma_8 \phantom{\Big]}$ & 
    	$\left< \sigma^{\mu}_{\,\,\, \tau} \omega^{\tau}_{\,\,\,\nu}  \right>$ & 
    	$(r_{\rm{H}}-r)^{-1}$ 
    \\
    \hline
	\end{tabular}
\end{center}
\caption{ Constraints on near-horizon behaviour of $\gamma_3$, $\gamma_4$, $\gamma_5$, $\gamma_6$, $\gamma_7$ and $\gamma_8$ from the non-relativistic incompressible Navier-Stokes scaling limit.}
\label{gamma345678-div}
\end{table}
Remarkably, it will turn out that these restrictions will also be sufficient to determine the near-horizon forms of $\gamma_1$ and $\gamma_2$, as those will appear as source terms in the equations for the radial flow of the other second order tensor transport coefficients. We will soon provide an additional argument to constrain the near-horizon behaviour of these transport coefficients independently by studying the scaling of the hypersurface curvature.

We now turn to the possible behaviour of second order scalar transport coefficients near the horizon. For this, we note that the rescaled pressure $P^{\textrm{non-rel}}$ at the horizon is not defined by the equation of state, but via incompressible Navier-Stokes equations as a function of the velocity fields only as in \eqref{pnr}. We therefore have an additional freedom which involves reabsorbing some (not all) higher derivative corrections to the Navier-Stokes equations via redefinition of $P^{\textrm{non-rel}}$.

For definiteness, let us consider the relativistic term $\delta_{6} \sigma_{\alpha\beta}\sigma^{\alpha\beta}$. If $\delta_6$ blows up like $(r_{\rm{H}}-r)^{-k}$ at the horizon, then it's contribution to Navier-Stokes equation will be 
\begin{equation}
\xi^{2-k}\delta_{6\, \rm{H}}\partial_i(\sigma_{jk}\sigma_{jk}) 
\end{equation}
and it's contribution to the incompressibility condition is
\begin{equation}
\xi^{3-k}\delta_{6\, \rm{H}}(\sigma_{ij}\sigma_{ij}) \partial_k v_k.
\end{equation} 
In order to have a leading order contribution which should be the same as Navier-Stokes, we require $k=1$. It might seem that by doing so we are changing the Navier-Stokes equation at the horizon, which we have said earlier to be disallowed. Nevertheless, we can readily see that if we redefine the pressure as below: 
\begin{equation}
\widetilde{P}^{\textrm{non-rel}} = P^{\textrm{non-rel}} + \delta_{6\,\textrm{H}}(\sigma_{jk}\sigma_{jk}),
\end{equation}
we can again use the incompressibility condition and the Navier-Stokes equations to show that:
\begin{equation}
\partial_j\partial_j\widetilde{P}^{\textrm{non-rel}} = - \rho_0 \partial_i\partial_j(v_i v_j) + \eta_{\textrm{H}}\partial_i\partial_j(\partial_i v_j+ \partial_j v_i ),
\end{equation}
\emph{i.e.} $\widetilde{P}^{\textrm{non-rel}}$ satisfies the same condition like $P^{\textrm{non-rel}}$ as in \eqref{pnr}. We thus have the same equations for the velocity fields as in non-relativistic incompressible Navier-Stokes after eliminating $\widetilde{P}^{\textrm{non-rel}}$. Thus, even if $\delta_{6}$ blows up like $(r_{\rm{H}}-r)^{-1}$ at the horizon, the horizon fluid still follows non-relativistic incompressible Navier-Stokes equations.

We can similarly constrain the behaviour of all second order scalar transport coefficients except $\delta_1$ and $\delta_2$ which involve the curvatures. The result is as in the Table \ref{delta34567-div}.
\begin{table}
\begin{center}
    \begin{tabular}{| c | c | c |}
    \hline
    Transport coefficient & Corresponding scalar & Allowed leading order near-horizon behaviour     \\
    \hline  
 	   
    	$\phantom{\Big]}\delta_3 \phantom{\Big]}$ & 
    	$\left( \nabla \cdot u\right)^2$ & 
    	$(r_{\rm{H}}-r)^{-1}$
    \\
    \hline
    	$\phantom{\Big]}\delta_4 \phantom{\Big]}$ & 
    	$ {\nabla_\bot}^\mu {\nabla_\bot}_\mu \ln s $ & 
    	$(r_{\rm{H}}-r)^{-3}$ 
    \\
    \hline
    	$\phantom{\Big]}\delta_5 \phantom{\Big]}$ & $ {\nabla_\bot}^\mu \ln s {\nabla_\bot}_\mu \ln s $ & 
    	$(r_{\rm{H}}-r)^{-7}$   
    	\\
    \hline
    	$\phantom{\Big]}\delta_6 \phantom{\Big]}$ & 
    	$\sigma^{\mu}_{\,\,\, \nu} \sigma^{\nu}_{\,\,\,\mu}  $ & 
    	$(r_{\rm{H}}-r)^{-1}$ 
    \\
    \hline
    	$\phantom{\Big]}\delta_7 \phantom{\Big]}$ & 
    	$ \omega^{\mu}_{\,\,\, \nu} \omega^{\nu}_{\,\,\,\mu}  $ & 
    	$(r_{\rm{H}}-r)^{-1} $
    \\
    \hline
    		\end{tabular}
\end{center}
\caption{Allowed leading order near-horizon behavior of $\delta_3$, $\delta_4$, $\delta_4$, $\delta_5$ and $\delta_7$ by the non-relativistic incompressible Navier-Stokes scaling limit.}
\label{delta34567-div}
\end{table}
Once again these restrictions on the near-horizon forms of other second order scalar transport coefficients will be sufficient to determine the near-horizon forms of $\delta_1$ and $\delta_2$, as these will be sourcing the radial flow of the above transport coefficients.

Let us now ask what could be the consistent scaling of the curvature of the metric in the frame of the infalling observer. Clearly these curvatures need to be small as some power of $\xi$ in order for a static equilibrium (about which we are expanding in $\xi$ expansion) to exist. In order to obtain the scaling of the Riemann curvature, we need to understand how the fluctuations of the metric scale with $\xi$. We can adopt Riemann normal coordinates 
for the infalling observer who is at say $\mathbf{x} = 0$ and $\tau =0$ and at a definite value of $\tilde{r}$. The hypersurface metric in the vicinity of this observer takes the form:
\begin{equation}
g_{\mu\nu} = \eta_{\mu\nu} - \frac{1}{3} R_{\mu\rho\nu\sigma}x^\rho x^\sigma + \mathcal{O}(\nabla^3).
\end{equation}
Let us call $g_{\mu\nu}- \eta_{\mu\nu}$ as $\delta g_{\mu\nu}$. Our results will require 
\begin{equation}
\delta g_{\tau\tau} = \mathcal{O}(\xi^3), \quad \delta g_{i\tau}= \mathcal{O}(\xi^2), \quad \delta g_{ij} = \mathcal{O}(\xi^3) \, 
\end{equation}
which will imply that
\begin{equation}
R_{ijkl} = \mathcal{O}(\xi^3) \quad \text{and} \quad
R_{i\tau j\tau} = \mathcal{O}(\xi^3).
\end{equation}
It is easy to see that with the above scaling the norm condition of $u^\mu$ is preserved, as up to $\mathcal{O}(\xi^2)$, the form of $u^\mu$ given by $(1, \xi v^i)/ \sqrt{1-\xi^2 v_j v_j}$ satisfies the norm with respect to the flat metric itself. It turns out that if the scaling of both $R_{i\tau j\tau}$ and $R_{ijkl}$ are $\mathcal{O}(\xi^3)$, the near-horizon behaviour of $\gamma_1$, $\gamma_2$, $\delta_1$ and $\delta_2$ that we will find in the next section do not change the fixed point given by non-relativistic incompressible Navier-Stokes equations at the horizon. In this case, the constraints on the near-horizon behavior of these transport coefficients is as in Table \ref{gamma12delta34-div}. It would be interesting to understand what determines the scaling of the curvature independently of our results for near horizon behavior of $\gamma_1$, $\gamma_2$, $\delta_1$ and $\delta_2$. We leave this for the future.
\begin{table}
\begin{center}
    \begin{tabular}{| c | c | c |}
    \hline
    Transport coefficient & Corresponding tensor & Near-horizon behavior should be weaker than     \\
    \hline  
 	   $\gamma_1$ & 
    	$\langle R^\mu_{\phantom{\mu}\nu}\rangle$ & 
    	$(r_{\rm{H}}-r)^{-1}$
    \\
    \hline
    	$\gamma_2$ & 
    	$\left< u^\alpha R^{\phantom{\alpha} \mu \phantom{\nu} \beta}_{\alpha \phantom{\mu} \nu \phantom{\beta}} u_\beta  \right>$ & 
    	$(r_{\rm{H}}-r)^{-1}$ 
    \\
    \hline
    		\end{tabular}
\end{center}
\begin{center}
    \begin{tabular}{| c | c | c |}
    \hline
    Transport coefficient & Corresponding scalar & Allowed leading order near-horizon behaviour     \\
    \hline  
 	   $\delta_1$ & $ R $ & 
    	$(r_{\rm{H}}-r)^{-2}$   
    	\\
    \hline
    	$\delta_2$ & 
    	$R_{\mu\nu}u^\mu u^\nu  $ & 
    	$(r_{\rm{H}}-r)^{-2}$ 
    \\
    \hline
    		\end{tabular}
\end{center}
\caption{Constraints on near-horizon behaviour of $\gamma_1$, $\gamma_2$, $\delta_1$ and $\delta_2$ from the non-relativistic incompressible Navier-Stokes scaling limit.}
\label{gamma12delta34-div}
\end{table}

Clearly, we can adopt our procedure to constrain the near-horizon behaviour of the higher order transport coefficients at any order in the derivative expansion. Our near-horizon scaling will fix unambiguously the allowed leading order behaviour of the transport coefficients even at higher orders in the derivative expansion. In the next section, we will provide evidence that these constraints will be sufficient to determine the near-horizon forms of all the transport coefficients independently of their boundary values, and thus we can solve the RG flow uniquely. Furthermore, the value of the transport coefficients at the boundary would be exactly those required for absence of naked singularities at the future horizon. Therefore these boundary values should coincide with the results which can be obtained from the traditional method of determining them from explicit construction of the bulk metric.

We end this section with some comments comparing our near horizon limit with that of \cite{Bredberg:2011jq}. Indeed our near-horizon limit is qualitatively similar to the latter, though we have applied the scaling of the radial and time coordinates and the fluid variables as in (\ref{reco}, \ref{reu}, \ref{tre}) to the RG flow while in the latter the same was applied to the bulk metric. The horizon fluid was also incompressible non-relativistic Navier-Stokes fluid with vanishing bulk viscosity and shear-viscosity equaling the membrane paradigm value  \cite{Bredberg:2011jq}. In our case, the value of the horizon shear-viscosity will turn out to be different, though $\eta/s$ will be same at all scales and equal to the universal value $1/4\pi$ (see the next section for details). The quantitative difference in the horizon shear viscosity will be because in \cite{Bredberg:2011jq} there was a rescaling of spatial coordinates involved to make the induced metric $\eta_{\mu\nu}$ everywhere on the cut-off hypersurface where the Dirichlet boundary condition was imposed on the metric fluctuations. We recover our quantitative result if we undo this rescaling of spatial coordinates in \cite{Bredberg:2011jq}. \footnote{Note the Fefferman-Graham and Eddington-Finkelstein slices $r= \text{constant}$ are the same far in the future (in the static final equilibrium geometry) up to trivial radial reparametrizations. So it is no surprise that the shear viscosity at the horizon turn out to be the same in both cases provided the units of measurement are the same. } This rescaling of spatial coordinates is non-singular at the horizon and hence do not change our results qualitatively. 

Our approach is however essential to define the incompressible non-relativistic Navier-Stokes fluid as a
fixed point of the rescaled RG flow in a rigorous manner. As already shown in the previous sections, we can define a unique first order RG flow without requiring to know the explicit metric up to trivial scale reparametrizations and numerical prefactors in the counter-terms. We will have sufficient evidence in the following section to show that even these numerical prefactors will be fixed uniquely by requiring that the fixed point of the rescaled RG flow will be the incompressible non-relativistic Navier-Stokes fluid at the horizon.

\section{Results for RG flow of transport coefficients}

In this section, we will solve the RG flow of the transport coefficients at the first and second orders in derivative expansion. We will follow the general algorithm of Section \ref{algo}, and then determine the transport coefficients uniquely using the physical principle of Section \ref{hf} that the fluid on the infra-red screen will follow the incompressible Navier-Stokes equation in  well defined scaling limit.
It will be clear that in our procedure we will need to go to higher orders to completely determine some lower order transport coefficients. In particular, we will see that the shear viscosity $\eta$ gets determined completely at second order.

Solving the RG flow explicitly up to second order will allow us to determine four of the five second order transport coefficients of the conformal fluid at the boundary. Our computations will reproduce the boundary values of these transport coefficients already known in the literature. We believe that the remaining second order transport coefficient will be determined by a higher order calculation.

\subsection{First order}

At the first order, we have only one tensor coefficient (see \eqref{ListOfFirstOrderSVT}), namely the shear viscosity $\eta$, which multiplies $-2\sigma^{\mu}_{\phantom{\mu}\nu}$ in the energy-momentum tensor\footnote{We recall again that the definition of $\sigma^{\mu}_{\phantom{\mu}\nu}$ varies in the literature, and so the $2$-factor in the shear viscosity term.} $t^{\mu}_{\phantom{\mu}\nu}$. As shown in Section \ref{algo}, in order to obtain the equation for $\eta$, we should first compute  $\sigma^{\mu\prime}_{\phantom{\mu}\nu}$. At the leading order, \emph{i.e.} first order, it should be proportional to $\sigma^{\mu}_{\phantom{\mu}\nu}$ itself as there is no other transverse traceless tensor at this order. We also recall from Section \ref{MetricElimination} this can be calculated readily from the zeroth order form of  $z^{\mu}_{\phantom{\mu}\nu}$, as given by \eqref{z-ZerothOrder} without knowing the metric explicitly. The result is as given in (the first order term of) (\ref{sigma-prime}). The equation for the shear viscosity follows from the traceless transverse part of the equation of  radial evolution of $t^{\mu}_{\phantom{\mu}\nu}$ as given by (\ref{EMTensorEoM-Tensor}), of which we retain the first order terms only. We obtain:
\begin{equation}
\eta^\prime = - \dfrac{r^{d-1}}{2} \epsilon \cdot \eta \,.
\end{equation}
The general solution of this first order ODE is:
\begin{equation}
\label{eta(r)}
\eta(r) = \dfrac{r_\textrm{H}^{2(d-1)}}{\left( r_\textrm{H}^d + r^d \right)^{2 \left(1 - 1/d \right) }} \eta_{\textrm{b}}
\, .
\end{equation}
The integration constant $\eta_\textrm{b}$, is actually the value of $\eta$ on the boundary.

As we have discussed in Section \ref{hf}, in order to have a sensible horizon fluid, we need $\eta$ to be finite at the horizon, and this indeed so for any value of $\eta_\textrm{b}$.  Thus from the first order RG flow equations we cannot determine the exact value of $\eta$ neither at the horizon nor at the boundary. Nevertheless, we will be able to so at the second order. We will see that unless $\eta_{\textrm{H}}$ is adjusted properly, the source term for one of the second order tensor transport coefficients will blow up at the horizon.

Comparing with the result for $s^\prime$ as given by \eqref{s,c_s-prime}, we readily see that:
\begin{equation}\label{eta-equals-s}
\Bigg(\frac{\eta}{s}\Bigg)^\prime = 0.
\end{equation}
This is the very well known result, first obtained in \cite{Iqbal:2008by}, that $\eta/s$ does not flow radially.

We now consider the vector projection of the equation for radial evolution of  $t^{\mu}_{\phantom{\mu}\nu}$ as given by \eqref{vfste}. Again, at the first order there is only one vector, and so there is a single $\beta^{(1)}_i$ parameter. We find that\footnote{For obvious reasons, we do not put an extra counting label} $\beta^{(1)}=0$. Thus the transverse contribution to $u^{\mu\prime}$ vanishes at the first order.

We are now in a position to determine the only scalar transport coefficient at the first order, namely the bulk viscosity $\zeta$ from . We use (\ref{EMTensorEoM-Scalar1}) of which we retain first order terms only. This gives the following equation for the bulk viscosity:
\begin{equation}
\zeta^\prime - \left( \dfrac{d-1}{r} (c_s^2+1) + r^{d-1} \left( P - \dfrac{1}{2} c_s^2 \epsilon \right)\right) \cdot \zeta = 0 \, .
\end{equation}
In deriving this identity we used \eqref{nabla-u-Example} together with \eqref{alpha_0}.
This homogeneous equation for $\zeta(r)$ can be solved analytically:
\begin{equation}
\label{zeta-Solution}
\zeta(r) \sim \dfrac{r^d \left( r_\textrm{H}^d + r^d \right)^{\frac{2}{d}-1}}{ \left( r_\textrm{H}^d - r^d \right)^{3}} \, .
\end{equation}
We recall from Section \ref{hf} that we need $\zeta$ to be constant at the horizon. Obviously, the only solution which is of this form is the trivial solution $\zeta(r)=0$.  Therefore, the bulk viscosity vanishes for all hypersurfaces in the foliation.

We can now determine the auxiliary variables $\lambda^{(1)}$ and $\alpha^{(1)}$ from the first order terms in \eqref{EMTensorEoM-Scalar2} and \eqref{NormCondition} respectively. We obtain $\lambda^{(1)} = \alpha^{(1)} =0$. Putting together these with $\beta^{(1)}=0$, we conclude that at the first order there both ${u^\mu}^\prime$ and $T^\prime$ receive no contributions:
\begin{equation}\label{fos}
({u^\mu}^\prime)^{(1)} = 0, \quad (T^\prime)^{(1)} = 0 \, .
\end{equation}
The above leads to some simplifications for the second order calculations.

\subsection{Second order}

We will now compute the radial flow of second order transport coefficients. We have seen from the general algorithm presented in Section \ref{algo}, which can be used at any order, that to find the $n$-th order transport coefficients, we need not solve for $\left( {u^\mu}^\prime \right)^{(n)}$ or $\left( T^\prime \right) ^{(n)}$ at the same $n$-th order. However, we do need to find $(u^{\mu\prime})^{(m)}$ and $(T')^{(m)}$ for all $m<n$, \emph{i.e.} we have to solve the RG flow equations completely at lower orders. As we will be interested in the second order transport coefficients only, which contains all the relevant physical information up to thus order, we will not solve for $({u^\mu}^\prime)^{(2)}$ or $(T^\prime)^{(2)}$, \emph{i.e.} the auxiliary variables at second order. These will be relevant, however, if one is about to calculate the third order transport coefficients.

\subsubsection{Tensor transport coefficients}

In order to obtain the second order transport coefficients, we should first calculate the radial derivatives of all the second order transverse traceless tensors and write them as a linear combination of the second order transverse traceless tensors. In other words, we should solve \eqref{Tensor_n-prime} at second order. This can be done as the explicit equilibrium form of $z^\mu_{\phantom{\mu}\nu}$ (given by \eqref{z-ZerothOrder}) is known. The explicit solution of this ``tensor mixing problem" is given in (\ref{T_i-prime}). 

Additionally we have to know the contributions coming from lower orders, as the second order terms in the expansion of $\left(  -2 \eta \sigma^{\mu}_{\phantom{\mu}\nu} \right)^\prime$. There are some simplifications coming from the result \eqref{fos}. 
First, following the arguments of Subsection \ref{algo}, $({u^\mu}^\prime)^{(1)}= 0$ implies that $\sigma^{\mu\prime}_{\phantom{\mu}\nu}$ is transverse and traceless at second order, \emph{i.e.} it's second order contributions can be written as a linear sum of second order transverse traceless tensors. The result of the calculation is in (\ref{sigma-prime}). In deriving this result it is sufficient to use ${u^\mu}^\prime = \alpha_0 u^\mu + \ldots$ with $\alpha_0$ given by \eqref{alpha_0} and the first order expansion of $z^\mu_{\phantom{\mu}\nu}$ as given by \eqref{z}\footnote{Notice that $t^\mu_{\phantom{\mu}\nu}$ is already known at the first order.}.
Second, the solution for $\eta$ as given in \eqref{eta(r)} can be expressed as $\eta(T,r)$ by eliminating $r_{\textrm{H}}$ in favour of $T$ using \eqref{T}. We, therefore, readily see that $(\eta^\prime)^{(1)} = (\partial \eta/ \partial T) (T^\prime)^{(1)}$ may contribute to the second order of $\left(  -2 \eta \sigma^{\mu}_{\phantom{\mu}\nu} \right)^\prime$. It, however, does not since $(T^\prime)^{(1)} = 0$.

Combining the projected equation for radial evolution of  $t^{\mu}_{\phantom{\mu}\nu}$ as given by (\ref{EMTensorEoM-Tensor}) with (\ref{T_i-prime}) and (\ref{sigma-prime}), and retaining the second order terms only, we arrive at the radial equations of motion for the second order tensor transport coefficients $\gamma_i(r)$'s (from now on we drop the ${}^{(2)}$ superscript):
\begin{eqnarray}
\label{gamma-EoM}
\gamma_1^\prime + \dfrac{r^{d-1}}{2} \left( - P + \dfrac{2}{d-1} \epsilon \right) \gamma_1 &=& 
											\dfrac{r}{d-2} \left( - P + \dfrac{2}{d-1} \epsilon \right)
\nonumber\\
\gamma_2^\prime + \dfrac{r^{d-1}}{2} \left( P + \dfrac{2d-3}{d-1} \epsilon \right) \gamma_2 &=& 
						- r^{d-1} \left( 2  \eta^2 + (P+\epsilon) \left( \gamma_1 - \dfrac{2 r^{2-d}}{d-2} \right) \right) 
\nonumber\\
\gamma_3^\prime + \dfrac{d-2}{2(d-1)} r^{d-1} \epsilon \gamma_3  &=& 2 \dfrac{d+1}{d-1} r^{d-1} \eta^2 + 
\nonumber \\
				+ \dfrac{r^{d-1}}{2} (P+\epsilon) \!\!\! && \!\!\!\!\! 
						\left( \left( c_s^2 + \dfrac{3}{d-1} \right) \left( \gamma_1 - \dfrac{2 r^{2-d}}{d-2} \right) 
						  + \left( c_s^2 + \dfrac{d}{d-1} \right) \gamma_2  \right) 
\nonumber\\
\gamma_4^\prime - r^{d-1} \left( P + \dfrac{d}{2(d-1)} \epsilon \right) \gamma_4 &=& 2  r^{d-1} c_s^2 \eta^2 + 
\nonumber\\
				+ \dfrac{ r^{d-1}}{2} (P+\epsilon) \!\!\! && \!\!\!\!\! 
				         \left( \left( c_s^2 - \dfrac{1}{d-1} \right) \left( \gamma_1 - \dfrac{2 r^{2-d}}{d-2} \right) +  
				             \left( c_s^2 - \dfrac{d-2}{d-1} \right) \gamma_2  \right) 
\nonumber\\
\gamma_5^\prime - \dfrac{r^{d-1}}{2} \left( 3 P + \dfrac{2d-1}{d-1} \epsilon \right) \gamma_5 &=& 
				2  r^{d-1} \left( \dfrac{\partial c_s^2}{\partial \ln s} -  c_s^4 \right) \eta^2 - 
\nonumber\\
        + \dfrac{r^{d-1}}{2} (P+\epsilon) \!\!\! && \!\!\!\!\! 
        			\left( \gamma_1 - \dfrac{2 r^{2-d}}{d-2} + \gamma_2 \right) 
              			\left( - 2 c_s^4 + \dfrac{\partial c_s^2}{\partial \ln s} + (c_s^2+1) \left( c_s^2 - \dfrac{1}{d-1} \right) \right) +
\nonumber\\   
     + r^{d-1} (P+\epsilon)  \!\!\! && \!\!\!\!\! 
     	\left( - \dfrac{d-3}{2(d-1)} (c_s^2+1)   \gamma_2 
    	 + \dfrac{1}{2} \left( c_s^2 + \dfrac{d+1}{d-1} \right) \gamma_4 \right)          
\nonumber \\
\gamma_6^\prime + \dfrac{r^{d-1}}{2} \left( P + \dfrac{2d-3}{d-1} \epsilon \right) \gamma_6 &=& 
			r^{d-1} \left( 2  \eta^2 + (P+\epsilon) \left( \gamma_1 - \dfrac{2 r^{2-d}}{d-2} \right) \right) 
\nonumber\\
\gamma_7^\prime - \dfrac{r^{d-1}}{2} \left( 3 P + \dfrac{2d-1}{d-1} \epsilon \right) \gamma_7 &=& 
			r^{d-1} \left( 2  \eta^2 - (P+\epsilon) \left( \gamma_1 - \dfrac{2 r^{2-d}}{d-2} - 2\gamma_2 \right) \right) 
\nonumber\\
\gamma_8^\prime - \dfrac{r^{d-1}}{2} \left( P + \dfrac{\epsilon}{d-1} \right) \gamma_8 &=& 
			2 r^{d-1} \left( 2  \eta^2 + (P+\epsilon) \left( \gamma_1 - \dfrac{2 r^{2-d}}{d-2} + \gamma_2 \right) \right)
\end{eqnarray}
In deriving these equations we have also used \eqref{d-eta-OVER-d-ln-s}.

Let us clarify our strategy for solving the $\gamma_i$'s equations. All eight equations are first order ODEs with non-vanishing sources. Dropping the sources on the right hand side, one receives homogeneous equations. Imposing the horizon criteria of the previous section, we may fix the integration constants, but, contrary to the bulk viscosity \eqref{zeta-Solution}, this will not necessarily set the corresponding functions to zero. This will only work provided the solutions of the underlying homogeneous equations violate the bound at the horizon that we summarized at Tables \ref{gamma345678-div} and \ref{gamma12delta34-div}. 

In what follows we will refer to the near horizon behaviour presented in these tables as \emph{regular} even if the function in question satisfies the bound while blowing up at $r=r_\textrm{H}$. 

As we will see momentarily, for $\gamma_2$ the related homogeneous solution goes to zero at the horizon, and so it seems that (at least) for this function there is no way to fix the integration constant, which in turn means that the boundary value of $\gamma_2$ could not be determined from this procedure. 

This argument, however, is only partially correct. The integration constant of $\gamma_2$, indeed, cannot be eliminated based only on the ``regularity" of $\gamma_2$ itself. Nevertheless, since this function appears also as a source in the $\gamma_3$ equation, the restriction on $\gamma_3$ in Table \ref{delta34567-div} \emph{may} impose an additional \emph{stronger} constraint on the horizon behaviour of  $\gamma_2$. Remarkably, this is precisely what happens. Even more surprisingly, the regularity of $\gamma_3$ fixes also the boundary value of $\eta$ that we left unfixed in \eqref{eta(r)}.

Let us start with $\gamma_1$. Substituting \eqref{e(r)-and-P(r)} we observe that the solution of the corresponding homogeneous equation diverges as $(r_\textrm{H}-r)^{-1}$ near the horizon. According to Table \ref{gamma12delta34-div} the Navier-Stokes limit does not allow us to exclude this mode. Notice, though, that $\gamma_1$ appears as a source in the RG flow equations of all the other $\gamma$-functions in \eqref{gamma-EoM}. A quick look at these equations reveals that this diverging mode necessarily leads to the near-horizon divergences that explicitly violate the bounds in Table \ref{gamma345678-div}. The easiest way to see this is from the $\gamma_6$ equation. Since $ P \sim (r_\textrm{H}-r)^{-1}$ near the horizon, $\gamma_6$ behaves there exactly as  $\gamma_1$ does. To conclude, unless $\gamma_1$ is set to a constant at $r=r_\textrm{H}$, we cannot satisfy the conditions of the previous section. Moreover, as we will see immediately, eliminating the ``singular" $(r_\textrm{H}-r)^{-1}$ mode is sufficient for maintaining the ``regular" near-horizon solutions for \emph{all} other $\gamma$'s.

Excluding the homogeneous $(r_\textrm{H}-r)^{-1}$ solution we arrive at the following unique solution:
\begin{equation}
\label{gamma_1}
\gamma_1(r) = \dfrac{4}{(d-2)(r_\textrm{H}^d - r^d)} \left( r_\textrm{H}^{d-2} \left( \dfrac{r_\textrm{H}^d + r^d}{2}  \right)^{4/d-1} - r^2 \right) \, . 
\end{equation}
For a reason unclear to us, this expression significantly simplifies if $d=4$. Furthermore, for any $d$ the function $\gamma_1(r)$ is monotonically decreasing as one moves towards the horizon, and its value there is:
 \begin{equation}
 \gamma_1(r=r_\textrm{H}) = \dfrac{2}{(d-2)} r_\textrm{H}^{2-d} \, . 
 \end{equation}
Substituting (\ref{gamma_1}) into the $\gamma_2$ equation results in a very long expression, and so we will focus mostly only the $d=4$ case. Independently of the integration constant, $C_2$,  the solution of the $\gamma_2$ equation goes to zero as $(r_\textrm{H}-r)$  near the horizon. Thus, at this stage we still cannot fix $\gamma_2$ completely, and consequently the boundary value of $\gamma_2$ at $r=0$ also remains undetermined. Overall $\gamma_2$ depends now on two free constants: $C_2$ and $\eta_\textrm{b}$.

Let us now consider the $\gamma_3$ ODE. Because $\epsilon $ takes a non-zero constant value at the horizon, while $P$ and $c_s^2$ blow up their as $(r_\textrm{H}-r)^{-1}$ and $(r_\textrm{H}-r)^{-2}$ respectively, $\gamma_3$ has to diverge as $(r_\textrm{H}-r)^{-2}$. The only way to avoid this, is to require that the combination which multiplies $c_s^2$: 
\begin{equation}
\label{gamma3Condition}
\gamma_1 - \dfrac{2 r^{2-d}}{d-2} + \gamma_2 
\end{equation}
behaves like $(r_\textrm{H}-r)^3$. This looks like a strong condition, since this expression has both $(r_\textrm{H}-r)$, $(r_\textrm{H}-r)^2$ and also two logarithmic terms, while we can modify only two constants $C_2$ and $\eta_\textrm{b}$. Amazingly, though, setting to zero only the $(r_\textrm{H}-r)$ and the $(r_\textrm{H}-r)\ln(r_\textrm{H}-r)$ terms renders (\ref{gamma3Condition}) exactly the right $(r_\textrm{H}-r)^3$ behaviour near $r=r_\textrm{H}$! The value of $\eta_{\textrm{b}}$ that does the job for $d=4$ is:
\begin{equation}
\label{eta-b}
\eta_{\textrm{b}} = \dfrac{2 \sqrt{2}}{r_\textrm{H}^3} \,
\end{equation}
and the final solution for $\gamma_2(r)$ is:
\begin{equation}
\label{gamma_2}
\gamma_2(r) = 2 r_\textrm{H}^2 \dfrac{r_\textrm{H}^4 - r^4}{(r_\textrm{H}^4 + r^4)^2} 
   \left( \dfrac{2 r_\textrm{H}^2 r^2}{(r_\textrm{H}^2 + r^2)^2} 
       + \ln \left( \dfrac{(r_\textrm{H}^2 + r^2)^2}{2(r_\textrm{H}^4 + r^4)}\right)  \right) \, . 
\end{equation}
Though it is possible to write down the result for $d \neq 4$, for general $d$ this (and all the other) expression(s) become extremely complicated, so we will restrict ourself in this subsection to the $d=4$ case, while presenting explicit solutions for the $\gamma$'s. At the same time, all the conclusions about the regular/singular behaviour  of these functions will be completely general.

With \eqref{eta-b} and \eqref{gamma_2}, the source on the right hand side of the $\gamma_3$ equation now approaches a constant non-zero value at $r=r_\textrm{H}$. However, the homogeneous solution of the $\gamma_3$ equation is regular (non-zero constant) at $r=r_\textrm{H}$. So, though we can guarantee that $\gamma_3$ does not blow up at the horizon by properly adjusting the values of $\eta$ and $\gamma_2$ there, we cannot still fix the integration constant in the $\gamma_3$ equation itself.

The situation almost repeats itself for $\gamma_4$, $\gamma_5$ and $\gamma_6$, as these functions also cannot be completely fixed just from the ``regularity" condition. The solution of the related homogeneous equation of $\gamma_4$ behaves like $(r_\textrm{H}-r)^{-2}$, while the source is finite similar to the source of the $\gamma_3$ equation. According to Table \ref{gamma345678-div} any $\gamma_4$ solution will, then be consistent with the non-relativistic limit of the previous section. Moreover, if $\gamma_4$ behaves like $(r_\textrm{H}-r)^{-2}$, then the source in the $\gamma_5$  equation diverges only as $(r_\textrm{H}-r)^{-5}$. Since the homogeneous $\gamma_5$ solution goes like $(r_\textrm{H}-r)^{-3}$ at $r=r_\textrm{H}$, we see that $\gamma_5$ also respects the bound of Table \ref{gamma345678-div} for an arbitrary value of the integration constant.

Before reconsidering the functions $\gamma_3$, $\gamma_4$, $\gamma_5$ and $\gamma_6$, let us address the $\gamma_7$ and $\gamma_8$ equations. The homogeneous solutions for the corresponding equations behave near the horizon as  $(r_\textrm{H}-r)^{-3}$ and \mbox{$(r_\textrm{H}-r)^{-1}$} respectively. Requiring that $\gamma_7$ and $\gamma_8$ gain finite values at $r=r_\textrm{H}$ one can uniquely fix these two functions. The final solutions are:
\begin{eqnarray}
\label{gamma7gamma8}
\gamma_7(r) &=& - 2 r_\textrm{H}^2  \dfrac{r_\textrm{H}^4 - r^4}{(r_\textrm{H}^4 + r^4)^2} 
   \left( 2 \dfrac{\left(r_\textrm{H}^8 + r_\textrm{H}^6 r^2 + 4 r_\textrm{H}^4 r^4 + r_\textrm{H}^2 r^6 + r^8\right)}{(r_\textrm{H}^2 + r^2)^4} 
       + \ln \left( \dfrac{(r_\textrm{H}^2 + r^2)^2}{2(r_\textrm{H}^4 + r^4)}\right)  \right) \, 
\nonumber \\
\gamma_8(r) &=& - 4 r_\textrm{H}^2  \dfrac{r_\textrm{H}^4 - r^4}{(r_\textrm{H}^4 + r^4)^2} 
   \left( 2 \dfrac{r_\textrm{H}^2 r^2}{(r_\textrm{H}^2 + r^2)^2} 
       + \ln \left( \dfrac{(r_\textrm{H}^2 + r^2)^2}{2(r_\textrm{H}^4 + r^4)}\right)  \right) \, .
\nonumber 
\end{eqnarray}

To summarize, using the finiteness of the second order tensor transport coefficients on the horizon, we were able to determine completely $\eta$ (left unfixed at the first order), $\gamma_1$, 
$\gamma_2$, $\gamma_7$ and $\gamma_8$. We strongly believe that the integration constant in the remaining four functions, $\gamma_3, \ldots,\gamma_6$,  will be determined at the higher orders. Our expectations are firmly supported by the results of this subsection. At the first order we managed to set the integration constant in the bulk viscosity expression, but in order to determine the shear viscosity we had to go to the next (second) order, where, again, requiring regularity we could determine exactly half of transport coefficients appearing at this order.

\begin{table}[t]
\begin{center}
    \begin{tabular}{| c | c | c | c |  }
    \hline
    \multirow{2}{*}{Transport coefficient} & Corresponding  & Behaviour/Value  & Value at    \\ 
     &  tensor  & near the horizon & the boundary     \\    
    \hline \hline 
 	   $\phantom{\Bigg()} \eta(r)$ & 
 	   $-2 \sigma^\mu_{\,\,\,\nu}$ & 
       $\left( \pi T_\textrm{b} \right)^3/2 \sqrt{2}$ & 
       $\left( \pi T_\textrm{b} \right)^3$ \\ 
    \hline
    	$\phantom{\Bigg()} \gamma_1(r)$ & 
    	$\left< R^\mu_{\phantom{\mu} \nu} \right>$ & 
    	$\dfrac{1}{2} \left( \pi T_\textrm{b} \right)^2$ & 
    	$\left( \pi T_\textrm{b} \right)^2$ 
    \\ 
    \hline
    	$\phantom{\Bigg()} \gamma_2(r)$ & 
    	$\left< u^\alpha R^{\phantom{\alpha} \mu \phantom{\nu} \beta}_{\alpha \phantom{\mu} \nu \phantom{\beta}} u_\beta  \right>$ & 
    	$\mathcal{O} \left( (r_\textrm{H}-r) \right)$  & 
    	$-\ln 2 \left( \pi T_\textrm{b} \right)^2$ \\
    \hline
    	$\phantom{\Bigg()} \gamma_3(r)$ & 
    	$\left( \nabla \cdot u\right) \sigma^\mu_{\,\, \nu}$ & 
    	$\mathcal{O} \left( (r_\textrm{H}-r) \right)$ & 
    	$- \dfrac{1}{3} (2 - \ln 2) \left( \pi T_\textrm{b} \right)^2$ 
    \\
    \hline
    	$\phantom{\Bigg()} \gamma_4(r)$ & 
    	$\left< {\nabla_\bot}^\mu {\nabla_\bot}_\nu \ln s \right>$ & 
    	$\mathcal{O} \left( (r_\textrm{H}-r)^{-1} \right)$ & 
    	$- \dfrac{1}{3} (2 - \ln 2) \left( \pi T_\textrm{b} \right)^2$  
    \\
    \hline
    	$\phantom{\Bigg()} \gamma_5(r)$ & $\left< {\nabla_\bot}^\mu \ln s {\nabla_\bot}_\nu \ln s \right>$ & 
    	$\mathcal{O} \left( (r_\textrm{H}-r)^{-3} \right)$ & 
    	$\dfrac{1}{9}(2 - \ln 2) \left( \pi T_\textrm{b} \right)^2$ 
    \\
    \hline
    	$\phantom{\Bigg()} \gamma_6(r)$ & 
    	$\left< \sigma^{\mu}_{\,\,\, \tau} \sigma^{\tau}_{\,\,\,\nu}  \right>$ & 
    	$\mathcal{O} \left( (r_\textrm{H}-r) \right)$ & 
    	$ ( C_6 + \ln 2 ) \left( \pi T_\textrm{b} \right)^2$ 
    \\
    \hline
    	$\phantom{\Bigg()} \gamma_7(r)$ & 
    	$\left< \omega^{\mu}_{\,\,\, \tau} \omega^{\tau}_{\,\,\,\nu}  \right>$ & 
    	$\mathcal{O} \left( (r_\textrm{H}-r) \right)$ & 
    	$- (2 - \ln 2) \left( \pi T_\textrm{b} \right)^2$ 
    \\
    \hline
    	$\phantom{\Bigg()} \gamma_8(r)$ & 
    	$\left< \sigma^{\mu}_{\,\,\, \tau} \omega^{\tau}_{\,\,\,\nu}  \right>$ & 
    	$\mathcal{O} \left( (r_\textrm{H}-r) \right)$ & 
    	$2 \ln 2 \left( \pi T_\textrm{b} \right)^2$ 
    \\
    \hline
	\end{tabular}
\end{center}
\caption{To compare the boundary values \protect\eqref{BoundaryConstants} we have to multiply them by $\kappa_\textrm{AdS}$, because we have rescaled the stress tensor in \protect\eqref{t-rescaling}. Here we present our results for the boundary values of the  tensor transport coefficients. The boundary temperature, $T_b$, is identified with the Hawking temperature, given by \protect\eqref{Hawking}. For $d=4$ it means that $T_\textrm{b}=\sqrt{2}/\pi r_\textrm{H}$. Our method allows to fix the boundary values of the shear viscosity and of four out of eight second order tensor transport coefficients, $\gamma_1$, $\gamma_2$, $\gamma_7$ and $\gamma_8$. The remaining four will be determined at higher orders. However, three of these four, $\gamma_3$, $\gamma_4$ and $\gamma_5$ can alternatively be fixed requiring the boundary fluid to be conformal. Our findings agree with \cite{Baier:2007ix},\cite{Bhattacharyya:2008jc} with the exception of $\gamma_6$ that has to be determined at higher orders.}
\label{ComparisonTable}
\end{table}

Since, the third order calculation is not available  for the moment, we can try to find the remaining four integration constants by imposing conformality constraint on the boundary fluid.

Indeed, we know that out eight transversal tensors only five are conformal, and we have already presented them in  \eqref{ListOfConformalTensors}. Notice that three tensors ${\mathcal{T}_3}^\mu_{\phantom{\mu}_\nu}$, ${\mathcal{T}_4}^\mu_{\phantom{\mu}_\nu}$ and ${\mathcal{T}_5}^\mu_{\phantom{\mu}_\nu}$ appear only in one conformal combination together with ${\mathcal{T}_2}^\mu_{\phantom{\mu}_\nu}$. Therefore, knowing the boundary value of $\gamma_2$ and requiring conformality, we can determine the boundary (and so the horizon) values of $\gamma_3$,  $\gamma_4$ and $\gamma_5$.\footnote{Notice that the boundary values of $\gamma_1$ and $\gamma_2$ in our calculation have automatically appeared in a correct combination to reproduce the first conformal coefficient in  \eqref{ListOfConformalTensors}.} The results are:
\begin{equation}
\label{gamma_3}
\gamma_3(r) = - \dfrac{2}{3}  \dfrac{r_\textrm{H}^2}{(r_\textrm{H}^4 + r^4)^2} 
   \left( 2 (r_\textrm{H}^4 + 5 r_\textrm{H}^2 r^2 + r^4)  \dfrac{r_\textrm{H}^2 - r^2}{r_\textrm{H}^2 + r^2} 
       +  \dfrac{(r_\textrm{H}^8 - 10 r_\textrm{H}^4 r^4 + r^8) }{r_\textrm{H}^4 - r^4} \ln \left( \dfrac{(r_\textrm{H}^2 + r^2)^2}{2(r_\textrm{H}^4 + r^4)}\right)  \right) 
\end{equation}
and
\begin{eqnarray}
\label{gamma_45}
{\gamma}_4(r) &=& - \dfrac{2}{3}   \dfrac{r_\textrm{H}^2}{(r_\textrm{H}^4 - r^4)(r_\textrm{H}^4 + r^4)^2} 
   \Bigg(  2 \left( r_\textrm{H}^8 - r_\textrm{H}^6 r^2 + 4 r_\textrm{H}^4 r^4 - r_\textrm{H}^2 r^6 + r^8 \right) +
\nonumber
\\
&& \qquad \qquad  \qquad \qquad \qquad \qquad \qquad \qquad \qquad  
	+ \left( r_\textrm{H}^8 + 6 r_\textrm{H}^4 r^4  + r^8 \right)  \ln \left( \dfrac{(r_\textrm{H}^2 + r^2)^2}{2(r_\textrm{H}^4 + r^4)} \right)  \Bigg)  \, ,
\\ 
{\gamma}_5(r) &=& - \dfrac{2}{9}   \dfrac{r_\textrm{H}^2}{(r_\textrm{H}^4 - r^4)^3(r_\textrm{H}^4 + r^4)^2} 
   \Bigg( 4 (r_\textrm{H}^4 + r^4)^4 - 2 (r_\textrm{H}^4 - r^4)^2 \left( r_\textrm{H}^4 + r_\textrm{H}^2 r^2 +  r^4 \right) \left( 3 r_\textrm{H}^4 - 2 r_\textrm{H}^2 r^2 + 3 r^4 \right) +
\nonumber\\
&& \qquad  \qquad \qquad \qquad \qquad \qquad \qquad  
       + \left( 16 r_\textrm{H}^4 r^4 (r_\textrm{H}^4 + r^4)^2  - (r_\textrm{H}^4 - r^4)^4 \right)  \ln \left( \dfrac{(r_\textrm{H}^2 + r^2)^2}{2(r_\textrm{H}^4 + r^4)}\right)  \Bigg)    \, .
\nonumber          
\end{eqnarray}
Remarkably, imposing conformality at the boundary reduces the horizon divergence of the functions $\gamma_4$ and $\gamma_5$ by exactly one power of $(r_\textrm{H} - r)$. They behave now as $\left(  r_\textrm{H} - r \right) ^{-1}$ and $\left(  r_\textrm{H} - r \right) ^{-3}$ respectively respeecting the conditions of Table \ref{gamma345678-div}. Analogously,  $\gamma_3$, vanishes at the horizon instead of approaching a constant non-zero value there. We have not found any deep physical explanation for this fact. Most likely it becomes obvious from the coupling of these functions to higher order tensor/scalar transport coefficients.

Finally, the integration constant in $\gamma_6$ cannot be determined by imposing the boundary fluid conformality, since the corresponding transverse tensor ${\mathcal{T}_2}^\mu_{\phantom{\mu}_\nu}$ is conformal by its own right and does not mix up with other tensor in \eqref{ListOfConformalTensors}. We may, of course, fix the integration constant by comparing the boundary value of $\gamma_6$ to the previously known results, but we prefer to leave the constant explicitly in the final expression for $\gamma_6$:\footnote{Contrary to the $\gamma_3$ example setting the ``right" boundary value of $\gamma_6$, see \eqref{BoundaryConstants}, does not result in an extra $(r_\textrm{H}- r)$ power in the near horizon behaviour of $\gamma_6$.}  
\begin{eqnarray}
\label{gamma6}
\gamma_6(r) &=& 2 r_\textrm{H}^2 \dfrac{r_\textrm{H}^4 - r^4}{(r_\textrm{H}^4 + r^4)^2} 
   \left( C_6 - \dfrac{2 r_\textrm{H}^2 r^2}{(r_\textrm{H}^4 + r^4)^2} 
       - \ln \left( \dfrac{(r_\textrm{H}^2 + r^2)^2}{2(r_\textrm{H}^4 + r^4)}\right)  \right) \, .
\end{eqnarray}
Again, in order to find $C_6$ one has to go the third (or higher) order in the derivative expansion.

Our results for the boundary values of the tensor transport coefficients perfectly match the known results of \cite{Baier:2007ix},\cite{Bhattacharyya:2008jc} that we have presented already in \eqref{BoundaryFluid} and \eqref{BoundaryConstants}. To make comparison clear we have to use the relation between $T_\textrm{b}=T_\textrm{Hawking}$ and $r_\textrm{H}$ shown in 
\eqref{Hawking}. We also have to reintroduce $\kappa_\textrm{AdS}$, because we have rescaled $t^\mu_{\phantom{\mu}\nu}$ as in \eqref{t-rescaling}. The results are presented in Table \ref{ComparisonTable}.

\subsubsection{Scalar transport coefficients}

To obtain the equations for the RG flow of the second order scalar transport coefficients, we should first solve the ``scalar mixing problem", namely to find the scalar analogue of the expansions \eqref{Tensor_n-prime} and \eqref{Vector_n-prime}. Unlike in the tensor expansion, though, our task now is a bit simpler, since we don't have to keep in mind possible contributions coming from the first order viscosity term. Our results for $\left( \mathcal{S}^{(2)}_i \right)^\prime$ are summarized in \eqref{S_i-prime}. Using these and retaining only second order terms in \eqref{EMTensorEoM-Scalar1}, we obtain the following RG flow equations for the second order scalar transport coefficients:
\begin{eqnarray}
\label{delta-ODEs}
\delta_1^\prime &+& \left( -\dfrac{d-1}{r} (c_s^2+1) + 
		r^{d-1} \left( - P + \dfrac{1}{2}\left( c_s^2 - \dfrac{1}{d-1}\right)  \epsilon \right)  \right) \delta_1 = 
\\ 
&&  \qquad \qquad
= \dfrac{r}{2(d-2)} \left( \left( c_s^2 + \dfrac{2d-5}{d-1} \right) P +
   				 \left( c_s^2 + \dfrac{d-3}{(d-1)^2}\right) \epsilon \right)
\nonumber \\   				  
\delta_2^\prime &+& \left( -\dfrac{d-1}{r} (c_s^2+1) + 
		\dfrac{r^{d-1}}{2} \left( - P + \left( c_s^2 + \dfrac{d-2}{d-1}\right)  \epsilon \right)  \right) \delta_2 = 
\nonumber \\ 
&&  \qquad
= \dfrac{r}{(d-1)(d-2)} \Big( \big( c_s^2 (d-1) + (d-4) \big) P +
   				 \big( c_s^2 + (d-2) \big) \epsilon \Big) +
   		r^{d-1}(P+\epsilon) \delta_1			 
\nonumber 
\end{eqnarray}
\begin{eqnarray}
\delta_3^\prime &+& \left( -\dfrac{d-1}{r} (c_s^2+1) + 
		r^{d-1} \left( - P + \dfrac{1}{2} \left( c_s^2 - 1 \right)  \epsilon \right)  \right) \delta_3 = 0
\nonumber \\
\delta_4^\prime &+& \left( -\dfrac{d-1}{r} (c_s^2+1) + 
		\dfrac{r^{d-1}}{2} \left( - 3 P + \left( c_s^2 - \dfrac{2}{d-1} \right)  \epsilon \right)  \right) \delta_4 = 
\nonumber \\		
&&  \qquad \qquad \qquad \qquad
= \dfrac{d-2}{2(d-1)} r^{d-1} (P+\epsilon) \left( \delta_2 + \dfrac{2 r^{2-d}}{d-2} \left(c_s^2 - \dfrac{{1}}{d-1}\right)\right)		
\nonumber \\
\delta_5^\prime &+& \left( -\dfrac{d-1}{r} (c_s^2+1) + 
		r^{d-1} \left( - 2 P + \dfrac{1}{2} \left( c_s^2 - \dfrac{2d-1}{d-1} \right)  \epsilon \right)  \right) \delta_5 = 
\nonumber \\		
&&  \quad 
= r^{d-1} (P+\epsilon) \left( \dfrac{d-2}{2(d-1)} (1+c_s^2) \left( \delta_2 + \dfrac{2 r^{2-d}}{d-2} \left(c_s^2 - \dfrac{{1}}{d-1}\right)\right)	+ \left( \dfrac{c_s^2}{2} + \dfrac{1}{d-1} \right) \delta_4 \right)	
\nonumber \\
\delta_6^\prime &+& \left( -\dfrac{d-1}{r} (c_s^2+1) + 
		\dfrac{r^{d-1}}{2} \left( c_s^2 + \dfrac{2d-3}{d-1} \right)  \epsilon   \right) \delta_6 = \dfrac{r^{d-1}}{2} (P+\epsilon) \left( \delta_2 + \dfrac{2 r^{2-d}}{d-2} \left(c_s^2 - \dfrac{{1}}{d-1} \right) \right)
\nonumber \\		
\delta_7^\prime &+& \left( -\dfrac{d-1}{r} (c_s^2+1) + 
		r^{d-1} \left( - P + \dfrac{1}{2} \left( c_s^2 - \dfrac{2d-1}{d-1} \right)  \epsilon   \right) \right) \delta_7 =
\nonumber \\		
&&  \qquad \qquad \qquad \qquad \qquad \qquad \qquad
		 =- \dfrac{3}{2} r^{d-1} (P+\epsilon) \left( \delta_2 + \dfrac{2 r^{2-d}}{d-2} \left(c_s^2 - \dfrac{{1}}{d-1} \right) \right)	
\nonumber				  				  
\end{eqnarray}
Similar to the tensor transport coefficients we will refer to $\delta$'s as ``regular" and ``singular" if they satisfy the incompressible Navier-Stokes criteria of Tables \ref{delta34567-div} and \ref{gamma12delta34-div}.

Before analysing the full non-homogeneous equations let us solve first the corresponding homogeneous ODEs. Those are obtained by dropping the right hand side of the equations. The calculation is straightforward even for $d \neq 4$. We summarize the results in the third column of Table \ref{HomogeneousDeltas}. We can learn from this table that for $\delta_1$, $\delta_3$, $\delta_4$ and  $\delta_7$ we have to eliminate the homogeneous solution in order to follow the rules of Tables \ref{gamma12delta34-div} and \ref{delta34567-div}. This will not yet be sufficient to guarantee the correct behaviour of these functions as we have also consider the sources of the related equations. To provide a complete answer we will have to address the near horizon behaviour of $\delta_2$. The only exception here is $\delta_3$ which has no source and so:
\begin{equation}
\delta_3 = 0
\end{equation}
is the only possible solution. As for the remaining two functions,  $\delta_5$ and $\delta_6$, we cannot fully fix these functions using the criteria of the previous section, because they do not contribute to the sources of the other ODEs and the solutions of their homogeneous equations are  not ``singular" near the horizon.

We are now in a position to study the $\delta_1$. Eliminating the $(r_\textrm{H}-r)^{-3}$ mode we may fix this function completely. For $d=4$\footnote{One can solve this equation straightforwardly for a general $d$, but exactly as for the tensor transport coefficients $\gamma$'s the final expression will look very lengthy.} the result is:
\begin{equation}
\delta_1(r) = - \dfrac{4}{9} r^2 
				\left[ \dfrac{3 r_\textrm{H}^2 + r^2}{(r_\textrm{H}^2 + r^2)^3} + 
					12 \dfrac{r_\textrm{H}^6 r^2 }{\left( r_\textrm{H}^4 - r^4 \right)^3} \left( \dfrac{\pi}{4} 
										- \arctan \left( \dfrac{r^2}{r_\textrm{H}^2}\right)\right)
				\right] \, .
\end{equation}
The solution goes like $(r_\textrm{H}-r)^{-2}$ near $r=r_\textrm{H}$ and we respect the bound of Table \ref{gamma12delta34-div}.

Next we consider  $\delta_2$. At the first glance, it looks that we cannot fix the integration constant in its ODE solution as the homogeneous mode goes like $(r_\textrm{H}-r)^{-2}$ and therefore, according to Table \ref{gamma12delta34-div}, there is no constrain on it. Notice, though, that $\delta_2$ appears as a source in most of the equations in \eqref{delta-ODEs}. Moreover, its contribution always comes in the following form:
\begin{equation}
\label{delta-source}
 (P+\epsilon) \left( \delta_2 + \dfrac{2 r^{2-d}}{d-2} \left(c_s^2 - \dfrac{{1}}{d-1} \right) \right)	\, .
\end{equation}
If we want to satisfy the bounds of Table \ref{delta34567-div}, this expression has to go as $(r_\textrm{H}-r)^{-2}$ near $r=r_\textrm{H}$. \footnote{For example, the left hand side of the $\delta_7$ equation looks like:
\begin{equation}
\nonumber
\delta_7^\prime - \dfrac{3}{r_\textrm{H}-r} \delta_7 \, ,
\end{equation}
near $r=r_\textrm{H}$. Therefore, the right hand side should go at most as $\left( r_\textrm{H}-r\right)^{-2}$.}
Recalling that $c_s^2 = 2 r_\textrm{H}^2/ (d(d-1)) \cdot (r_\textrm{H}-r)^{-2} + \ldots$, we see that the $\delta_2$ integration constant has to be fixed by requiring that the leading $(r_\textrm{H}-r)^{-2}$ term in the expansion of  $\delta_2$ cancels the same order term coming from $c_s^2$. This is the only way to assure that \eqref{delta-source} goes like $(r_\textrm{H}-r)^{-1}$, which subsequently guarantees that all the $\delta_i$'s for $i=4, \ldots, 7$ are ``regular" at the horizon, namely that they behave there like $(r_\textrm{H}-r)^{-1}$ in conjunction with Table \ref{delta34567-div}. For $d=4$ we find:
\begin{eqnarray}
\delta_2(r) &=& - \dfrac{4}{9} \dfrac{r^2}{\left( r_\textrm{H}^4 - r^4 \right)^3 \left( r_\textrm{H}^4 + r^4 \right)} 
				\Bigg[ \left( r_\textrm{H}^2 - r^2 \right)^2 \left( 6 r_\textrm{H}^{8} - 22 r_\textrm{H}^6 r^2 - 11 r_\textrm{H}^4 r^4 + 2 r_\textrm{H}^2 r^6 + r^8 \right) + 
\nonumber \\				
&& \qquad \qquad \qquad \qquad \qquad \qquad
			+ 24 r_\textrm{H}^6 r^2 \left( r_\textrm{H}^4 + r^4 \right)  \left( \dfrac{\pi}{4} 
										- \arctan \left( \dfrac{r^2}{r_\textrm{H}^2}\right)\right) \, .
				\Bigg]
\end{eqnarray}

To conclude, the differential equations for $\delta_1$ and $\delta_3$ can be solved unambiguously by eliminating the ``singular" $(r_\textrm{H}-r)^{-3}$ modes. The horizon behaviour of $\delta_4$, $\delta_5$, $\delta_6$, $\delta_7$ determines $\delta_2$ completely. Next, the ``regularity" fixes uniquely 
$\delta_4$ and $\delta_7$, since their homogeneous modes violate the bounds of the previous section. As for the remaining two functions, $\delta_5$ and $\delta_6$, we cannot eliminate the homogeneous modes as they respect the criteria of Table \ref{delta34567-div}. In order to fix the integration constants of these functions one has to go to the next order in the derivative expansion, exactly like in the $\gamma_6$ case discussed above. Importantly, contrary to the tensor transport coefficients, here we cannot fix the integration constants by imposing conformality on the boundary, because all the $\delta$'s vanish there independently of the horizon behaviour. This, of course, is of no surprise since the boundary conformal fluid has no (first and higher orders) scalars in its energy momentum tensor. To see that all the $\delta$'s do indeed vanish at $r=0$, notice that for small $r$ one finds from \eqref{delta-ODEs} that $\delta_i^\prime - d \cdot \delta_i = \mathcal{O}\left( r^{d+1} \right)$ for $i=3, \ldots, 7$ and so $\delta_i \sim r^d$ near $r=0$.\footnote{$\delta_1(r)$ and $\delta_2(r)$ vanish instead as $r^2$.}

We summarize the results of this subsection in Table \ref{HomogeneousDeltas}. We will not report the explicit solutions for $\delta_4(r)$, $\delta_5(r)$,  $\delta_6(r)$ and  $\delta_7(r)$, since the final expressions are very lengthy, while the prime goal of our discussion was only to demonstrate that one can fix the integration constants in all $\delta$'s, except $\delta_5$ and $\delta_6$, already at the second order.

\begin{table}[t]
\begin{center}
\begin{tabular}{| c | c | c | c | c | c |}
    \hline
 	The  transport & The   & The homogeneous & The full solution         & Boundary  & Solution fixed
    \\
 	coefficient    & scalar             & solution        & behaviour & value   & uniquely
    \\
    \hline \hline
    $ \delta_1(r)$ & $R$   & $ r^d \dfrac{(r_\textrm{H}^d+r^d)^{-1+4/d}}{(r_\textrm{H}^d-r^d)^3}$ 
    							    & $\mathcal{O}((r_\textrm{H}-r)^{-2})$ & \phantom{\Bigg( } 0 \phantom{\Bigg)} & Yes \\
    \hline							     
    $\delta_2(r)$ &  $u_\mu R^{\mu}_{\,\, \nu} u^\nu$  & $r^d \dfrac{(r_\textrm{H}^d+r^d)^{-2+4/d}}{(r_\textrm{H}^d-r^d)^2}$ 
    								& $\mathcal{O}((r_\textrm{H}-r)^{-1})$ & \phantom{\Bigg( } 0 \phantom{\Bigg)} & Yes \\
    \hline								 
    $\delta_3(r)$ &  $\left( \nabla \cdot u\right)^2$  & $r^d \dfrac{(r_\textrm{H}^d+r^d)}{(r_\textrm{H}^d-r^d)^3}$
        							& $0$ & \phantom{\Bigg( } 0 \phantom{\Bigg)} & Yes \\ 
    \hline								 
    $\delta_4(r)$ &  ${\nabla_\bot}^\mu {\nabla_\bot}_\mu \ln s$  & $r^d \dfrac{(r_\textrm{H}^d+r^d)^{-2+8/d}}{(r_\textrm{H}^d-r^d)^4}$ 
    								& $\mathcal{O}((r_\textrm{H}-r)^{-1})$ & \phantom{\Bigg( } 0 \phantom{\Bigg)} & Yes \\ 
    \hline								 
    $\delta_5(r)$ &  ${\nabla_\bot}^\mu \ln s {\nabla_\bot}_\mu \ln s $  & $r^d \dfrac{(r_\textrm{H}^d+r^d)^{1+4/d}}{(r_\textrm{H}^d-r^d)^5}$ 
    								& $\mathcal{O}((r_\textrm{H}-r)^{-5})$ & \phantom{\Bigg( } 0 \phantom{\Bigg)} & No \\ 
    \hline								 
    $\delta_6(r)$ &  $\sigma^\mu_{\,\, \nu} \sigma^{\nu}_{\,\,  \mu}$  & $r^d \dfrac{(r_\textrm{H}^d+r^d)^{-3+4/d}}{(r_\textrm{H}^d-r^d)}$ 
        							& $\mathcal{O}((r_\textrm{H}-r)^{-1})$ & \phantom{\Bigg( } 0 \phantom{\Bigg)} & No \\ 
    \hline								 
    $\delta_7(r)$ &  $\omega^\mu_{\,\, \nu} \omega^{\nu}_{\,\, \mu}$  & $r^d \dfrac{(r_\textrm{H}^d+r^d)^{3}}{(r_\textrm{H}^d-r^d)^3}$   
        							& $\mathcal{O}((r_\textrm{H}-r)^{-1})$ & \phantom{\Bigg( } 0 \phantom{\Bigg)} & Yes \\ 
    \hline
	\end{tabular}	
\end{center}
\caption{Here we sum up the results for the second order scalar transport coefficients. The homogeneous solutions of $\delta_1$, $\delta_3$, $\delta_4$ and $\delta_7$ do not satisfy the bound of Tables \protect\ref{delta34567-div} and \protect\ref{gamma12delta34-div}. We have to set to zero these modes in order to get incompressible non-relativistic Navier-Stokes fluid on the horizon. This uniquely determines these transport coefficients. Furthermore, to guarantee that the final solutions for  $\delta_4$ and $\delta_7$ still satisfy the bound one has to adjust properly the integration constant in $\delta_2(r)$. The same form of $\delta_2(r)$ leads to  $\delta_5$ and $\delta_6$ that also satisfy the bound of Table \protect\ref{delta34567-div}, but we cannot determine these functions unambiguously. This might be done at higher orders. Finally, all $\delta$'s automatically vanish on the boundary in accordance with the boundary fluid conformal symmetry.}
\label{HomogeneousDeltas}
\end{table}

\section{Discussion}

In the context of fluid/gravity correspondence, we have shown here that we can construct a holographic RG flow such that spacetime emerges from the flow of physical data, namely transport coefficients and auxiliary variables that parametrize the flow of the hydrodynamic variables. The RG flow is genuinely a system of first order differential equations, meaning we do not require to impose any boundary condition even implicitly at the cut-off hypersurface, which represents the scale where all physical variables are evaluated. The RG flow is uniquely determined by requiring that the fluid on the infra-red holographic screen, which coincides with the late-time horizon, follows incompressible Navier-Stokes equations with precise forcing terms. The bulk metric, which can be reconstructed from the explicit solution of the RG flow, should have a regular future horizon, order by order in the derivative expansion.

The main lesson we can learn from our results is that the fluid/gravity limit probably gives a unique definition of holographic RG flow which corresponds to an explicit emergence of space-time. Interpreting this holographic RG flow in terms of field theory is beyond the scope of the present work, however the RG flow, from the point of view of gravity, is defined uniquely up to trivial scale reparametrizations, \emph{i.e.} redefinition of the radial coordinate as a function of itself. We have seen explicitly that we need to choose Fefferman-Graham foliation of spacetime to be able to construct the RG flow without knowing the bulk metric explicitly, indeed our arguments presented here can be readily generalized. We have also argued that the form of the renormalized energy-momentum tensor which do not require any implicit boundary condition at the cut-off hypersurface, is fixed up to numerical constants in the counter-terms. We will soon say more about this. It can also be expected that quite generally the physical matter on the infra-red screen, where the Fefferman-Graham foliation terminates, should follow forced incompressible Navier-Stokes' equations, and this should determine the RG flow uniquely. 

We conjecture that holographic RG in the fluid/gravity limit should be sufficient to determine all numerical constants in the counter-terms which renormalise the hypersurface energy-momentum tensor. Clearly, if this is true, the fluid/gravity limit indeed determines holographic RG completely, because the renormalization by our arguments should be valid universally, \emph{i.e.} for all states in the universal sector of the conformal field theory holographically described by Einstein's equations.

Our conjecture is based on the simple observation that adjusting properly only two integration constants (coming from $\eta$ and $\gamma_2$) we were able to set to zero \emph{four} divergent terms in the source of the $\gamma_3$ differential equation (see the discussion following \eqref{gamma3Condition}). Similarly, the correct choice of $\delta_2$ insured that the other $\delta$'s satisfied the bound imposed by the non-relativistic incompressible Navier-Stokes limit as given in Table \ref{delta34567-div}. With a different numerical coefficient of the second order counter-term (\emph{i.e.} the Einstein tensor) these cancellations would not have happened. This coefficient was also necessary to cancel ultraviolet divergence, hence at the second order we could not have chosen the numerical coefficient in any other way. At higher orders in derivatives, however, most of the counter-term coefficients will not be fixed by the UV finiteness (see Section \ref{Section-Counterterms}). We believe that in this situation the horizon limit itself should be sufficient to determine these coefficients.

We would like to test our conjecture by studying holographic RG flow of the perfect fluid energy-momentum tensor which describes a wide class of stationary black holes \cite{Bhattacharyya:2007vs,Caldarelli:2012cm}. It turns out that the boundary fluid which describes these exact black hole solutions has a non-trivial vorticity, therefore there will be corrections to infinite orders in the derivative expansion. However, for these special solutions it is possible to re-sum these infinite corrections. We would like to do a RG flow analysis of these solutions using the methodology of this paper. This would allow us to explicitly see if (i) we indeed have a perfect fluid at the horizon in the scaling limit even after considering the corrections to all orders in derivatives, \footnote{note that in absence of dissipation the shear is zero, therefore Navier-Stokes equations become just Euler equations} and (ii) whether all the counterterm coefficients also get determined from restrictions on the near-horizon forms of the transport coefficients imposed by the horizon limit.

In another line of development, we should extend our construction of holographic RG flow in hydrodynamic limit by including bulk vector fields, scalars and fermions. Holographic duals of charged superfluid hydrodynamics can be reproduces by coupling gravity to charged scalars and gauge fields \cite{Banerjee:2008th,Bhattacharya:2011eea}. Similarly holographic duals of supersymmetric versions of hydrodynamics have also been considered in the literature by coupling fermions to gravity \cite{Policastro:2008cx,Hoyos:2012dh,Erdmenger:2013thg}. We should consider the holographic RG flow of superfluid charged hydrodynamics in such cases, including the role of goldstinos. We should thus investigate whether the fluid/gravity limit uniquely determines the construction of holographic RG flow for general classical theories of gravity.

Furthermore, we can also investigate the holographic RG flow along similar lines away from the hydrodynamic limit. We can expect that holographic RG flow will lead us to new understanding of quasi-normal modes and their non-linear dynamics. In order to do this, we need phenomenological parametrization of the energy-momentum tensor away from the hydrodynamic limit and also generalization of phenomenological equations which include dynamics of non-hydrodynamic variables. The holographic RG flow of the generalized phenomenological coefficients and all physical variables, including non-hydrodynamic ones, should lead to reconstruction of spacetime. It has been shown earlier that such phenomenological equations can indeed be constructed and in certain situations can even be derived from Einstein's gravity systematically \cite{Iyer:2009in, Iyer:2011qc, Iyer:2011ak}. However the calculation of non-hydrodynamic phenomenological coefficients and in fact the phenomenological equations themselves are very complicated \footnote{The complexity comes from the fact that we need to re-sum all local time derivatives in the amplitude expansion (where the expansion parameter is the ratio of non-hydrodynamic shear stress to the local pressure), because the non-hydrodynamic variables vary slowly in space but decay fast in time close to thermal equilibrium.}. It could be expected that as infra-red dynamics is simple, in fact purely hydrodynamic, the holographic RG flow should lead us to extrapolate a simpler structure of the generalized phenomenological equations. This will require us to generalize our Ansatz for the RG flow as given in Section 3 incorporating the non-hydrodynamic degrees of freedom.

It will also be interesting to consider the holographic RG flow of fluids in non-relativistic versions of holography. In the non-relativistic limit, sometimes there is an enlarged symmetry group \cite{Bagchi:2009my, Mukhopadhyay:2009db, Berkeley:2012kz} which can be realized covariantly \cite{Mukhopadhyay:2009db}, meaning that the equations of fluid mechanics transform covariantly under such transformations. It may be expected that holographic RG will give us new insights about role of symmetries in the dynamics particularly at long times, as these should be related to possible nature of the horizons in the related classical gravity theories.  In particular, as the holographic RG flow reconstructs the metric from RG flow of the fluid, if we can understand how the RG flow preserves the symmetries of the fluid, we would be able to understand the properties of the dual bulk spacetimes in the non-relativistic limit better. 

Finally, we should extend our methods to construct the holographic RG flow of correlation functions, \emph{i.e.} the holographic analogue of Callan-Symanzik equations of quantum field theory. Since our method applies in very general geometries, we should be able to construct the RG flow of correlation functions not only in equilibrium but also in non-equilibrium situations. Recently, holographic prescriptions have been developed for calculating non-equilibrium two-point Schwinger-Keldysh propagators \cite{Banerjee:2012uq, Mukhopadhyay:2012hv}. In the context of holographic RG flow, understanding of the right behaviour of the correlation functions at the horizon should be enough to reproduce the results for the boundary correlators obtained via these prescriptions. Furthermore, we should be able to independently verify our results for holographic RG flow of transport coefficients which could also be obtained from thermal multi-point correlation functions of the energy-momentum tensor.

The above realization may open a door to new understanding of fully developed turbulence via holographic RG flow.\footnote{For similar views see \cite{Eling:2010vr}.} In particular, the structure functions of turbulence are limits of multi-point non-equilibrium correlation functions of the velocity fields. The late-time behaviour of these structure functions should be governed by horizon dynamics in the holographic set-up. Furthermore, as the infra-red dynamics in the holographic RG flow is forced incompressible Navier-Stokes equations, we can expect that the structure functions in the holographic infra-red screen, should be the same as in realistic turbulence at very late times.  Thus it might be possible to derive the anomalous scaling of structure functions of fully developed turbulence using holographic RG flow. At a more modest level, holographic RG flow can give insights into the mechanism of energy cascade in turbulence.

\acknowledgments{We thank Vishnu Jejjala for collaboration in the initial stage of this work. SK is supported in part by the ANR grant 08-JCJC-0001-0 and the ERC Starting Grants 240210 - String-QCD-BH and 259133 -- ObservableString. The research of AM is supported by the LABEX P2IO, the ANR contract 05-BLAN-NT09-573739, the ERC Advanced Grant 226371 and the ITN programme PITN-GA-2009-237920. A part of this work was done when AM was visiting the CERN theory group for three months. AM also thanks the Neils Bohr Institute for giving an opportunity to give a seminar on this work prior to publication. We are grateful to G. Policastro, M. Petropoulos, Y. Oz and N. Obers for useful discussions.}

\appendix

\section{Various identities}

\label{VariousUsefulIdentities}

The following identities do not rely on the zero order equations of motion and are valid at any number of dimensions:
\begin{equation}
\label{NablaU}
\nabla^\mu u_\nu = \sigma^\mu_{\,\, \nu} +  \omega^\mu_{\,\, \nu} 
                  + \dfrac{\nabla \cdot u}{d-1} \Delta^\mu_{\,\, \nu}  - u^\mu D u_\nu \, .
\nonumber\\
\end{equation}
Contracting the free indices we get (here $\sigma^2=\sigma^\mu_{\,\, \nu} \sigma^\nu_{\,\, \mu}$ 
and $\omega^2=\omega^\mu_{\,\, \nu} \omega^\nu_{\,\, \mu}$):\footnote{See the remark following \eqref{ListOfSecondOrderScalars}.}
\begin{eqnarray}
\nabla^\mu u_\nu \nabla^\nu u_\mu &=& \sigma^2 + \omega^2 
                       + \dfrac{\left(\nabla \cdot u \right)^2}{d-1}                 
\\
\nabla^\mu u_\nu \nabla_\mu u^\nu &=& \sigma^2 - \omega^2 
                       + \dfrac{\left(\nabla \cdot u \right)^2}{d-1} - D u^\mu D u_\mu \, .
\nonumber
\end{eqnarray}
Using 
\begin{equation}
\label{NablaNablaR}
\left[ \nabla_\alpha, \nabla_\beta \right] u_\gamma = u^\delta R_{\delta \gamma \beta \alpha} 
\end{equation}
we find \eqref{D-Nabla-u}.

Let us also write down tensor equations that follow from (\ref{NablaU}) and (\ref{NablaNablaR}):
\begin{eqnarray}
\left< \nabla^\mu u_\tau \nabla^\tau u_\nu \right> &=& 
                             \left< \sigma^{\mu }_{\phantom{\nu} \tau} \sigma^{\tau}_{\phantom{\mu}  \nu }\right> +
                             \left< \omega^{\mu }_{\phantom{\nu} \tau} \omega^{\tau}_{\phantom{\mu}  \nu }\right> +
                             \dfrac{2}{d-1} \left( \nabla \cdot u \right) \sigma^{\mu}_{\phantom{\mu} \nu}
\nonumber\\
\left< \nabla^\mu u^\tau \nabla_\nu u_\tau \right> &=& 
                              \left< \sigma^{\mu }_{\phantom{\nu} \tau} \sigma^{\tau}_{\phantom{\mu}  \nu }\right>  -
                             \left< \omega^{\mu }_{\phantom{\nu} \tau} \omega^{\tau}_{\phantom{\mu}  \nu }\right>  -
                             2 \left< \sigma^{\mu }_{\phantom{\nu} \tau} \omega^{\tau}_{\phantom{\mu}  \nu }\right>  +
                             \dfrac{2}{d-1}  \left( \nabla \cdot u \right) \sigma^{\mu}_{\phantom{\mu} \nu}
\\                             
\left< \nabla^\tau u^\mu \nabla_\tau u_\nu \right> &=& 
                              \left< \sigma^{\mu }_{\phantom{\nu} \tau} \sigma^{\tau}_{\phantom{\mu}  \nu }\right>  -
                             \left< \omega^{\mu }_{\phantom{\nu} \tau} \omega^{\tau}_{\phantom{\mu}  \nu }\right>  +
                             2 \left< \sigma^{\mu }_{\phantom{\nu} \tau} \omega^{\tau}_{\phantom{\mu}  \nu }\right>  +
                             \dfrac{2}{d-1}  \left( \nabla \cdot u \right) \sigma^{\mu}_{\phantom{\mu} \nu} -
                             \left< D u^\mu D u_\nu \right>
\nonumber\\
D \left< \nabla^\mu u_\nu \right>  &=&
	   u^\alpha R^{\,\,\, \left< \mu \right. \,\,\, \beta}_{\alpha \,\,\, \left. \nu \right>} u_\beta -
                              \left< \sigma^{\mu }_{\phantom{\nu} \tau} \sigma^{\tau}_{\phantom{\mu}  \nu }\right>   -
                             \left< \omega^{\mu }_{\phantom{\nu} \tau} \omega^{\tau}_{\phantom{\mu}  \nu }\right>  -
                             \dfrac{2}{d-1}  \left( \nabla \cdot u \right) \sigma^{\mu}_{\phantom{\mu} \nu} +
                             \left<{\nabla_\bot}^\mu D u_\nu \right>  \, .
\nonumber                                  
\end{eqnarray}
The last equation can be used to find 
$\left< D \sigma^{\mu}_{\phantom{\mu} \nu} \right>$:\footnote{Notice that by definition $\sigma^{\mu}_{\phantom{\mu}\nu}= \left< \sigma^{\mu}_{\phantom{\mu}\nu} \right>$ but 
$D \sigma^{\mu}_{\phantom{\mu}\nu} \neq \left< D \sigma^{\mu}_{\phantom{\mu}\nu} \right>$.}
\begin{equation}
\left< D \sigma^{\mu}_{\phantom{\mu} \nu} \right> =
\left< u^\alpha R^{\phantom{\alpha} \mu \phantom{\nu} \beta}_{\alpha  \phantom{\mu} \nu \phantom{\beta}} u_\beta  \right> -
                             \left< \sigma^{\mu }_{\phantom{\nu} \tau} \sigma^{\tau}_{\phantom{\mu}  \nu }\right> -
                             \left< \omega^{\mu }_{\phantom{\nu} \tau} \omega^{\tau}_{\phantom{\mu}  \nu }\right> -
                             \dfrac{2}{d-1}  \left( \nabla \cdot u \right) \sigma^{\mu}_{\phantom{\mu} \nu} +
                             \left<{\nabla_\bot}^\mu  D u_\nu \right>  +
                             \left< D u^\mu D u_\nu \right>  \, .
\end{equation}

\section{Hydrodynamic scalar mixing}
\label{sm}

At the second order in space-time derivatives we have \emph{seven} scalars listed in \eqref{ListOfSecondOrderScalars}.
The $r$-derivatives of these scalars satisfy the following identities (dropping the third order terms): 
\begin{eqnarray}
\label{S_i-prime}
r^{1-d} \mathcal{S}_1^\prime &=& - \dfrac{\epsilon}{d-1} \mathcal{S}_1 - (P+\epsilon) \mathcal{S}_2
\nonumber\\
r^{1-d} \mathcal{S}_2^\prime &=& \dfrac{1}{2} \left( P + \dfrac{d-3}{d-1} \epsilon \right) \mathcal{S}_2
							- \dfrac{1}{2} \left( P+\epsilon \right) 
							     \left( \dfrac{d-2}{d-1} \left( {\mathcal{S}}_4 + \left( c_s^2 + 1\right) {\mathcal{S}}_5 \right) + 
							     \mathcal{S}_6 - 3 \mathcal{S}_7
							  \right) 
\nonumber\\
r^{1-d} \mathcal{S}_3^\prime &=& - \dfrac{\epsilon}{d-1} \mathcal{S}_3
\nonumber\\
r^{1-d} \mathcal{S}_4^\prime &=& -\dfrac{1}{2} \left( P + \dfrac{d+1}{d-1} \epsilon \right) \mathcal{S}_4 - 
                            \left( P+\epsilon \right) \left( \dfrac{c_s^2}{2} + \dfrac{1}{d-1} \right) \mathcal{S}_5
\nonumber\\
r^{1-d} \mathcal{S}_5^\prime &=& - \left( P + \dfrac{d}{d+1} \epsilon \right) \mathcal{S}_5
\nonumber\\
r^{1-d} \mathcal{S}_6^\prime &=& \left( P + \dfrac{d-2}{d-1} \epsilon \right) \mathcal{S}_6
\nonumber\\
r^{1-d} \mathcal{S}_7^\prime &=& -\left( P + \dfrac{d}{d-1} \epsilon \right) \mathcal{S}_7
\end{eqnarray}
To derive these equations we followed the steps outlined in the main text together with various identities from Appendix \ref{VariousUsefulIdentities}.

\section{Hydrodynamic tensor mixing}

We also have \emph{eight} tensors listed in \eqref{ListOfSecondOrderTensors}. Following the rules described above and the identities of Appendix \ref{VariousUsefulIdentities} we obtain the following  $r$-derivatives of the transport coefficients (for convenience we omit here the tensor indices in ${\mathcal{T}_i}^\mu_{\phantom{\mu}\nu}$):
\begin{eqnarray}
\label{T_i-prime}
\dfrac{r^{1-d}}{P+\epsilon} \mathcal{T}_1^\prime &=& - \dfrac{\epsilon}{P+\epsilon} \cdot \dfrac{\mathcal{T}_1}{d-1} + \mathcal{T}_2
						- \dfrac{1}{2} \left( c_s^2 + \dfrac{3}{d-1} \right) \mathcal{T}_3 
						- \dfrac{1}{2} \left( c_s^2 - \dfrac{1}{d-1} \right) \mathcal{T}_4
\nonumber \\						
&& \qquad				+ \dfrac{1}{2} \left( 2 c_s^4 - \dfrac{\partial c_s^2}{\partial \ln s} 
											- (c_s^2+1) \left( c_s^2 - \dfrac{1}{d-1} \right) \right) \mathcal{T}_5
						- \mathcal{T}_6 + \mathcal{T}_7 - 2 \mathcal{T}_8 
\nonumber\\
\dfrac{r^{1-d}}{P+\epsilon} \mathcal{T}_2^\prime &=&  \dfrac{P+\dfrac{d-2}{d-1}\epsilon}{P+\epsilon} \cdot \mathcal{T}_2
						- \dfrac{1}{2} \left( c_s^2 + \dfrac{d}{d-1} \right) \mathcal{T}_3 
						- \dfrac{1}{2} \left( c_s^2 - \dfrac{d-2}{d-1} \right) \mathcal{T}_4
\nonumber \\						
&& \qquad				+ \dfrac{1}{2} \left( 2 c_s^4 - \dfrac{\partial c_s^2}{\partial \ln s} 
											- (c_s^2+1) \left( c_s^2 - \dfrac{d-2}{d-1} \right) \right) \mathcal{T}_5
						- 2 \mathcal{T}_7 - 2 \mathcal{T}_8 
\nonumber\\
r^{1-d} \mathcal{T}_3^\prime &=& \dfrac{1}{2} \left( P + \dfrac{d-3}{d-1} \epsilon \right) \mathcal{T}_3
\nonumber\\
r^{1-d} \mathcal{T}_4^\prime &=& - \dfrac{1}{2} \left( P + \dfrac{d+1}{d-1} \epsilon \right) \mathcal{T}_4 - 
                             \dfrac{1}{2} \left( c_s^2 + \dfrac{d+1}{d-1} \right) \left( P+\epsilon \right) \mathcal{T}_5
\nonumber
\end{eqnarray}
\begin{eqnarray}
r^{1-d} \mathcal{T}_5^\prime &=& - \left( P + \dfrac{d-2}{d-1} \epsilon \right) \mathcal{T}_5
\nonumber\\
r^{1-d} \mathcal{T}_6^\prime &=& \left( P + \dfrac{d-2}{d-1} \epsilon \right) \mathcal{T}_6
\nonumber\\
r^{1-d} \mathcal{T}_7^\prime &=& - \left( P + \dfrac{d}{d-1} \epsilon \right) \mathcal{T}_7
\nonumber\\
r^{1-d} \mathcal{T}_8^\prime &=& - \dfrac{ \epsilon}{d-1} \cdot \mathcal{T}_8
\end{eqnarray}
Apart from the derivative of the second order tensor transport coefficients we have to consider the contribution of the shear viscosity term. We must calculate ${\sigma^\mu_{\,\, \nu}}^\prime$ including second order terms:
\begin{eqnarray}
\label{sigma-prime}
\dfrac{r^{1-d}}{\eta} {\sigma^\mu_{\phantom{\mu} \nu}}^\prime &=&   \dfrac{1}{2} \left( P + \dfrac{d-2}{d-1} \epsilon \right) \dfrac{\sigma^\mu_{\phantom{\mu} \nu}}{\eta}
- {\mathcal{T}_2}^\mu_{\phantom{\mu}\nu} + 
      \left( \dfrac{2}{d-1} + \dfrac{1}{\eta} \dfrac{\partial \eta}{\partial \ln s}\right) {\mathcal{T}_3}^\mu_{\phantom{\mu}\nu} +
      c_s^2 {\mathcal{T}_4}^\mu_{\phantom{\mu}\nu} + 
\\
&&  \qquad \qquad
	  +  \left( -c_s^4 + \dfrac{\partial c_s^2}{\partial \ln s}\right) {\mathcal{T}_5}^\mu_{\phantom{\mu}\nu} +
      {\mathcal{T}_6}^\mu_{\phantom{\mu}\nu} + {\mathcal{T}_7}^\mu_{\phantom{\mu}\nu} + 2 {\mathcal{T}_8}^\mu_{\phantom{\mu}\nu} \, . 
\nonumber      
\end{eqnarray}
The term $\partial \eta/\partial \ln s$ can be computed using:
\begin{equation}
\label{d-eta-OVER-d-ln-s}
\dfrac{\partial \eta}{\partial \ln s} = \dfrac{\partial \eta}{\partial r_\textrm{H}} \left( \dfrac{\partial \ln \, s}{\partial r_\textrm{H}} \right)^{-1} = \eta \, ,
\end{equation}
where the last identity follows from \eqref{eta(r)} and \eqref{s(r)-T(r)}. It is consistent with the fact that $\eta/s$ is constant along the RG flow as in \eqref{eta-equals-s}.

\bibliographystyle{utphys}
\bibliography{RGflowFinal}

\end{document}